\documentclass[a4paper,11pt]{article}

\usepackage{jheppub_mod} 
\usepackage[T1]{fontenc} 

\usepackage{amsmath,booktabs}
\usepackage{graphicx}
\usepackage{xspace}
\usepackage{color}
\pagecolor{white}
\usepackage[dvipsnames]{xcolor}
\usepackage{url}
\usepackage[utf8]{inputenc}
\usepackage{epstopdf}
\usepackage{hyperref}
\usepackage{multirow}
\usepackage{physics}
\usepackage{arydshln}
\usepackage[makeroom]{cancel}

\newcommand{\N}{\mathcal{N}}
\newcommand{\Order}{\mathcal{O}}
\newcommand{\beq}{\begin{equation}}
\newcommand{\eeq}{\end{equation}}
\newcommand{\beqa}{\begin{eqnarray}}
\newcommand{\eeqa}{\end{eqnarray}}
\newcommand{\diff}{\text{d}}

\newcommand{\MeV}{\,\text{MeV}}
\newcommand{\fm}{\,\text{fm}}
\newcommand{\Br}{\text{Br}}
\newcommand{\F}{\mathcal{F}}
\newcommand{\mN}{m_N}
\newcommand{\rr}{\mathbf r}
\newcommand{\jj}{\mathbf j}
\newcommand{\rrhat}{\hat{\mathbf r}}
\newcommand{\pp}{\mathbf p}
\newcommand{\kk}{\mathbf k}
\allowdisplaybreaks[1]

\DeclareMathOperator{\sgn}{sgn}

\title{Uncertainty quantification for $\boldsymbol{\mu\to e}$ conversion in nuclei: charge distributions}

\author[a]{Frederic No\"el}
\author[a]{and Martin Hoferichter}

\affiliation[a]{Albert Einstein Center for Fundamental Physics, Institute for Theoretical Physics, University of Bern, Sidlerstrasse 5, 3012 Bern, Switzerland}

\emailAdd{noel@itp.unibe.ch}
\emailAdd{hoferichter@itp.unibe.ch}

\abstract{Predicting the rate for $\mu\to e$ conversion in nuclei for a given set of effective operators mediating the violation of lepton flavor symmetry crucially depends on hadronic and nuclear matrix elements. In particular, the uncertainties inherent in this non-perturbative input limit the discriminating power that can be achieved among   operators by studying different target isotopes. In order to quantify the associated uncertainties, as a first step, we go back to nuclear charge densities and propagate the uncertainties from electron scattering data for a range of isotopes relevant for $\mu\to e$ conversion in nuclei, including $^{40,48}$Ca, $^{48,50}$Ti, and $^{27}$Al. We provide as central results  Fourier--Bessel expansions of the corresponding charge distributions  with complete covariance matrices, accounting for Coulomb-distortion effects in a self-consistent manner throughout the calculation. As an application, we evaluate the overlap integrals for $\mu\to e$ conversion mediated by dipole operators. In combination with modern ab-initio methods, our results will allow for the evaluation of general $\mu\to e$ conversion rates with quantified uncertainties.
}

\begin{document}

\maketitle
	
\section{Introduction}
\label{sec:intro}

Charged lepton flavor violation (LFV) constitutes an intriguing avenue to search for physics beyond the Standard Model (BSM). On the one hand, it is known from neutrino oscillations that leptons can change flavor, but on the other, the induced rates for charged leptons are tiny, suppressed by the small neutrino masses to $\Order(10^{-50})$, rendering any SM contribution unobservable for the foreseeable future. Any observation of charged LFV thus amounts to a clear BSM signal~\cite{Petcov:1976ff,Marciano:1977wx,Marciano:1977cj,Lee:1977qz,Lee:1977tib}. The most prominent process is arguably $\mu\to e\gamma$, for which a limit of $\Br[\mu\to e\gamma] < 4.2\times 10^{-13}$~\cite{MEG:2016leq} (here and below given at $90\%$ confidence level) has been obtained, while $\mu\to 3e$, $\Br[\mu\to 3e]<1.0\times 10^{-12}$~\cite{SINDRUM:1987nra}, gives complementary limits mainly for pure leptonic LFV operators. Both limits are set to improve over the next years at the MEG II~\cite{MEGII:2018kmf} and Mu3e~\cite{Mu3e:2020gyw} experiments, with potential further improvements thanks to the HIMB project~\cite{Aiba:2021bxe}. The third main channel to search for $\mu\to e$ transitions is $\mu\to e$ conversion in nuclei, 
\beq
 \Br[\mu\to e, \text{Ti}] < 6.1 \times 10^{-13} \quad \text{\cite{Wintz:1998rp}},\qquad 
 \Br[\mu\to e, \text{Au}] < 7 \times 10^{-13} \quad \text{\cite{SINDRUMII:2006dvw}},
 \label{muelimits}
\eeq
adding complementary sensitivity especially to LFV operators involving quark fields.\footnote{Conventionally, these limits are normalized to muon capture~\cite{Suzuki:1987jf}. Reference~\cite{Wintz:1998rp} supersedes the earlier limit $\Br[\mu\to e, \text{Ti}]<4.3\times 10^{-12}$~\cite{SINDRUMII:1993gxf} as the final result from the SINDRUM-II experiment.} Major advances in $\mu\to e$ conversion sensitivity are expected in the future from   
 Mu2e~\cite{Mu2e:2014fns} and COMET~\cite{COMET:2018auw}, projecting improvements by up to four orders of magnitude over Eq.~\eqref{muelimits}. 
 
Traditionally, $\mu\to e$ conversion in nuclei has been analyzed in terms of so-called overlap integrals~\cite{Kitano:2002mt,Borrel:2024ylg}, which amount to a convolution of the bound-state wave function of the initial-state muon, the final-state electron wave function, and nuclear densities corresponding to the respective underlying operator. Such a description captures the leading, coherently enhanced responses, e.g., for dipole, scalar, and vector operators, but in view of the expected experimental developments also subleading contributions have been studied recently~\cite{Cirigliano:2017azj,Davidson:2017nrp,Rule:2021oxe,Cirigliano:2022ekw,Hoferichter:2022mna,Haxton:2022piv,Noel:2024}. A key point in the analysis becomes to what extent different BSM scenarios can be distinguished, both from $\mu\to e$ conversion alone and in combination with limits for the purely leptonic processes, and this task can be systematically addressed in terms of effective operators for the underlying LFV interaction~\cite{Cirigliano:2009bz,Petrov:2013vka,Crivellin:2013hpa,Crivellin:2014cta,Hazard:2016fnc,Crivellin:2017rmk,Davidson:2018kud,Davidson:2020hkf,Davidson:2022nnl,Ardu:2023yyw}. However, in order to be able to disentangle subleading responses and to assess the discriminatory power in a robust manner, it is critical that both the uncertainties in the hadronic matrix elements, converting the effective operators from quarks and gluons to hadronic degrees of freedom, and in the nuclear responses, accounting for the overlap with the nuclear wave function, be quantified and controlled. In this context, one of the weakest points concerns the neutron responses, since, in contrast to charge form factors accessible in electron--nucleus scattering, nuclear weak form factors, giving direct, model-independent access to the neutron distribution, have only recently been measured in parity-violating electron scattering (PVES) for a few selected nuclei and momentum transfers~\cite{PREX:2021umo,Qweak:2021ijt,CREX:2022kgg}. One potential way around this limitation was pioneered in Ref.~\cite{Hagen:2015yea}, by combining modern ab-initio methods for $^{48}$Ca with experimental information on the charge radius to extract improved constraints on the neutron distribution, and a similar strategy was pursued in Ref.~\cite{Payne:2019wvy} for $^{40}$Ar. In Ref.~\cite{prep} we will show that such an approach also applies to $\mu\to e$ overlap integrals, by means of a correlation analysis with experimentally better determined charge distributions, which allows one to quantify and reduce the uncertainties associated with the neutron responses. 

However, such an analysis crucially rests on reliable charge distributions with quantified uncertainties. Unfortunately, the widely used Fourier--Bessel (FB) parameterizations from Ref.~\cite{DeVries:1987atn} do not provide uncertainties at all, leading to the unsatisfactory situation that uncertainties can at best be guessed from nuclear charge radii or model-dependent fits. For this reason, we went back to the original data for electron scattering off a few selected isotopes most relevant for $\mu\to e$ conversion, to propagate the uncertainties profiting from modern statistical techniques and computational resources that were not available when the data were originally taken. To make matters worse, the data themselves were often poorly documented, with minimal information on systematic uncertainties, only available in unpublished PhD theses, or even lost as private communication. We therefore spent considerable effort locating the original data wherever possible, collecting as complete a data base as seems to be possible for $^{40,48}$Ca~\cite{Croissiaux:1965,Bellicard:1967,Frosch:1968zz,Eisenstein:1969,Sinha:1973zz,Sick:1979bkt,Emrich:1983,Emrich:1983xb}, $^{48,50}$Ti~\cite{Engfer:1966,Theissen:1967,Frosch:1968zz,Heisenberg:1971rw,Romberg:1971hbw,Heisenberg:1972zza,Selig:1985,Selig:1988ket}, and $^{27}$Al~\cite{Stovall:1967yyp,Lombard:1967skb,Bentz:1970,Li:1970zza,Lapikas:1973njq,Li:1974vj,Singhal:1977wvn,Singhal:1982,Dolbilkin:1983,Ryan:1983zz}. The latter two targets are directly motivated by the current limit from Ref.~\cite{Wintz:1998rp} and future ones from Mu2e and COMET, respectively, while calcium is required as an intermediate step relative to which the titanium measurements were performed. Moreover, charge distributions for $^{27}$Al~\cite{Qweak:2021ijt} and $^{48}$Ca~\cite{CREX:2022kgg} are also of direct interest for the analysis of PVES, providing further motivation to focus on this set of isotopes. 

The computationally most intensive part of the fits to electron-scattering data originates from the consideration of Coulomb-distortion effects, especially, when repeating the fits many times to obtain reliable uncertainty estimates. The formalism to account for such Coulomb effects in terms of phase shifts obtained via solving the Dirac equation is, in principle, well known in the literature, see, e.g., Refs.~\cite{Yennie:1954zz,Tuan:1968,Uberall:1971,Dreher:1974pqw,Merle:1976,Heisenberg:1981mq,Donnelly:1984rg}, but a concise description including the subtleties critical for an efficient numerical implementation is not easily available.   
In the main part of this paper we will therefore review the formalism for electron scattering, see Sec.~\ref{sec:formalism}, and present our key results for the charge distributions, see Sec.~\ref{sec:charge}, supplemented by detailed appendices on the multipole decomposition (App.~\ref{app:PWBA}), phase-shift formalism (App.~\ref{app:Dirac}), fit details and strategy (Apps.~\ref{app:fit_details} and~\ref{app:fit_strat}), FB parameterizations (App.~\ref{app:parameterizations}), and data sets (App.~\ref{app:data}).  As an application, we calculate the overlap integrals for $\mu\to e$ dipole operators in Sec.~\ref{sec:dipole}, as they can be formulated directly in terms of charge distributions, see Eq.~\eqref{eq:D_1,D_2}, whose uncertainties we are now in the position to propagate. Summary and outlook to future applications, in combination with ab-initio techniques, are provided in Sec.~\ref{sec:summary}. Our results for the charge distributions, including the complete covariance matrices, are made available as a supplementary \texttt{python} notebook.

\section{Formalism for electron scattering}
\label{sec:formalism}

\subsection{Cross section and charge distribution for plane waves}

The differential cross section for the scattering of electrons off nuclei, 
\beq
e^-(k)+\N(p) \to e^-(k')+\N(p'),
\eeq
with momentum transfer $\bar q_\mu=k'_\mu-k_\mu=p_\mu-p'_\mu$,
is often written in the form
\beq
\dv{\sigma}{\Omega}=\bigg(\dv{\sigma}{\Omega}\bigg)_\text{Mott}\times\frac{E'_e}{E_e}\times \big|F(q,\theta)\big|^2,
\eeq
where the first factor denotes the Mott cross section~\cite{Mott:1929,Mott:1932}
\beq
\label{Mott_cross_section}
\bigg(\dv{\sigma}{\Omega}\bigg)_\text{Mott}=\frac{\alpha^2}{4E^2_e}\frac{\cos^2\frac{\theta}{2}}{\sin^4\frac{\theta}{2}},
\eeq
the second the recoil correction
\beq
\frac{E'_e}{E_e}=\bigg(1+\frac{2E_e}{M}\sin^2\frac{\theta}{2}\bigg)^{-1},
\eeq
and the third the nuclear form factor $F(q,\theta)$. The scattering angle $\theta$ and incoming/outgoing electron energy $E_e$/$E'_e$ are defined in the laboratory frame, and $q=|\bar{\mathbf q}|$ denotes the modulus of the three-momentum transfer, which, up to relativistic corrections, is related by 
\beq 
\label{def_q}
q^2=-{\bar q}^2=4E_e E'_e\sin^2\frac{\theta}{2}>0,
\eeq
so that $q=2E_e\sin\tfrac{\theta}{2}$ up to $1/M$ corrections.
$M$ denotes the mass of the target nucleus, and $\alpha=e^2/(4\pi)$ the fine-structure constant. Information on the charge distribution of the nucleus is contained in $F(q,\theta)$.  In the plane-wave Born approximation (PWBA), the form factor can be further decomposed into longitudinal, $F_L(q)$, and transverse, $F_T(q)$, components
\beq
\label{angular_separation}
\big|F(q,\theta)\big|^2=\big|F_L(q)|^2+\bigg(\frac{1}{2}+\tan^2\frac{\theta}{2}\bigg)\big|F_T(q)|^2,
\eeq
both of which can be further expanded into pieces with definite angular momentum $L$
\beq
\label{FL:FT_multipoles}
\big|F_L(q)\big|^2=\sum_{L\,\text{even}\, \leq 2J}\big|ZF_L^\text{ch}(q)\big|^2,\qquad 
\big|F_T(q)\big|^2=\sum_{L\,\text{odd}\, \leq 2J}\big|F_L^\text{mag}(q)\big|^2,
\eeq
where $J$ is the spin of the nucleus.\footnote{We keep the slightly unfortunate convention to denote the angular momentum components of the longitudinal form factor $F_L$ again by $L$.} For instance, for the nucleon only a single multipole contributes, related to the usual Sachs form factors~\cite{Ernst:1960zza}\footnote{The decomposition of the matrix element of the electromagnetic current $j^\mu_\text{em}$ into Dirac and Pauli form factors is $\langle N(p')|j^\mu_\text{em}|N(p)\rangle=\bar u(p')\big[F_1(t)\gamma^\mu-\frac{i}{2\mN}\sigma^{\mu\nu}\bar q_\nu F_2(t)\big]u(p)$, with $t=\bar q^2$ and nucleon mass $\mN$.} 
\beq
G_E(t)=F_1(t)-\frac{t}{4\mN^2}F_2(t),\qquad G_M(t)=F_1(t)+F_2(t),
\eeq
by
\beq
\label{free_nucleon}
F_L(q)=\bigg(1-\frac{q^2}{8\mN^2}\bigg)G_E(-q^2),\qquad F_T(q)=\frac{q}{\sqrt{2}\,\mN}G_M(-q^2).
\eeq
The 
charge distribution $\rho(r)=\rho_0(r)$ is now related to the Fourier transform of the $L=0$ component of $F_L(q)$
\beq
ZF_0^\text{ch}(q)=\int \diff^3r \, \rho(r) e^{-i{\mathbf q}\cdot{\mathbf r}} = 4\pi\int\diff r\, r^2\, j_0(q r) \rho(r), \label{eq:Fch<->rho0}
\eeq
with inversion
\beq
\rho(r)=\int\frac{\diff^3 q}{(2\pi)^3} ZF_0^\text{ch}(q)e^{i{\mathbf q}\cdot{\mathbf r}}=\frac{1}{2\pi^2}\int\diff q\, q^2\,j_0(qr)ZF_0^\text{ch}(q).
\eeq
In fact, generalizations of these relations apply to all angular momentum components, conventionally written as
\begin{align}
\label{rho_L_j_LL}
    ZF_L^\text{ch}(q)&= 4\pi\int\diff r\, r^2\, j_L(q r) \rho_L(r), & 
    \rho_L(r)&=\frac{1}{2\pi^2}\int\diff q\, q^2\,j_L(qr)ZF_L^\text{ch}(q),\notag\\
 F_L^\text{mag}(q)&= 4\pi\int\diff r\, r^2\, j_L(q r) j_{LL}(r), & 
    j_{LL}(r)&=\frac{1}{2\pi^2}\int\diff q\, q^2\,j_L(qr)F_L^\text{mag}(q),
\end{align}
where the inversions follow from the orthogonality properties of the spherical Bessel functions $j_L(x)$. The quantities $\rho_L(r)$ correspond to higher-order terms in an expansion of the charge distribution in spherical harmonics, and similarly the $j_{LL}(r)$ occur in the expansion of the current. Ultimately, for the $\mu\to e$ application we are mainly interested in the spherically symmetric charge distribution $\rho(r)$, but for some nuclei of interest, most notably $^{27}$Al with spin $J=5/2$, also the subleading form factors need to be considered. In the PWBA, longitudinal and transverse contributions can be separated by the angular dependence~\eqref{angular_separation}, with measurements close to $\theta=180^\circ$ particularly sensitive to the magnetic corrections. In practice, we will see that the impact of contributions beyond $L=0$ for the extraction of $\rho(r)$ is small compared to the uncertainties from the measured cross sections, as we will study in the case of $^{27}$Al using estimates for the $L>0$ multipoles from both shell-model and ab-initio calculations.  More details on the decomposition of the PWBA form factors into nuclear multipoles are provided in App.~\ref{app:PWBA}. 

\subsection{Coulomb distortion}

In practice, an analysis based on expressions using the PWBA is not sufficient for a realistic description of the measured cross sections. This is because the very presence of the Coulomb field of the nucleus distorts the electron incoming and outgoing plane waves, and the corresponding corrections, especially in the vicinity of form factor minima, are substantial. In order to extract the charge distribution from the data, one thus needs a systematic way to obtain a self-consistent description of the cross section. 
This may be achieved by employing the so-called phase-shift model, which deduces the scattering cross section from the phase shifts of the numerically determined electron wave functions, obtained by solving the Dirac equation for a charge potential generated by the Coulomb field of the nucleus. While this procedure, which is equivalent to the distorted-wave Born approximation (DWBA), is in principle well known in the literature, for the practical implementation techniques to accelerate the partial-wave convergence are required, and the Coulomb part of the wave function needs to be evaluated in a stable manner. Accordingly, evaluating the cross section many times for a self-consistent, model-independent fit of the charge distribution in terms of FB coefficients as well as a thorough uncertainty analysis was computationally very expensive at the time when most electron scattering data were taken. For spin-zero nuclei, propagating a FB parameterization of $\rho(r)$ in the phase-shift analysis is indeed the major part of the calculation. In this case, the cross section can be written as
\begin{equation}
\dv{\sigma}{\Omega}=\big|A_\text{s}(\theta)\big|^2+\big|A_\text{sf}(\theta)\big|^2 
\overset{m_e=0}{=} \Big(1 + \tan^2\frac{\theta}{2}\Big) \big|A_\text{s}(\theta)\big|^2, \label{eq:crosssection_phaseshift}
\end{equation}
with 
\begin{align}
    A_\text{s}(\theta) &= \frac{1}{2 i k_e} \sum_{\kappa>0} \kappa \qty[ e^{2i\delta_\kappa} P_\kappa(\cos\theta) + e^{2i\delta_{-\kappa}} P_{\kappa-1}(\cos\theta)], \notag \\
  A_\text{sf}(\theta) &= -\frac{\sin\theta}{2 i k_e} \sum_{\kappa>0} \qty[ e^{2i\delta_\kappa} P'_\kappa(\cos\theta) - e^{2i\delta_{-\kappa}} P'_{\kappa-1}(\cos\theta)],
    \label{eq:phase_shift_amplitude}
\end{align}
where $P_\kappa$ are the Legendre polynomials, $\delta_{\kappa}$ the phase shifts obtained from the solution of the Dirac equation (fulfilling $\delta_{\kappa}=\delta_{-\kappa}$ in the $m_e=0$ limit), and $k_e=|\kk|$ denotes the electron momentum.
Coulomb-distortion effects can also be evaluated when contributions with higher $L$ are present, we refer to App.~\ref{app:Dirac} for a review of the full DWBA in this case.

\section{Charge distributions}
\label{sec:charge}

\subsection{Fourier--Bessel parameterization and fit strategy}
\label{sec:FB_fit}

We parameterize the charge densities in a model-independent way in terms of a FB series~\cite{Dreher:1974pqw}. This series expansion takes the form
\begin{align}
    \rho (r) &= \begin{cases} \sum^N_{n=1} a_n j_0(q_n r), &  r\leq R\\ 0,  & r > R  \end{cases} \quad \qq{with} \quad q_n=\frac{\pi n}{R}, \label{eq:FB}
\end{align}
where $N$ denotes the number of terms in the series, the cutoff radius $R>0$ is related to the range of the charge distribution, and ${a_n \in \mathbb{R}}$ are the parameters. The quantities $q_n$ are chosen in such a way that 
${j_0(q_n R)=0}$ $\forall n$, ensuring that the charge density $\rho(r)$ is continuous at ${r=R}$.

The parameterization~\eqref{eq:FB} was established in Ref.~\cite{Dreher:1974pqw} as a model-independent method to describe charge densities, with a more direct relation to the available data than, e.g., for sums 
of Gaussians~\cite{Sick:1974suq}.
Already in Ref.~\cite{Dreher:1974pqw}, the main motivation to introduce the FB series was to
extract charge densities from elastic electron--nucleus scattering in a model-independent way, including also the extraction of meaningful uncertainty estimates. 
Unfortunately, the strategies to derive such measures of uncertainty 
were not employed or at least not documented in 
subsequent reviews such as Ref.~\cite{DeVries:1987atn}. Furthermore, due to computational limitations, at the time the strategy involved the use of a toy model to transfer the cross-section data to form-factor pseudodata, which could then be fit directly using the FB parameterization. In this toy model, the charge density was described as a sum of $\delta$ functions, under the assumption that for a sufficient density of $\delta$ functions, the result would become indistinguishable from the physical charge density, see, e.g., Ref.~\cite{Friedrich:1972iz}. Within this approach, the phase shifts can be calculated analytically, by solving the Dirac equation stepwise between each $\delta$ function, which improved the speed of the calculation considerably, however, at the expense of introducing new systematic errors related to the convergence of the fit. 
Since with modern computational resources and optimizations to the code for the phase-shift solution 
a direct fit using the FB parameterization becomes feasible, we fit directly to the cross-section data, avoiding the previous detour and the associated systematic uncertainties. 
 More details on our implementation of the FB parameterization and analytical expressions of derived quantities are provided in App.~\ref{app:FBparam}.\footnote{Existing public codes for the solution of the Dirac equation are typically optimized for different ranges in $(E_e,\theta)$ and/or for specific analytic choices of the potential~\cite{Salvat:2005,Salvat:2019}.}
 
In the fit, we account for the statistical and systematic uncertainties from the measurements, including their correlations, wherever possible. Moreover, we employ a change of variables in which the conservation of the total charge becomes manifest, see App.~\ref{subsec:norm}. Recoil corrections are included 
by changing from the laboratory system (Lab) to the center-of-mass system (CMS), then performing the numerical calculation of the phase shifts, and finally transforming back into the Lab system, see App.~\ref{app:recoil}. 

Finally, to ensure robust control of statistical and systematic uncertainties, we employ the following fit strategies to scan over different combinations of $N$ and $R$ parameters, while fitting the FB coefficients $a_n$ to the elastic scattering data of the considered nuclei, both with and without additional constraints from muonic atoms using Barrett moments, see App.~\ref{subsec:barrett} and Eq.~\eqref{eq:Barrett}. 
The range of these fits is then narrowed down to a selection with acceptable statistical quality and asymptotic behavior, to form the basis for our central solutions as well as statistical and systematic error bands. We divide the fitting strategy into the following steps:
\begin{enumerate} 
    \item Fit for a large grid of $N$ and $R$ pairs, beyond the expected range of validity of the FB expansion. \label{item:base}
    \item Use the statistical fit quality of all solutions in the grid to define a reasonable range of $N$ and $R$ pairs. \label{item:sen}
    \item Redo the fits in the deduced range including the constraints to prevent unphysical oscillations, see App.~\ref{subsec:oscillations}. \label{item:pen}
    \item Select fits that still display sufficient statistical quality and acceptable asymptotic behavior, see App.~\ref{subsec:asymp}. \label{item:sel}
    \item Select a central solution based on statistical fit quality, favoring lower values of $N$ and $R$. \label{item:central}
    \item Deduce upper and lower systematic uncertainty bands based on the remaining solutions, see App.~\ref{subsec:bands}. \label{item:band}
    \item Redo steps~\ref{item:pen}--\ref{item:band} including the constraint from the Barrett moments. \label{item:redo}
\end{enumerate}
Following this strategy, we end up with two solutions for each nucleus, enabling cross checks of the consistency of electron scattering data and muonic-atom constraints. In each case, the fit is set up in such a way that we first allow for a wide range of values of $N$ and $R$, to ensure that no admissible fits be overlooked. The  
maximal values are related to the largest momentum transfer for which measurements are available, essentially, because the charge form factor near the momentum transfer $q_n$ is determined by the coefficient $a_n$ in the FB series, see App.~\ref{app:FBparam}, leading to an initial guess of suitable ranges. The subsequent steps then ensure that an overparameterization of the data be avoided, by giving precedence to lower values of $N$ and $R$ for comparable fit quality. More details of the fit strategy, intermediate results, and the implementation of the various constraints are provided in Apps.~\ref{app:fit_details} and \ref{app:fit_strat}.
In the following sections, we present our key results for the charge densities of the different nuclei, with tables of the resulting parameter sets, including uncertainties and correlations, given in App.~\ref{app:parameterizations}. All results are also included in the supplementary \texttt{python} notebook. 

\begin{figure}[t]
    \centering
    \includegraphics[width=0.85\linewidth]{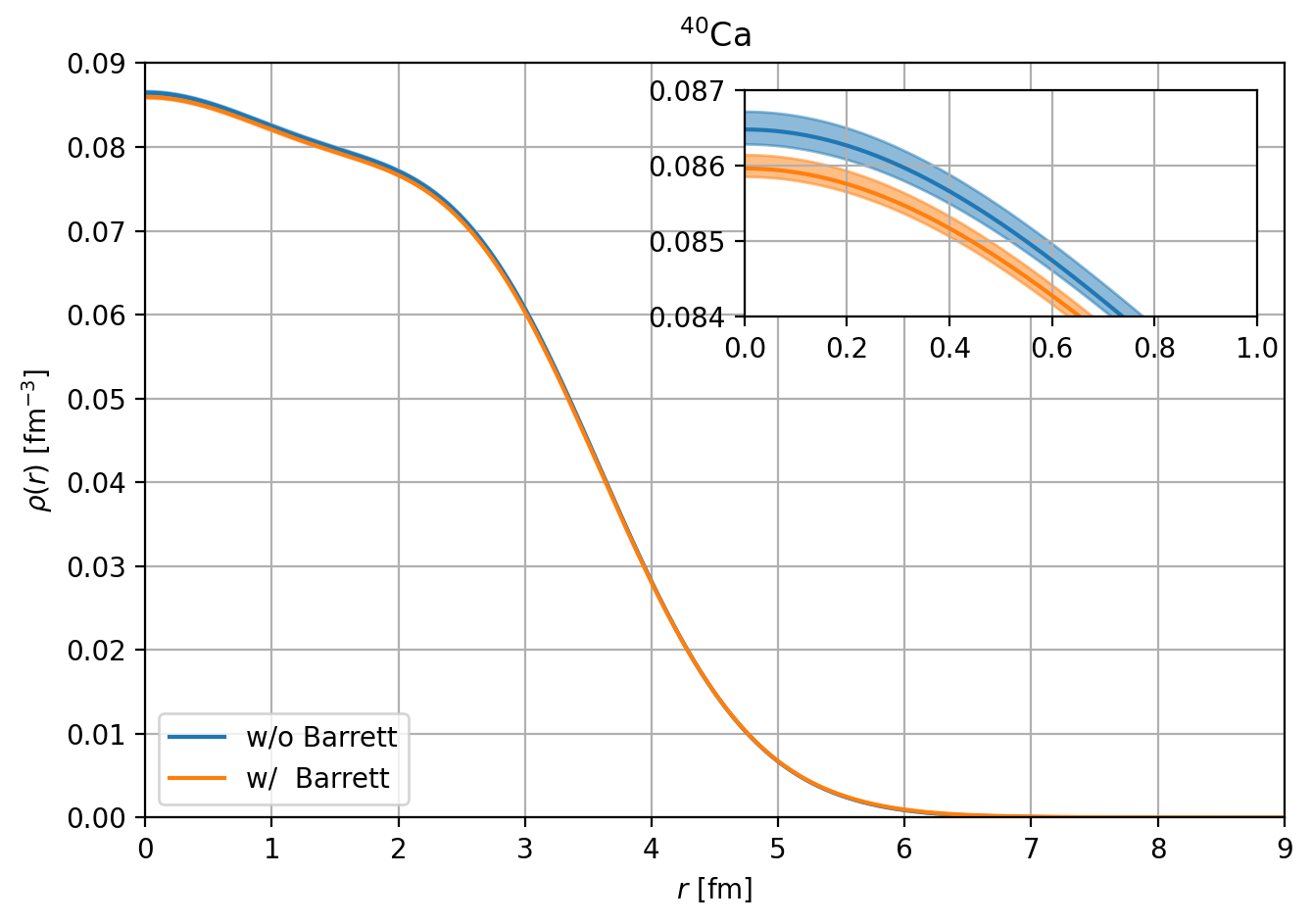}
    \includegraphics[width=0.85\linewidth]{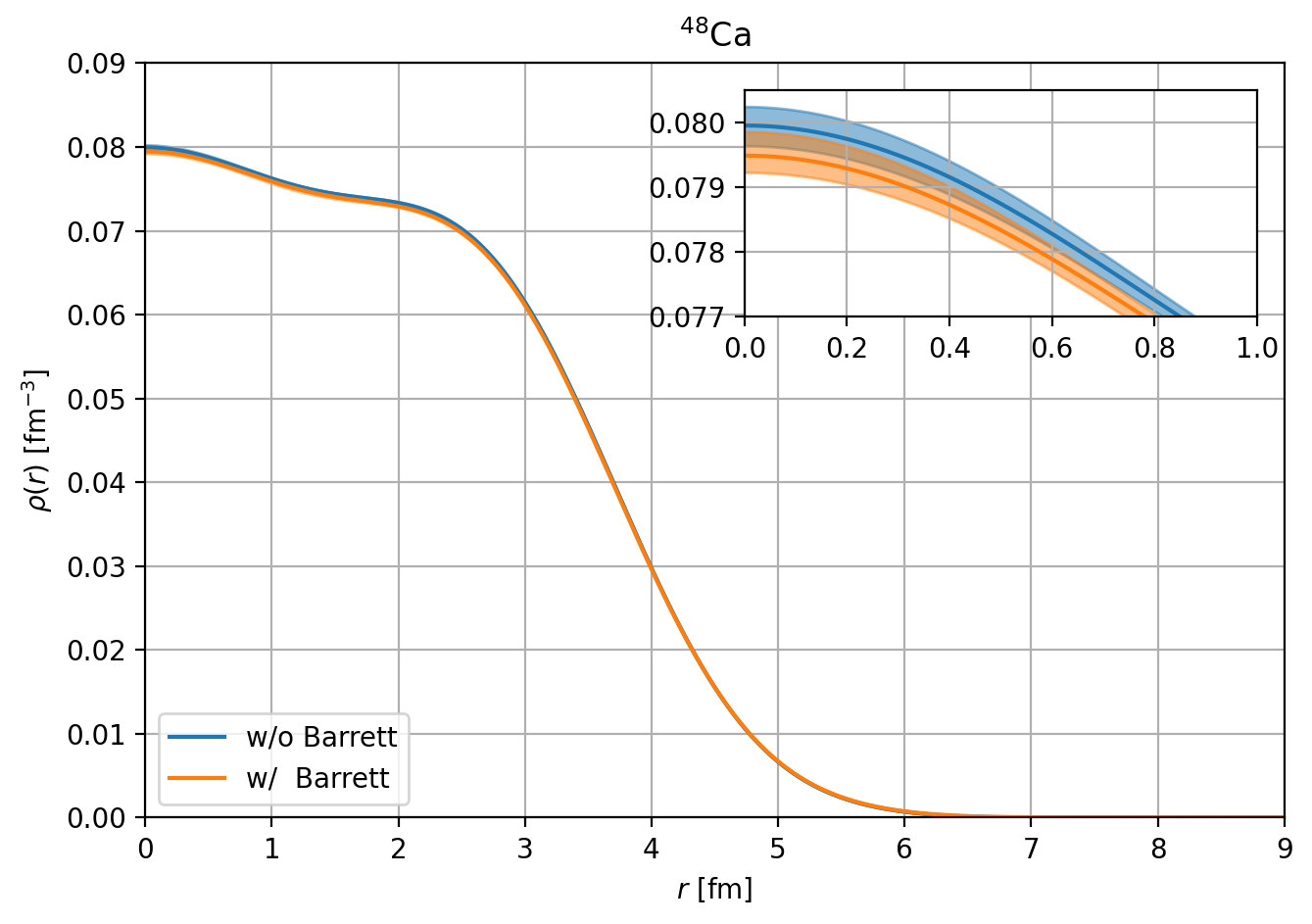}
    \caption{Charge densities of $^{40}$Ca and $^{48}$Ca. Two variants are shown, with (orange) or without (blue) the constraints from Barrett moments.}
    \label{fig:Ca_best}
\end{figure}

\subsection{Calcium}

We start our discussion with two calcium isotopes, $^{40,48}$Ca, for the following reasons. First, both nuclei have spin $J=0$, so that the simplest form of the phase-shift model as given in Eq.~\eqref{eq:crosssection_phaseshift} applies. 
Second, these isotopes are of interest either as reference point for titanium, as discussed in the subsequent section, or directly for phenomenological applications, e.g., in the context of PVES or neutrinoless double-$\beta$ decay. Third, the data situation is exceptionally good, at least compared to other isotopes of interest for $\mu\to e$ conversion.

We spent considerable effort going through the literature for electron scattering off calcium, including Refs.~\cite{Croissiaux:1965,Bellicard:1967,Frosch:1968zz,Eisenstein:1969,Sinha:1973zz,Sick:1979bkt,Emrich:1983,Emrich:1983xb}. Unfortunately, most data sets are poorly documented, in many cases making it impossible to even access the original data, let alone retrieving information on uncertainties and correlations. Moreover, later surveys such as Ref.~\cite{Angeli:1999} heavily criticized especially early data sets, discarding a fair number of them due to suspected systematic effects. By these standards, the documentation in the PhD thesis~\cite{Emrich:1983} is exceptionally good, including complete data tables and a discussion of statistical and systematic uncertainties. In addition, data from Ref.~\cite{Sick:1979bkt} are made explicit, while only provided as a figure in the original reference.  Taken together, these data sets then cover a wide range of momentum transfers, clearly superseding earlier measurements. We checked for possible tensions by comparing to the data from Ref.~\cite{Frosch:1968zz}, a point to which we will return in the context of the charge distributions for titanium, while for the calcium fits the data provided in Ref.~\cite{Emrich:1983} were obviously superior. Since this reference is also not easily accessible, we provide the data as used in our fits in App.~\ref{app:data}.

The final results of our fits are shown in Fig.~\ref{fig:Ca_best}, where we display the extracted charge distributions for $^{40}$Ca and $^{48}$Ca including uncertainty bands containing statistical and systematic components. For $^{40}$Ca one can see in the close-up that a tension between the electron-scattering-only fits and the variant including Barrett-moment constraints emerges for $r\to 0$, while for $^{48}$Ca a more significant tension arises at larger distances. We return to this point in the context of charge radii in Sec.~\ref{sec:charge} and provide parameterizations for both solutions in App.~\ref{app:parameterizations}.

\subsection{Titanium}

For titanium the data situation is considerably worse than for the calcium isotopes discussed in the preceding section, see Refs.~\cite{Engfer:1966,Theissen:1967,Frosch:1968zz,Heisenberg:1971rw,Romberg:1971hbw,Heisenberg:1972zza,Selig:1985,Selig:1988ket}. In particular, the only reference from which the original data for the most abundant isotope $^{48}$Ti could be retrieved is Ref.~\cite{Frosch:1968zz}, in which a measurement relative to $^{48}$Ca was performed. Unfortunately, Ref.~\cite{Frosch:1968zz} is among the works whose calcium results were criticized later due to inconsistencies with subsequent measurements, and we confirm the prevalence of such inconsistencies in comparison to the $^{40}$Ca data of Ref.~\cite{Emrich:1983}.
However, the $^{40}$Ca data in Ref.~\cite{Frosch:1968zz} required an absolute calibration, involving a number of systematic effects that are absent in a relative measurement. Indeed, comparing the measurement of $^{48}$Ca from Ref.~\cite{Frosch:1968zz} (taken relative to $^{40}$Ca) with Ref.~\cite{Emrich:1983}, the situation improves considerably, suggesting that the relative measurement of $^{48}$Ti should also be less affected by the systematics observed for the absolute $^{40}$Ca measurement. Accordingly, our $^{48}$Ti fits are based on the relative measurement from Ref.~\cite{Frosch:1968zz}, together with the $^{48}$Ca results from the previous section. 

A similar strategy was followed in Refs.~\cite{Selig:1985,Selig:1988ket}, in which their $^{48}$Ti data set is compared to the relative measurement of Ref.~\cite{Frosch:1968zz}, together with independent input for $^{48}$Ca. While we were successful in retrieving the PhD thesis~\cite{Selig:1985}, the original data are not contained therein, so  that, unfortunately, the results from this measurement appear to be lost, too, leaving Ref.~\cite{Frosch:1968zz} as the only remaining source to extract the charge distribution for $^{48}$Ti. Still, the analysis presented in Ref.~\cite{Selig:1985} finds consistency with Ref.~\cite{Frosch:1968zz}, increasing confidence that at least the relative measurement for $^{48}$Ti therein should be reliable.    

\begin{figure}[t!]
    \centering
    \includegraphics[width=0.85\linewidth]{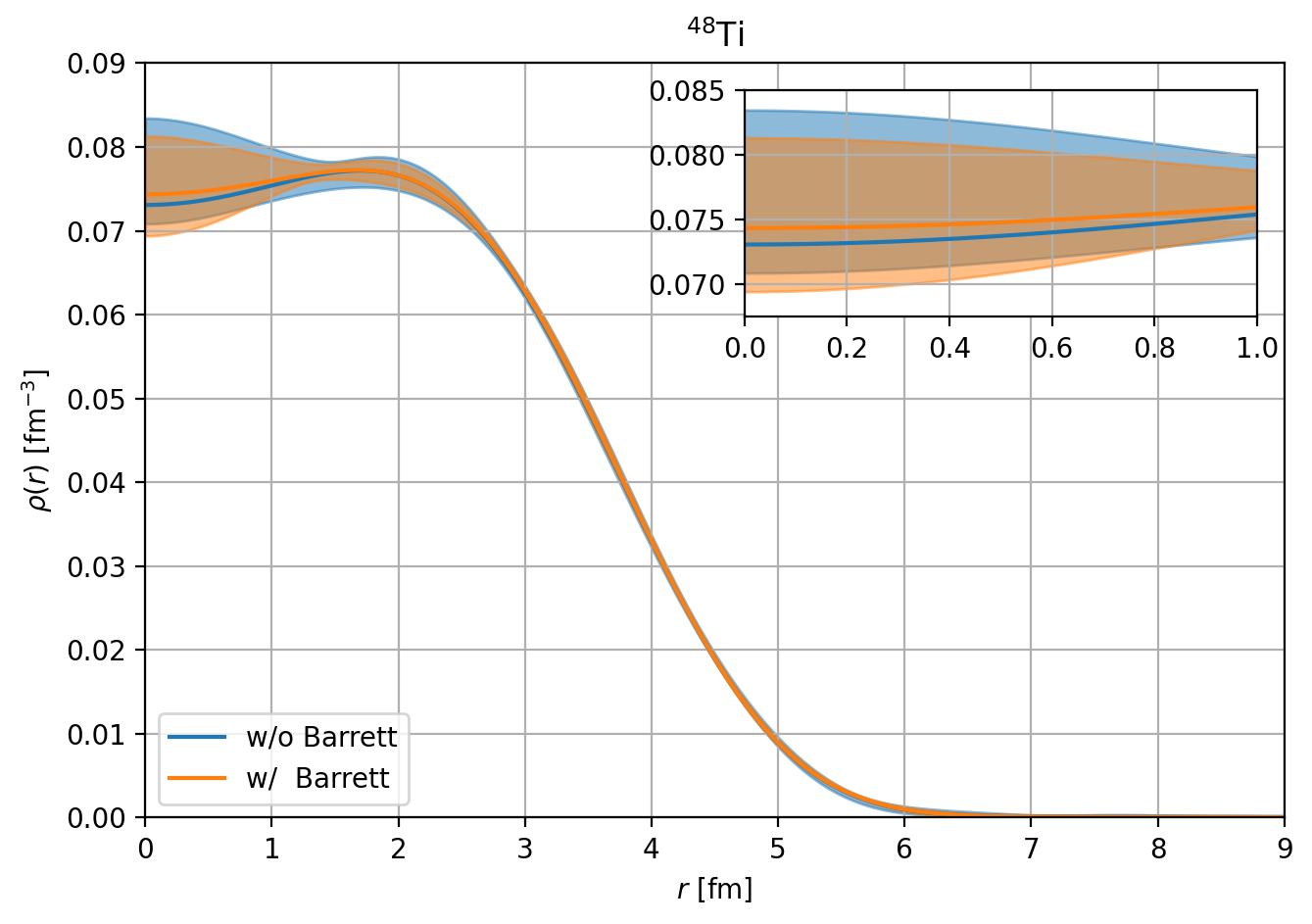}
    \includegraphics[width=0.85\linewidth]{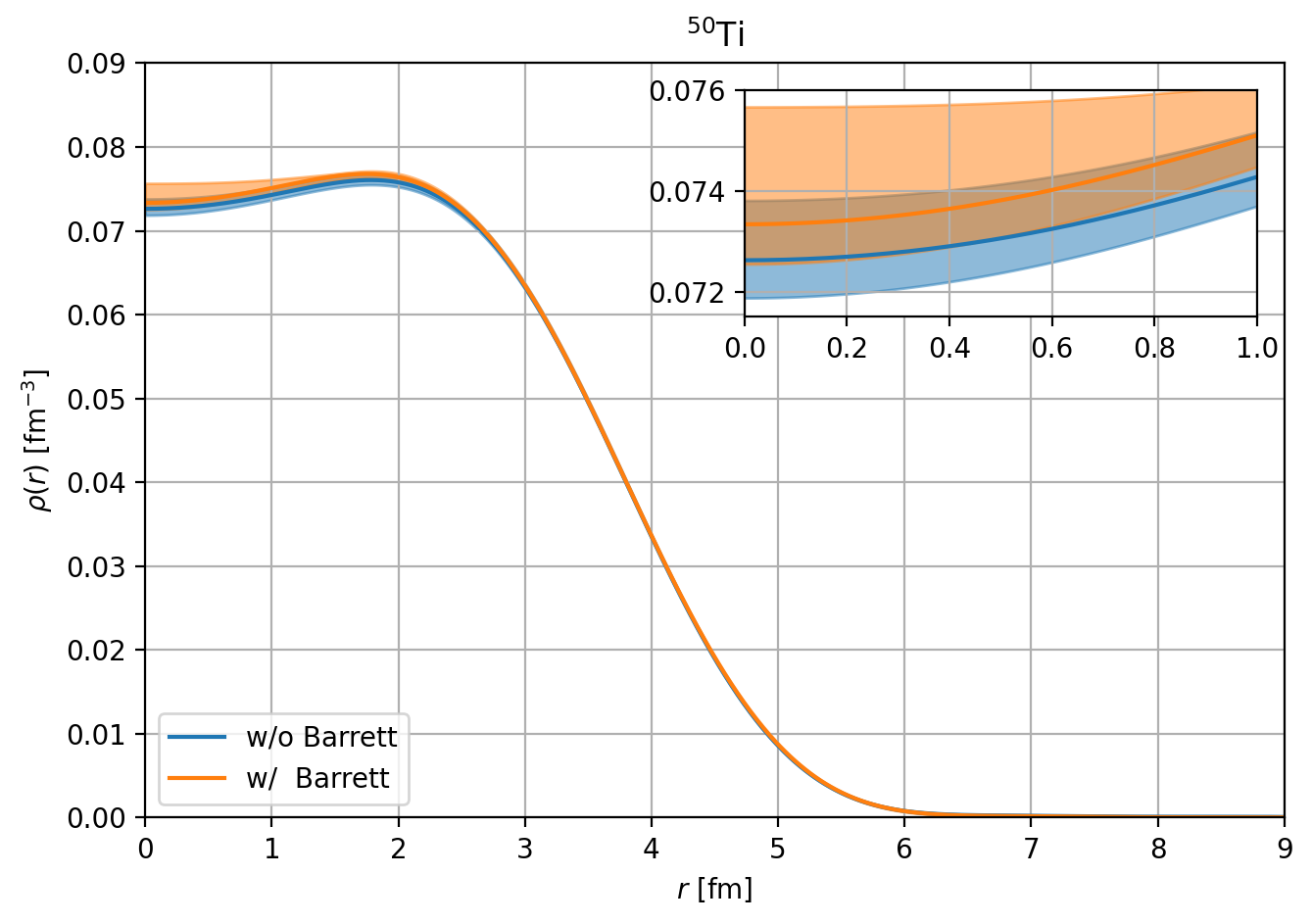}
    \caption{Charge densities of $^{48}$Ti and $^{50}$Ti. Two variants are shown, with (orange) or without (blue) the constraints from Barrett moments.}
    \label{fig:Ti_best}
\end{figure}

Similar difference measurements, relative to $^{48}$Ti, are available for $^{46}$Ti and $^{50}$Ti~\cite{Heisenberg:1972zza}, i.e., for the analysis of these charge distributions we could, in principle, use our previous results for $^{48}$Ti as input. Unfortunately, the quality of these difference measurements is borderline, and for $^{46}$Ti we were not able to extract  meaningful uncertainty estimates. This conclusion is actually in line with the discussion of charge radii already given in Ref.~\cite{Heisenberg:1972zza}, for which a tension to the previous measurement from Ref.~\cite{Romberg:1971hbw} is observed, and indeed the isotope shift for $^{46}$Ti compared to $^{48}$Ti is compatible with zero instead of being slightly positive as found in spectroscopy measurements. More details on our fits for $^{46,48}$Ti are discussed in App.~\ref{app:Ti4648}, the conclusion being that we only quote results for the case of $^{50}$Ti. Apart from a more consistent data situation compared to $^{46}$Ti, also a (qualitative) cross check against Ref.~\cite{Selig:1985} becomes possible, increasing confidence that the extracted charge distribution for $^{50}$Ti is still useful despite the issues described in App.~\ref{app:Ti4648}. 

For the remaining naturally occurring titanium isotopes, $^{47}$Ti ($J=5/2$) and $^{49}$Ti ($J=7/2$), we only found measurements of the magnetic part of the interaction in Refs.~\cite{Likhachev:1976db,Platchkov:1981ni}, but not the main Coulomb contribution. The necessity of having reasonable control over the magnetic part of the cross section for a robust extraction of the charge distribution in case of $J>0$, see $^{27}$Al in the next section, would complicate the extraction in either case, but since not even information on the Coulomb part could be retrieved, we do not see a way to extract charge distributions for these isotopes from electron-scattering data at present. 

Our final results for $^{48}$Ti and $^{50}$Ti are shown in Fig.~\ref{fig:Ti_best}. 
The bands for the extracted charge distributions encompass statistical and systematic components, including the uncertainties propagated from the respective reference points as determined in our previous fits. The uncertainties are appreciably larger than for $^{40,48}$Ca, as a direct consequence of the tenuous data situation. Accordingly, in this case no significant tension with the Barrett moments arises. 

\begin{figure}[t]
    \centering
    \includegraphics[width=0.85\linewidth]{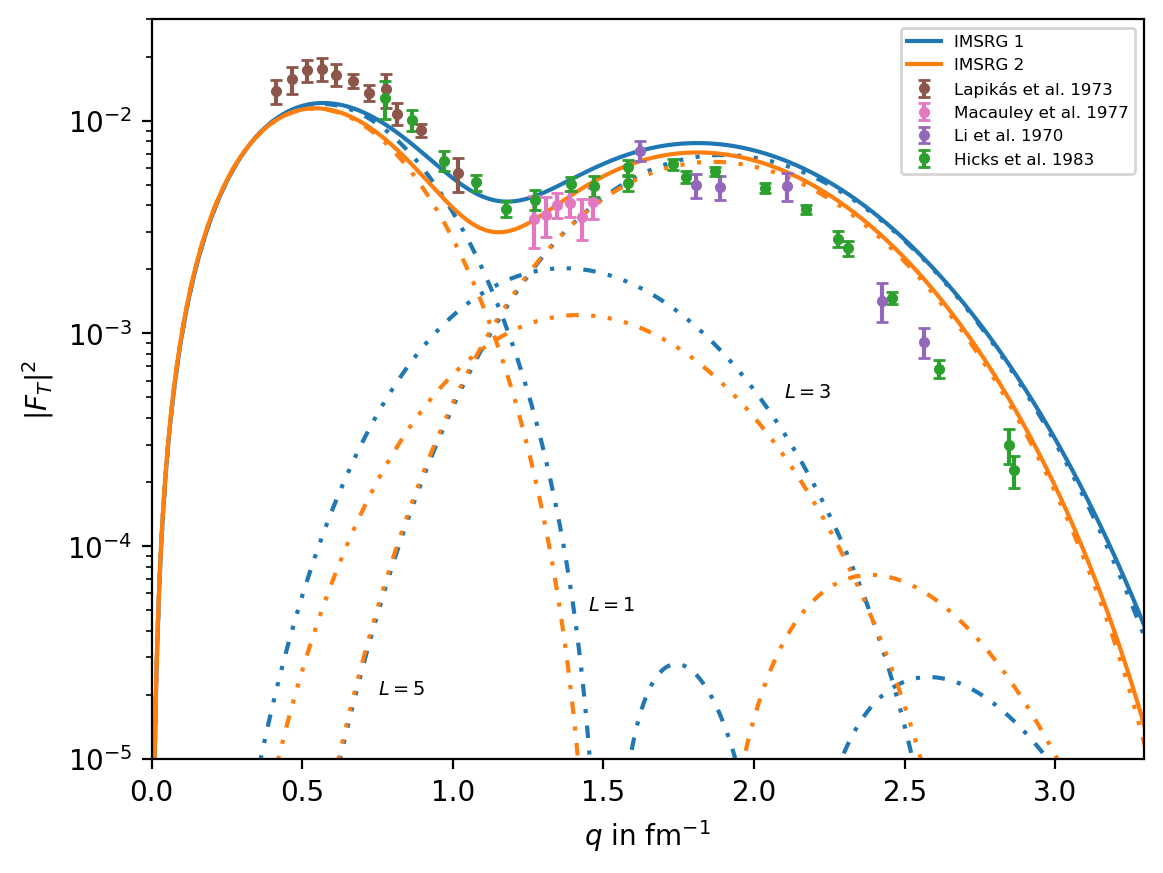}
    \caption{Results for the  $^{27}$Al transverse form factor $F_T$ in comparison to the data given in Ref.~\cite{Donnelly:1984rg}, digitized and converted to our normalization convention. The data sets are Lapik\'as et al.~1973~\cite{Lapikas:1973njq}, Li et al.~1970~\cite{Li:1970zza}, Macauley et al.~1977 (quoted in Ref.~\cite{Singhal:1982}), and Hicks et al.~1983 (``private communication''). In comparison, we show results from IMSRG calculations for two chiral Hamiltonians, as determined from the $\Sigma'_L$, $\Delta_L$ multipoles for $L=1,3,5$ (dot-dashed curves, with the total given by the solid lines). IMSRG~1 and IMSRG~2 refer to two chiral Hamiltonians, $\Delta$N2LO$_\text{GO}$~\cite{Jiang:2020the} and N2LO$_\text{sat}$~\cite{Ekstrom:2014dxa}, respectively.}
    \label{fig:FT_Al}
\end{figure}

\subsection{Aluminum}

For aluminum, we considered the scattering data taken in Refs.~\cite{Stovall:1967yyp,Lombard:1967skb,Bentz:1970,Li:1970zza,Lapikas:1973njq,Li:1974vj,Singhal:1977wvn,Singhal:1982,Dolbilkin:1983,Ryan:1983zz}. While, in principle, a fair number of measurements exists, only few references documented the measured cross sections, so that in many cases at best a digitization of results on logarithmic plots would be possible.  Even more disappointingly, the most precise scattering data off $^{27}$Al ever taken, forming the basis for the FB fits given in   
Ref.~\cite{DeVries:1987atn}, are not even published as a PhD thesis, but appear to be lost as ``private communication.'' Given these restrictions, it is evident that the resulting uncertainties for $^{27}$Al will again be more sizable than for our calcium fits.

Ultimately, we were able to track down the data from Refs.~\cite{Lombard:1967skb,Li:1974vj,Dolbilkin:1983}, where Ref.~\cite{Li:1974vj} shows the best coverage of momentum transfer.  
Reference~\cite{Dolbilkin:1983} would allow one to add a few data points in the very low-energy region, but due to internal inconsistencies (likely reflecting an incomplete report of systematic uncertainties) as well as minor impact on the fit (since the total charge is conserved by construction), there was at best marginal gain by including these points in the fit. Likewise, the impact of the data from Ref.~\cite{Lombard:1967skb} proved negligible, given that the same range in momentum transfer was covered in Ref.~\cite{Li:1974vj} with much higher precision. Finally, we considered additional measurements in Ref.~\cite{Li:1974vj} at scattering angle $135^\circ$ scanned over energy, but again the impact proved minimal due to the much higher contamination from higher multipoles compared to the main data set. Taken together with the significant cost of evaluating the phase-shift model for further energy values in the fit, we therefore concentrated on the main scans from Ref.~\cite{Li:1974vj} taken at $E=\{250,500\}\MeV$.

\begin{figure}[tp]
    \centering
    \includegraphics[width=0.85\linewidth]{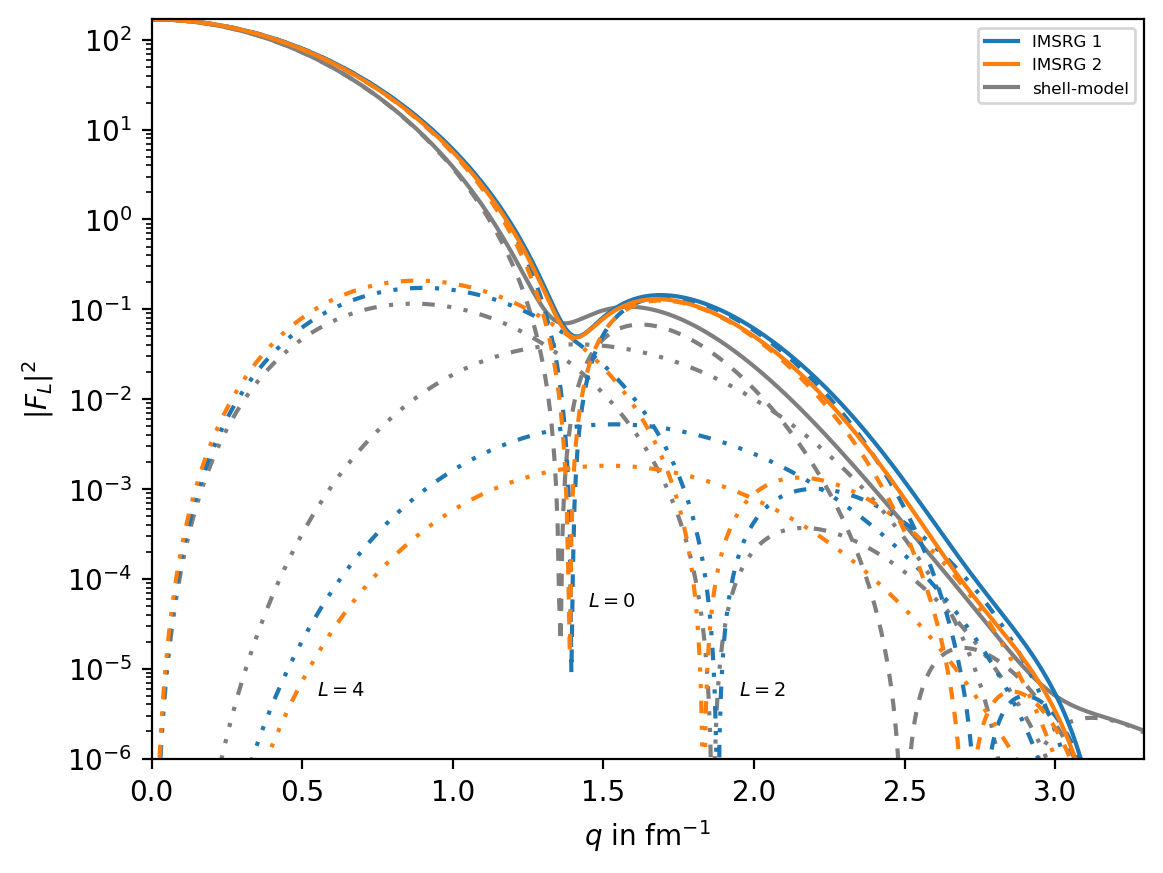}
    \caption{Results for the $^{27}$Al longitudinal form factor $F_L$. IMSRG~1 and IMSRG~2 refer to the same two chiral Hamiltonians as for $F_T$, with $F_L$ determined from the $M_L$, $\Phi_L''$ multipoles for $L=0,2,4$ (dot-dashed curves, with the total given by the solid lines).  Furthermore, we show results from the shell-model calculation from Ref.~\cite{Hoferichter:2022mna}, based on Refs.~\cite{Brown:2006gx,Caurier:2004gf,Otsuka:2018bqq,Caurier:1999}.}
    \label{fig:FL_Al}
\end{figure}

A key point in the fit for $^{27}$Al concerns the question how to account for the non-zero spin $J=5/2$, adding a new layer of complexity compared to the $J=0$ nuclei considered in the previous sections. There are two main effects that complicate the extraction of the charge distribution. First, higher Coulomb multipoles, see Eqs.~\eqref{FL:FT_multipoles} and~\eqref{rho_L_j_LL}, contribute, filling out the minima that arise in the leading spherically symmetric charge form factor at $L=0$. Second, magnetic contributions to the cross section, see Eq.~\eqref{angular_separation}, need to be taken into account before conclusions on the charge distribution become possible. We address these corrections using estimates from the nuclear shell-model~\cite{Hoferichter:2022mna} (for the higher Coulomb multipoles) and ab-initio calculations using the in-medium similarity renormalization group (IMSRG)~\cite{Stroberg:2016ung,Stroberg:2019bch,Stroberg:2019mxo,prep} (for both classes of corrections).\footnote{At the current level of precision it is not necessary to include the tiny corrections from two-body currents (suppressed to next-to-next-to-next-to-leading order in the chiral counting for the charge operator~\cite{Kolling:2009iq,Pastore:2011ip,Krebs:2019aka}).} 
We benchmarked the accuracy of these corrections by comparing to the elastic magnetic form factor as compiled in Ref.~\cite{Donnelly:1984rg} using input from Refs.~\cite{Li:1970zza,Lapikas:1973njq,Singhal:1982}  (as well as further ``private communication'' sources), see Fig.~\ref{fig:FT_Al}. These data sets contain measurements beyond the $180^\circ$ limit, in which only the magnetic contribution survives, so that the interpretation again requires assumptions on the treatment of the Coulomb contribution as well. Especially in view of these caveats, we observe that the IMSRG results, shown for two different chiral Hamiltonians, provide a very decent description of the data, justifying their use to subtract contributions beyond the $L=0$ Coulomb multipole. We also show the $L>0$ corrections to the longitudinal form factor, see Fig.~\ref{fig:FL_Al}, comparing the IMSRG results to nuclear shell-model calculations. Some differences are visible in the $L=4$ contribution, leading to a slightly faster filling of the minimum of the $L=0$ PWBA form factor in the case of the shell-model result. 

\begin{figure}[ptb]
    \centering
    \includegraphics[width=0.85\linewidth]{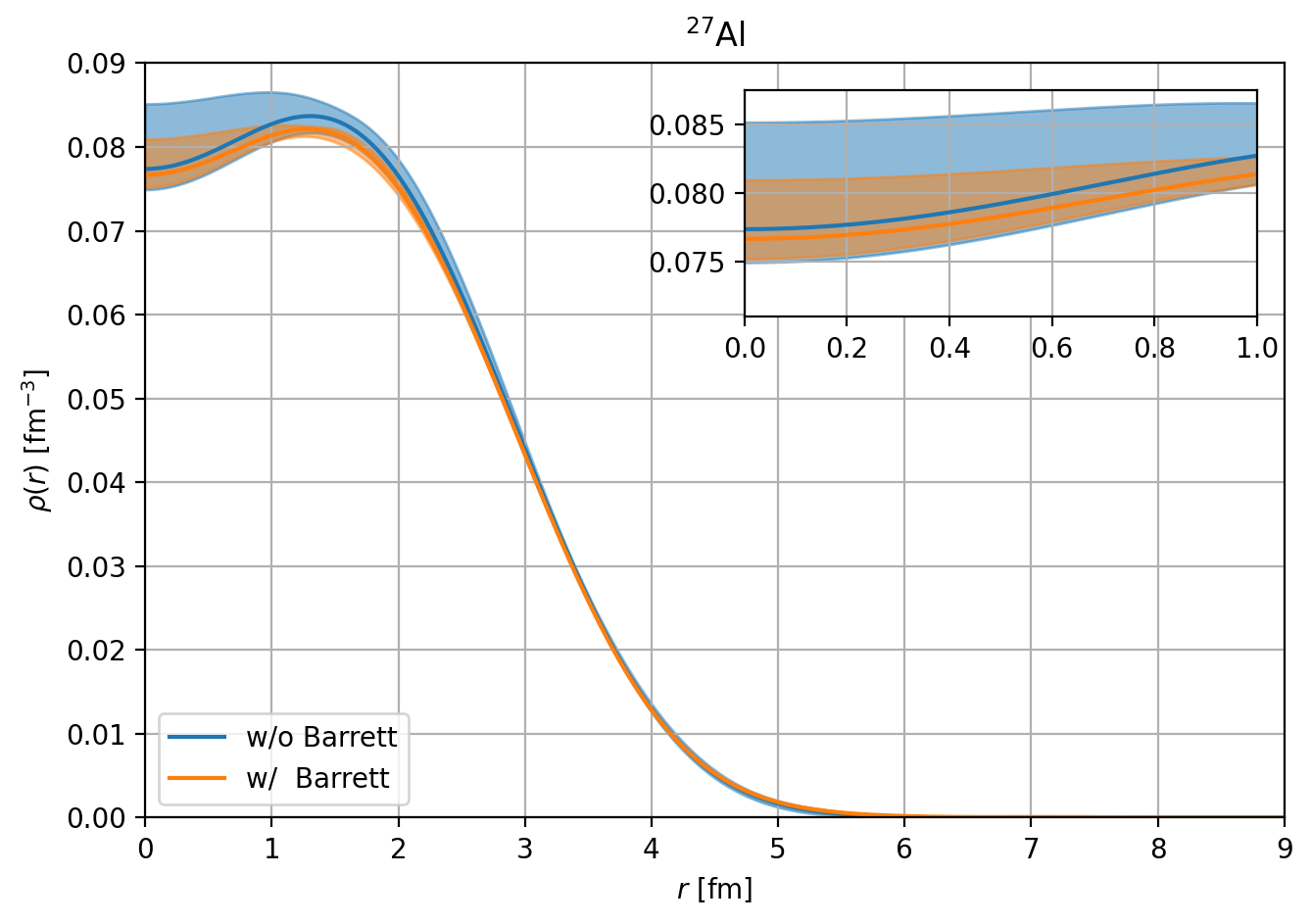}
    \caption{Charge density of $^{27}$Al ($L=0$). Two variants are shown, with (orange) or without (blue) the constraints from Barrett moments.}
    \label{fig:Al_best}
\end{figure}

Given that a purely data-driven determination of all the additional multipoles seems out of reach, we remove any data points that are clearly dominated by these corrections and only keep those points fulfilling the condition 
\begin{equation}
\label{cond}
 {\qty(\dv{\sigma}{\Omega})_{\text{data}}\!-\qty(\dv{\sigma}{\Omega})_{L>0} > \Delta\qty(\dv{\sigma}{\Omega})_{\text{data}}},
\end{equation}
i.e., we include data points in the fit as long as the remainder after subtracting the $L>0$ corrections is not consistent with zero. As input for $\qty(\dv{\sigma}{\Omega})_{L>0}$ we use the arithmetic mean of the two IMSRG calculations, while the spread between them (and to the shell-model calculation) can serve as an estimate of the uncertainty. In view of the sizable errors of the cross-section data themselves, we assume that for the data points that fulfill condition~\eqref{cond} the remaining systematic uncertainty is subdominant,  or at least sufficiently covered by the scale factor. Figure \ref{fig:Al_best} shows the extracted $L=0$ charge density for $^{27}$Al, including uncertainty bands that comprise both statistical and systematic components.

\subsection{Charge radii and discussion}
\label{sec:charge_radii}

\begin{table}[tp]
    \centering 
    \renewcommand{\arraystretch}{1.3}
    \begin{tabular}{cllll} 
        \toprule 
        & \multicolumn{2}{c}{$\sqrt{\expval{r^2}}$ [fm]} & \multicolumn{2}{c}{$\expval{r^k e^{-\alpha r}}$ [fm${}^{k}$]}\\ 
         Nucleus & Our fit & Refs.~\cite{DeVries:1987atn,Angeli:2013epw} & Our fit & Ref.~\cite{Fricke:1995zz}\\ 
         \midrule 
         \multirow{2}{*}{$^{27}$Al} 
           & $2.996(11)\mqty{(43)[44]\\[-2mm](\substack{+26\\-33})[35]}$ & $3.035(2)$ & $8.32(6)\mqty{(22)[23]\\[-2mm](\substack{+14\\-17})[18]}$ & \\ 
           & $3.063(3)\mqty{(30)[31]\\[-2mm](\substack{+0\\-1})[3]}$ & $3.0610(31)$ & $8.66(1)\mqty{(15)[15]\\[-2mm](\substack{+0\\-0})[1]}$ & $8.662(10)$ \medskip\\
         \multirow{2}{*}{$^{40}$Ca}  
           & $3.452(3)\mqty{(8)[9]\\[-2mm](\substack{+1\\-9})[10]}$ & $3.450(10)$ & $10.637(18)\mqty{(43)[47]\\[-2mm](\substack{+4\\-47})[50]}$ &  \\ 
           & $3.4771(17)\mqty{(17)[24]\\[-2mm](\substack{+0\\-5})[17]}$ & $3.4776(19)$ & $10.767(8)\mqty{(8)[11]\\[-2mm](\substack{+0\\-2})[8]}$ & $10.776(10)$\medskip\\ 
         \multirow{2}{*}{$^{48}$Ca}  
          & $3.4499(29)\mqty{(31)[42]\\[-2mm](\substack{+42\\-52})[60]}$ & $3.451(9)$ & $10.645(16)\mqty{(17)[23]\\[-2mm](\substack{+22\\-26})[31]}$ &  \\ 
          & $3.475(2)\mqty{(10)[10]\\[-2mm](\substack{+0\\-3})[4]}$ & $3.4771(20)$ & $10.772(9)\mqty{(46)[47]\\[-2mm](\substack{+1\\-10})[14]}$ & $10.781(10)$\medskip\\ 
         \multirow{2}{*}{$^{48}$Ti}  
           & $3.62(3)\mqty{(8)[8]\\[-2mm](\substack{+2\\-3})[4]}$ & $3.597(1)$ & $11.50(14)\mqty{(39)[41]\\[-2mm](\substack{+9\\-15})[21]}$ &  \\ 
           & $3.596(3)\mqty{(57)[57]\\[-2mm](\substack{+1\\-1})[3]}$ & $3.5921(17)$ & $11.39(1)\mqty{(28)[28]\\[-2mm](\substack{+0\\-0})[1]}$ & $11.388(11)$\medskip\\
        \hdashline\vspace{-1.25\bigskipamount}\\
         \multirow{2}{*}{$^{50}$Ti}
           & $3.612(16)\mqty{(42)[45]\\[-2mm](\substack{+19\\-48})[51]}$ & $3.572(2)$ & $11.45(8)\mqty{(21)[23]\\[-2mm](\substack{+9\\-23})[25]}$ & \\
           & $3.572(3)\mqty{(9)[9]\\[-2mm](\substack{+0\\-2})[3]}$ & $3.5704(22)$ & $11.254(12)\mqty{(43)[45]\\[-2mm](\substack{+0\\-2})[12]}$ & $11.256(10)$ \\
         \bottomrule
    \renewcommand{\arraystretch}{1.0}
    \end{tabular}
    \caption{Extracted charge radii $\sqrt{\expval{r^2}}$ and Barrett moments $\expval{r^k e^{-\alpha r}}$. For each isotope, the upper/lower entries refer to the fits excluding/including the Barrett moments as constraint, with input values from Table~\ref{tab:barrett} repeated in the last column. 
    Moreover, for each entry, the common first error refers to the statistical uncertainty, while we show two variants of the systematic (and total, in square brackets) uncertainty, the upper one being propagated from the charge distribution, the lower one directly from the charge radii corresponding to the set of FB fits on which the systematic uncertainty band for the charge distribution is based. 
    For comparison, we also quote the charge radii given in Ref.~\cite{DeVries:1987atn} (upper) and Ref.~\cite{Angeli:2013epw} (lower). 
    Note that, while the radii given in Ref.~\cite{Angeli:2013epw} are derived from spectroscopy measurements, for Ref.~\cite{DeVries:1987atn} it is not always clear if spectroscopy constraints are included, but at least for $^{48,50}$Ti Ref.~\cite{Selig:1985} confirms that this is the case. The results for $^{50}$Ti are separated by a dashed line, to indicate that the systematic uncertainties are harder to quantify than for the other isotopes, for the reasons spelled out in App.~\ref{app:Ti4648}.}
    \label{tab:rch_result}
\end{table}

To compare our results to previous work, it is instructive to consider the resulting charge radii, see Table~\ref{tab:rch_result}, where we compare our determinations (with and without constraints from Barrett moments) to the ones from Refs.~\cite{DeVries:1987atn,Angeli:2013epw}. The radii quoted in Ref.~\cite{Angeli:2013epw} are derived from spectroscopy measurements. Accordingly, these values agree well with our fits including Barrett moments, where the additional uncertainties in our case originate from the systematics of the combined fit. For $^{27}$Al and $^{40,48}$Ca the determinations from Ref.~\cite{DeVries:1987atn} are based on electron scattering data, so that the resulting values largely agree with our fits excluding the Barrett-moment constraint. For $^{27}$Al we do see a sizable difference, mainly due to the fact that we did not have access to the ``private communication'' data on which the FB  fit from Ref.~\cite{DeVries:1987atn} is based. Comparing with our uncertainty estimates for $^{40}$Ca, however, for which the data situation is exceptional in comparison, we believe that the uncertainty quoted in Ref.~\cite{DeVries:1987atn} for $\sqrt{\expval{r^2}}$ is unrealistic, even if data of similar quality as for $^{40}$Ca had been available at the time. Finally, for $^{48,50}$Ti we do not have a direct comparison of electron-scattering-only determinations, since the FB fit from Ref.~\cite{DeVries:1987atn} already includes spectroscopic information, as confirmed by comparing with the FB fit in Ref.~\cite{Selig:1985}.   

For each quoted charge radius, we show two variants of estimating the systematic uncertainty. The upper numbers are derived by propagating our systematic uncertainty band for the charge distribution, based on the final systematic covariance matrix for the FB parameters. In comparison, the lower numbers are deduced directly from the spread observed among the individual FB fits from which the systematic covariance matrix is constructed. In most cases, we observe reasonable agreement among the two estimates, with a few notable exceptions. First, the fits including Barrett moments essentially impose the spectroscopic value of the charge radius, in such a way that the spread among the individual fits is small. Accordingly, for these fits the systematic error propagated from the charge distribution tends to overestimate the true uncertainty (and potentially sizably so, as for $^{27}$Al and $^{48}$Ti). We emphasize, however, that this effect is special for the charge radius, and will become less pronounced the weaker a given quantity is correlated with the charge radius.  Second, for $^{48}$Ti also in the fit without Barrett moments the systematic uncertainties in the charge distribution appear overestimated, tracing back to the construction as an envelope of the individual fits. In this case, constructing a covariance matrix that encompasses all acceptable individual fits entails a large systematic uncertainty. This behavior is ultimately caused by the poor data quality, e.g., for $^{40,48}$Ca the different ways of estimating systematic errors show a higher degree of consistency.

Comparing the spectroscopy values from Ref.~\cite{Angeli:2013epw} with our electron-scattering-only fits, we can also quantify potential tensions between the two approaches. For $^{27}$Al, we do see a deficit, but only at the level of $1.5\sigma$, due to the sizable systematic uncertainty from the electron-scattering determination. Accordingly, the $7\sigma$ difference between Refs.~\cite{DeVries:1987atn,Angeli:2013epw} very likely originates from underestimated uncertainties in Ref.~\cite{DeVries:1987atn}.
For $^{40,48}$Ca the tensions are more significant, particularly for $^{48}$Ca. That is, in both cases the central values come out almost identical for both isotopes, in agreement with Refs.~\cite{DeVries:1987atn,Angeli:2013epw}, but we find that the systematic uncertainties for $^{48}$Ca are reduced compared to $^{40}$Ca by almost a factor $2$, increasing the tension in the charge radii from $2.5\sigma$ for $^{40}$Ca to $4.3\sigma$ for $^{48}$Ca. Interestingly, the systematic uncertainties hardly increase in the fits including Barrett moments, since the fit is able to accommodate a larger charge radius by modifications of the charge distribution at larger distances, see App.~\ref{app:fit_strat}. Finally, for $^{48,50}$Ti we find charge radii larger than the spectroscopic values, but in these cases the systematic uncertainties are large enough that no significant tension can be inferred.

In view of the tensions for the calcium fits, one could worry about the impact of higher-order radiative corrections. Such two-photon effects, which are, in principle, sensitive to the possible excitations of the intermediate-state nucleus, go under the name of dispersive corrections. They become most relevant in the vicinity of the minima, contributing further to the filling of the zeros of the PWBA form factor. Such corrections have been studied in the literature, see, e.g., Refs.~\cite{Bottino:1972cbl,Friar:1972af,Knoll:1974zz,Friar:1974bn,Mercer:1974zz,DeForest:1975mz,Hachenberg:1975zz,Friar:1976yro,Ravenhall:1976zz,Mercer:1977up,Friar:1979ak}, but phenomenological estimates of their impact remain uncertain and model dependent. In Ref.~\cite{Friar:1979ak} an upper bound on the impact of dispersive corrections on the charged radius is formulated in terms of
\begin{equation}
\label{disp}
 \Delta_\text{disp} \sqrt{\expval{r^2}}\simeq -\frac{3\sigma_{-1}}{4\pi Z \sqrt{\expval{r^2}}}\simeq -\frac{3a_0}{2\pi r_0}\simeq -7\times 10^{-3}\fm,
\end{equation}
with phenomenological parameters $\sqrt{\expval{r^2}}\simeq r_0 A^{1/3}$, $\sigma_{-1}\simeq a_0 A^{4/3}$, $r_0\simeq 1.1\fm$, $a_0\simeq 0.016\fm^2$ to try and estimate polarization effects. More realistic calculations tend to give smaller corrections~\cite{Friar:1979ak,Friar:1975pfp}, so that while Eq.~\eqref{disp} slightly moves the $^{40,48}$Ca radii from electron scattering towards the spectroscopy ones, one would suspect that the net effect is likely too small to make up the deficit. However, while in Ref.~\cite{Friar:1979ak} it was concluded that the experimental evidence for dispersive corrections was scant, 
the calcium data are precise enough that their impact could potentially be detected.  
To improve our analysis along these lines, dedicated nuclear-structure calculations of polarization corrections would be required, to obtain a more realistic assessment of the impact of dispersive corrections beyond rough estimates such as Eq.~\eqref{disp}.

\begin{table}[tp]
    \centering 
    \renewcommand{\arraystretch}{1.3}
    \begin{tabular}{cllllll} 
        \toprule
         &  \multicolumn{4}{c}{This work} & Ref.~\cite{Kitano:2002mt} & Ref.~\cite{DeVries:1987atn} \\ 
         Nucleus & \multicolumn{1}{c}{$D_1$} & \multicolumn{1}{c}{$D_2$}  & \multicolumn{1}{c}{$D$} & \multicolumn{1}{c}{$\Delta D$} & \multicolumn{2}{c}{$D$} \\ 
         \midrule
         \multirow{2}{*}{$^{27}$Al}  
          & $0.03651$ & $0.03668$ & $0.03652$ & $(14)\mqty{(55)[57]\\[-2mm](\substack{+41\\-31})[44]}$ & \multirow{2}{*}{$0.0362$} & \multirow{2}{*}{$0.03608$$^{*}$}\\ 
          & $0.03586$ & $0.03602$ & $0.03587$ & $(2)\mqty{(24)[24]\\[-2mm](\substack{+1\\-3})[4]}$ & \medskip\\ 
         \multirow{2}{*}{$^{40}$Ca} 
          & $0.07584$ & $0.07619$ & $0.07587$ & $(10)\mqty{(22)[24]\\[-2mm](\substack{+22\\-2})[24]}$ & \multirow{2}{*}{$0.0761$} & \multirow{2}{*}{$0.07595$$^{*}$} \\ 
          & $0.075275$ & $0.075625$ & $0.075311$ & $(31)\mqty{(33)[45]\\[-2mm](\substack{+14\\-10})[34]}$ & \medskip\\ 
         \multirow{2}{*}{$^{48}$Ca}  
          & $0.07526$ & $0.07561$ & $0.07532$ & $(9)\mqty{(10)[13]\\[-2mm](\substack{+10\\-10})[13]}$ & & \multirow{2}{*}{$0.07531$$^{*}$}\\ 
          & $0.07474$ & $0.07508$ & $0.07479$ & $(2)\mqty{(10)[10]\\[-2mm](\substack{+1\\-6})[7]}$ & & \medskip\\ 
         \multirow{2}{*}{$^{48}$Ti}   
          & $0.0860$ & $0.0864$ & $0.0860$ & $(5)\mqty{(16)[17]\\[-2mm](\substack{+5\\-3})[7]}$ & \multirow{2}{*}{$0.0864$} & \multirow{2}{*}{$0.08627$$^{*}$}\\ 
          & $0.0863$ & $0.0867$ & $0.0864$ & $(1)\mqty{(9)[9]\\[-2mm](\substack{+1\\-1})[1]}$ & \medskip\\
        \hdashline\vspace{-1.25\bigskipamount}\\
         \multirow{2}{*}{$^{50}$Ti} 
          & $0.0862$ & $0.0866$   & $0.0863$ & $(3)\mqty{(9)[9]\\[-2mm](\substack{+8\\-3})[9]}$ & & \multirow{2}{*}{$0.08705$$^{*}$}\\ 
          & $0.08696$ & $0.08736$ & $0.08702$ & $(12)\mqty{(28)[31]\\[-2mm](\substack{+1\\-12})[17]}$ & & \medskip\\ 
         \bottomrule
    \renewcommand{\arraystretch}{1.0}
    \end{tabular}
    \caption{Resulting dipole overlap integrals in units of $m_\mu^{5/2}$ in comparison to Ref.~\cite{Kitano:2002mt} and values calculated by us based on the parameterizations from Ref.~\cite{DeVries:1987atn}. The values for $D$ are obtained assuming ${m_e=0}$ and neglecting recoil corrections, in line with the conventions from Ref.~\cite{Kitano:2002mt}. $D_{1,2}$ correspond to Eq.~\eqref{eq:D_1,D_2}, with ${m_e \neq 0}$ and including recoil effects. The upper/lower values use the charge distributions from the fits excluding/including Barrett moments. In each case, the systematic uncertainties are quoted in the two variants introduced in Table~\ref{tab:rch_result}. The differences in the uncertainties $\Delta D$ between $D$, $D_1$, and $D_2$ are below the level quoted here.}
    \label{tab:D_result}
\end{table}

\section{Dipole operators}
\label{sec:dipole}

The decay rate for $\mu\to e$ conversion in nuclei is conventionally expressed in terms of the overlap integrals introduced in Ref.~\cite{Kitano:2002mt}. To illustrate how uncertainties from the nuclear charge densities can be propagated to these overlap integrals, we consider the case of a dipole interaction, which at the Lagrangian level is described by an effective operator
\begin{equation}
 {\mathcal L}_\text{dipole}=\sum_{Y=L,R}m_\mu C_Y^D\bar\mu\sigma^{\mu\nu}P_Y e F_{\mu\nu}+\text{h.c.},
\end{equation}
where we followed the convention from Ref.~\cite{Kitano:2002mt} to count the operator as  dimension-$6$ with a chirality flip mediated by the muon mass, and the $C_Y^D$ denote the corresponding Wilson coefficients. Since this operator still leads to a long-range interaction, the response is determined by the charge form factor of the nucleus, and, accordingly, the respective overlap integral follows directly from the charge distribution without the need for any further nuclear corrections. Explicitly, the definition reads~\cite{Kitano:2002mt}
\begin{align}
    D_1 &= -\frac{4}{\sqrt{2}} m_\mu \int_0^\infty \diff r \,E(r)\, \Big[ g^{{e}}_{-1}(r) \, f^{\mu}_{-1}(r) + f^{{e}}_{-1}(r) \, g^{\mu}_{-1}(r) \Big], \nonumber\\
    D_2 &= +\frac{4}{\sqrt{2}} m_\mu \int_0^\infty \diff r \,E(r)\, \Big[ f^{{e}}_{+1}(r) \, f^{\mu}_{-1}(r) - g^{{e}}_{+1}(r) \, g^{\mu}_{-1}(r) \Big], \label{eq:D_1,D_2}
\end{align}
where $E(r)$ denotes the electric field and $f_\kappa^{\ell}(r)$, $g_\kappa^{\ell}(r)$, ${\ell=e,\mu}$, the muon and electron wave functions obtained by solving the Dirac equation (see footnote \ref{footnote:1/r} regarding normalization conventions). In the limit $m_e=0$ it follows ${g^{{e}}_{+1} = f^{{e}}_{-1}}$ and ${f^{{e}}_{+1} = -g^{{e}}_{-1}}$ and thus ${D_1=D_2\equiv D}$, which is the limit considered in Ref.~\cite{Kitano:2002mt}. The electric field $E(r)$ follows from a given charge distribution $\rho(r)$ by virtue of Eq.~\eqref{eq:V_FB_E_FB}, while the quantum numbers of the lepton wave functions, $\kappa=\mp1$, are determined by the requirement that the initial-state muon occupy the $1s$ ground state. The total energy 
\begin{align}
    E_e&=\frac{(M + m_\mu + E^{\mu}_b)^2 - M^2 + m_e^2}{2 (M + m_\mu + E^{\mu}_b)} \simeq m_\mu+E^\mu_b - \frac{m_\mu^2}{2M} + \dots 
\end{align}
follows from the mass of the nucleus $M$ including recoil effects~\cite{Czarnecki:2011mx} and the binding energy $E^{\mu}_{b}<0$ via the ground-state solution. Going beyond the $m_e=0$ limit, the decay rate changes according to 
\begin{align}
\sum_{Y=L,R}\big|C_Y^D D\big|^2&\to \frac{1}{2} \Big(\big|\big(C_L^D+C_R^D\big) ~ D_1\big|^2+\big|\big(C_L^D-C_R^D\big) ~ D_2\big|^2\Big)\notag\\
&=\Big(\big|C_L^D\big|^2+\big|C_R^D\big|^2\Big)\frac{D_1^2+D_2^2}{2}
+2\Re\Big[C_L^D \big(C_R^D\big)^*\Big]\frac{D_1^2-D_2^2}{2},
\end{align}
and similarly when including scalar and vector operators. This formula accounts for the interference of electrons with $\kappa=\mp 1$, lifting the separation of the two chiralities in the $m_e=0$ limit.

The uncertainty propagation from $E(r)$ is straightforward and can, in principle, be carried out analytically. However, to include also the uncertainty propagation from changes of the wave functions it is easiest to calculate the derivatives with respect to the FB parameters numerically. Using the FB parameters and their covariances provided in App.~\ref{app:parameterizations}, we obtain the results summarized in Table~\ref{tab:D_result}, where we also include the second variant to estimate systematic uncertainties based on the individual FB fits on which the construction of the systematic covariance matrix is based, in analogy to Table~\ref{tab:rch_result}. 
First, we can check that our results are consistent with Ref.~\cite{Kitano:2002mt}, in which the FB parameterizations from Ref.~\cite{DeVries:1987atn} were employed, see the comparison of the last two columns. For all cases,
we observe a small deviation $\Delta D=1\times 10^{-4}$, less than the effect of keeping $m_e$, as indicated by the difference between $D_1$ and $D_2$. In the last column, we have also produced reference values for the charge distributions from  Ref.~\cite{DeVries:1987atn} for the remaining isotopes. As main results, we can now provide robust uncertainty estimates for $D$, directly propagated from the charge distributions. For $^{27}$Al, this uncertainty is reduced appreciably once the Barrett-moment constraint for the charge radius is imposed, with a final value 
\beq
D\big({}^{27}\text{Al}\big)=0.0359(2) 
\eeq
slightly lower than in Ref.~\cite{Kitano:2002mt}, but consistent within uncertainties (the systematic error could be estimated more aggressively by looking at the individual FB fits). For $^{40,48}$Ca, the tensions in the charge distributions are reflected at the $(2\text{--}3)\sigma$ level for $D$, with best values including Barrett moments of 
\beq
D\big({}^{40}\text{Ca}\big)=0.07531(5),\qquad 
D\big({}^{48}\text{Ca}\big)=0.07479(10). 
\eeq
Finally, for titanium we obtain 
\beq
D\big({}^{48}\text{Ti}\big)=0.0864(1),\qquad 
D\big({}^{50}\text{Ti}\big)=0.0870(3), 
\eeq
where for $^{48}$Ti we quote the systematic error based on the individual FB fits to avoid the overestimation discussed in Sec.~\ref{sec:charge_radii}. On the contrary, for $^{50}$Ti the uncertainty might be slightly underestimated 
as a result of the rather indirect way its charge distribution needs to be extracted from the data. 

\section{Summary and outlook}
\label{sec:summary}

In this work we started a program to derive robust uncertainty quantifications for the nuclear responses required for the interpretation of searches for $\mu\to e$ conversion in nuclei. As first step, one needs to be able to propagate the uncertainties from nuclear charge distributions, both because these determine the leptonic wave functions via the solution of the Dirac equation and via their impact on overlap integrals that characterize the conversion rate for a given LFV effective operator.  To this end, we revisited the extraction of charge distributions from electron--nucleus scattering for a set of isotopes of immediate interest for $\mu\to e$ conversion---$^{40,48}$Ca, $^{48,50}$Ti, and $^{27}$Al---to extend existing parameterizations by uncertainty estimates.

Having spent considerable effort to try and access the original literature, we have to conclude that in many cases not even the data for the measured cross sections were documented, let alone information on systematic uncertainties and correlations. Accordingly, in some cases the data underlying commonly used parameterizations appear to have been lost, emphasizing the need for sustainable data preservation efforts. The situation is most dire for $^{27}$Al, in which case the main source used in Ref.~\cite{DeVries:1987atn} was never published besides the quoted ``private communication.'' Accordingly, the precision of the still accessible  cross-sections measurements is inferior, reflected by substantial uncertainties in the derived charge distribution. Similarly, the latest data for $^{48,50}$Ti are only available in the form of Fourier--Bessel parameters, but not in terms of the actual cross sections. To derive uncertainty estimates, one therefore again has to rely on previous, less precise measurements, taken relative to $^{48}$Ca. In the case of calcium, we were able to access the original data in a PhD thesis, so that for $^{40,48}$Ca by far the best constraints on the charge distributions could be extracted.

For each isotope we performed a comprehensive set of fits, propagating a Fourier--Bessel expansion of the charge distribution via a phase-shift calculation to obtain cross sections including Coulomb-distortion effects. In particular, we studied the dependence on the range of the potential and the number of expansion parameters, to determine the amount of information that can actually be extracted from the available data, avoiding overfitting and strong theory assumptions on the shape of the form factor. In contrast, previous compilations often fit many more parameters than can be unambiguously determined from the data, thus subject to strong additional constraints that enforce a well-behaved solution. 
For $^{27}$Al and $^{48,50}$Ti, the systematic uncertainties from the truncations of the expansion proved substantial, while for $^{40,48}$Ca we observed much better convergence. Propagating the uncertainties to charge radii, we found that in the former cases determinations from muonic spectroscopy are much more precise, while for the latter a similar level of precision can be achieved. In fact, for $^{40}$Ca and especially for $^{48}$Ca a tension between electron scattering and spectroscopy was observed, motivating an investigation of the potential impact of higher-order radiative corrections. 

As key results, we provide sets of Fourier--Bessel parameterizations for all charge distributions (made available in a supplementary \texttt{python} notebook), both for the fits to electron scattering alone and for variants including  spectroscopic constraints. The application to overlap integrals for dipole operators mediating the LFV interaction illustrates the application to $\mu\to e$ conversion, in particular, how the uncertainties from charge distributions propagate. In the dipole case, this transition is immediate, since the response is determined by the nuclear charge form factor, while for other operators we anticipate a productive interplay with modern ab-initio nuclear-structure calculations. Some first results were already used in this work to estimate the $L\geq 1$ multipoles for $^{27}$Al, and, similarly, the required nuclear corrections for other LFV effective operators can be evaluated, using the measured charge distributions and radii as benchmarks~\cite{prep}.   
In this way, using ab-initio methods will help quantify the uncertainties in the nuclear responses, which are difficult to assess in more phenomenological approaches. In particular, uncertainties for a general set of overlap integrals can be derived, including correlations among them, which is critical to assess the discriminatory power of the measured rates regarding the underlying LFV mechanism.  While forming the basis for such future work, our analysis also shows that a remeasurement of electron scattering off $^{27}$Al would be well motivated, given the choice of target material for both Mu2e and COMET.

\acknowledgments
We thank  
Matthias Heinz for sharing IMSRG calculations for $L\geq 1$ multipoles in $^{27}$Al (anticipating results from Ref.~\cite{prep}) and for valuable comments on the manuscript, as well as T.~William Donnelly for correspondence on Ref.~\cite{Donnelly:1984rg}. We further thank Sonia Bacca and Luca Doria~\cite{Emrich:1983}, Franziska Hagelstein~\cite{Merle:1976}, and Jordy de Vries~\cite{Selig:1985} for heroic efforts tracking down unpublished PhD theses in the archives at JGU Mainz and the University of Amsterdam. Financial support by the Swiss National Science Foundation (Project Nos.\ PCEFP2\_181117 and TMCG-2\_213690) is gratefully acknowledged. 

\appendix

\addtocontents{toc}{\protect\setcounter{tocdepth}{1}}

\section{Plane-wave Born approximation and multipole decomposition}
\label{app:PWBA}

In the limit of plane waves for the initial- and final-state electron wave functions $\psi_{i}$ and $\psi_f$, the cross section for elastic electron--nucleus scattering becomes
\begin{align}
\label{elastic_cross_section_PWBA}
    \dv{\sigma}{\Omega} & =\bigg(\dv{\sigma}{\Omega}\bigg)_\text{Mott}\times\frac{E'_e}{E_e}\times  \bigg[\sum_{L} \big|Z F^\text{ch}_L(q)\big|^2 + \bigg(\frac{1}{2}+\tan^2\frac{\theta}{2}\bigg) \sum_{L} \big|F^\text{mag}_L(q)\big|^2 \bigg], 
\end{align}
see Sec.~\ref{sec:formalism} for the definitions, in particular, Eq.~\eqref{rho_L_j_LL} for the relation of the angular-momentum components of the charge and magnetic form factors to the distributions $\rho_L(r)$ and $j_ {LL}(r)$. In general, the transverse cross section also involves an electric part, which, however, vanishes for elastic processes due to time-reversal invariance~\cite{DeForest:1966ycn,Donnelly:1976fs}. Parity conservation then implies that only even (odd) partial waves contribute to the Coulomb  (magnetic) multipoles, respectively.  

For the practical calculation, it is important to match the decomposition~\eqref{elastic_cross_section_PWBA} onto the standard conventions for the nuclear multipoles~\cite{Serot:1978vj,Donnelly:1978tz,Donnelly:1979ezn,Serot:1979yk,Donnelly:1984rg,Walecka:1995mi}, see also the case of dark-matter--nucleus~\cite{Fitzpatrick:2012ix,Anand:2013yka,Klos:2013rwa,Hoferichter:2016nvd,Hoferichter:2018acd} and neutrino--nucleus~\cite{Hoferichter:2020osn} scattering for similar matching relations. Using the notation of Ref.~\cite{Serot:1978vj}, we find the relations 
\begin{align}
\label{F_L_multipoles}
    Z F^\text{ch}_L(q) &=  \bigg(1 - \frac{\expval{r_E^2}^p}{6} q^2 - \frac{q^2}{8 \mN^2}\bigg) \F^{M_{L}}_p(q) - \frac{\expval{r_E^2}^n}{6} q^2 \F^{M_{L}}_n(q) \notag\\
    &\qquad+ \frac{1 + 2 \kappa_p}{4 \mN^2} q^2 \F^{\Phi^{\prime\prime}_{L}}_p(q) + \frac{ 2 \kappa_n}{4 \mN^2} q^2 \F^{\Phi^{\prime\prime}_{L}}_n(q) + \mathcal{O}(q^4),  \nonumber\\
    F^\text{mag}_L(q) &=  \frac{-i q}{\mN} \bigg(\F^{\Delta_{L}}_p(q) - \frac{1 + \kappa_p}{2} \F^{\Sigma^{\prime}_{L}}_p(q) - \frac{\kappa_n}{2} \F^{\Sigma^{\prime}_{L}}_n(q) \bigg) + \mathcal{O} (q^3), 
\end{align}
where the various multipoles $\F_N^{A_L}$, $A\in\{M,\Phi^{\prime\prime},\Delta,\Sigma^\prime,\ldots\}$ are separated for proton ($p$) and neutron ($n$), normalized for the $M$-response to the charge $Z=\F_p^{M_0}(0)$ and neutron number $N=\F_n^{M_0}(0)$, respectively. Similarly, the normalizations $\F^{\Phi^{\prime\prime}_{0}}_N(0)$ can be related to spin-orbit corrections to nuclear charge radii~\cite{Hoferichter:2020osn,Bertozzi:1972jff}
\beq
\langle r^2\rangle_\text{so}=-\frac{3}{2\mN^2Z}\Big(\big(1+2\kappa_p\big) \F^{\Phi''_0}_p(0)
+2\kappa_n\F^{\Phi''_0}_n(0)\Big),
\eeq
the ones for $\Sigma^\prime$ to spin-expectation values $\langle \mathbf{S}_N\rangle$\footnote{$\langle \mathbf{S}_N\rangle=1/2$ for a free proton or neutron. This normalization for the $\Sigma'$ response differs from Ref.~\cite{Hoferichter:2020osn} according to  $\F_N^{\Sigma'_L}\big|_\text{this work}=2\sqrt{\frac{4\pi}{2J+1}}\F_N^{\Sigma'_L}\big|_\text{\cite{Hoferichter:2020osn}}$.}
\beq
\F^{\Sigma^\prime_{1}}_N(0)=2\sqrt{\frac{2}{3}}\sqrt{\frac{J+1}{J}} \langle \mathbf{S}_N\rangle,
\eeq
and the $\Delta$ response is related to the reduced matrix element of the angular-momentum operator
\beq
\F_N^{\Delta_1}(0)=-\frac{1}{\sqrt{6}}\sqrt{\frac{J+1}{J}} \langle \mathbf{L}_N\rangle.
\eeq
For a free nucleon, for which $\F^{\Phi''_0}_N(0)=\F^{\Delta_1}_N(0)=0$, the expansion~\eqref{F_L_multipoles} reproduces Eq.~\eqref{free_nucleon} up to the indicated order in $q^2$. Moreover, the multipoles in Eq.~\eqref{F_L_multipoles} are typically the most convenient objects that can be calculated with nuclear-structure techniques.

\section{Dirac equation and phase-shift formalism}
\label{app:Dirac}

The following remarks closely follow the conventions and deliberations of Ref.~\cite{Uberall:1971}, while some adjustments for consistency and readability were made. Crosschecks were performed and further inputs taken in particular from Refs.~\cite{Rose:1961,Tuan:1968,Heisenberg:1981mq,Kitano:2002mt}. Note that while some of the equations are quoted with explicit $m_e$ dependence, in practice we set $m_e=0$ in the calculations, except for the dipole overlap integrals in $\mu \to e$ conversion, which was the only situation in which such mass corrections proved relevant. 

\subsection{Dirac equation}

The electron wave functions in a spherically symmetrical potential can be parameterized as\footnote{Note that, historically, usually $g_\kappa$ and $f_\kappa$ are defined with the factor of $1/r$ included, which is then removed in a second line of definitions. We skip this intermediate step and directly separate a factor $1/r$.\label{footnote:1/r}}
\begin{align}
    \psi_\kappa^\mu(\rr)= \frac{1}{r} \mqty(g_\kappa(r) \phi_\kappa^\mu(\rrhat) \\ i f_\kappa(r) \phi_{-\kappa}^\mu(\rrhat)),
\end{align}
with $\phi_\kappa^\mu$ the two-component Pauli spinors given by 
\begin{align}
    \phi_\kappa^\mu(\rrhat) = \sum_{m,m'=\pm\tfrac{1}{2}} \braket{lm,\tfrac{1}{2}m'}{j\mu} Y_{lm}(\rrhat) \chi_{m'},
\end{align}
where $r=|\rr|$ and $\kappa$ and $\mu$ are labels referring to the angular momentum degree of freedom. We have
\begin{align}
    j &= \abs{\kappa} - \frac{1}{2} ,&
    l &= \begin{cases} \kappa & , \quad \kappa>0 \\ -\kappa-1 & , \quad \kappa<0 \end{cases} ,& \mu &= j_z, 
\end{align}
with $j$ the total spin of the system and $l$ the angular momentum (we will also need $\bar l(\kappa)\equiv  l(-\kappa)$). The angular wave functions fulfill the orthogonality relation
\begin{equation}
\label{orthogonality}
    \int\diff\Omega \big[\phi_\kappa^\mu(\rrhat)\big]^*\phi_{\kappa'}^{\mu'}(\rrhat)=\delta_{\kappa\kappa'}\delta_{\mu\mu'}.
\end{equation}

In our convention the radial Dirac equation is given by the following coupled differential equation
\begin{align}
    \begin{pmatrix}g'_\kappa(r) \\ f'_\kappa(r)\end{pmatrix} = \begin{pmatrix}-\frac{\kappa}{r} & E_e-V(r)+m_e) \\ -(E_e-V(r)-m_e) & \frac{\kappa}{r}\end{pmatrix} \begin{pmatrix}g_\kappa(r) \\f_\kappa(r)\end{pmatrix},
\end{align}
with $m_e$, $E_e$ mass and energy of the electron and $V(r)$ the potential. One may use numerical methods to solve the differential equation starting at the origin to find solutions for $g_\kappa$ and $f_\kappa$. 
We employ either an explicit \textsc{Runge--Kutta} method of order 8~\cite{DOP853:1993} (called \texttt{'DOP853'}) or a backwards differentiation formula (BDF)~\cite{LSODA:1983} (called \texttt{'LSODA'}), both from the \texttt{scipy} routine \texttt{scipy.integrate.solve\_ivp}~\cite{Virtanen:2019joe}. 
The second method is significantly faster as it is a wrapper of the \texttt{fortran} solver from \texttt{ODEPACK}~\cite{ODEPACKS:1982}. 
The initial conditions at $r\to0$ can be given by the analytical solutions of the differential equation in the limit $r\to0$. For ${V(r\to0)=V_0=\text{const}}$ these are given by
\begin{align}
    g_\kappa(r)&= c ~ \begin{cases} -\sqrt{\frac{\bar{E}_e'}{\bar{E}_e}}\, r  j_{\kappa}(k_0 r) & ,\quad \kappa>0 \\ r  j_{-\kappa-1}(k_0 r) & ,\quad \kappa<0 \end{cases} ,\qquad  
   & f_\kappa(r)&= c ~ \begin{cases} -r  j_{\kappa-1}(k_0 r) & ,\quad \kappa>0 \\ - \sqrt{\frac{\bar{E}_e}{\bar{E}_e'}}\, r  j_{-\kappa}(k_0 r) & ,\quad \kappa<0 \end{cases}, 
\end{align}
where  
\begin{equation}    
k_0 =\sqrt{(E_e-V_0)^2-m_e^2} =\sqrt{\bar{E_e}\bar{E}_e'},\qquad 
    \bar{E}_e=E_e-V_0-m_e,\qquad 
    \bar{E}_e'=E_e-V_0+m_e,
\end{equation}
$j_\kappa(x)$ denotes the spherical Bessel function, and $c$ is a global normalization factor. For bound-state solutions this normalization is determined by 
\beq
\int\dd[3]{r} \big[\psi_\kappa^\mu(\rr)\big]^*\psi_{\kappa'}^{\mu'}(\rr)=\delta_{\kappa\kappa'}\delta_{\mu\mu'},
\eeq
so that, after separating the angular components via Eq.~\eqref{orthogonality}, 
\beq
\int_0^\infty \diff r \big(|g_\kappa(r)|^2+|f_\kappa(r)|^2\big) = 1.
\eeq
Similarly, one has for continuum solutions\footnote{The factor $2\pi$ is a matter of convention, e.g., Ref.~\cite{Uberall:1971} uses a normalization without. We chose this normalization in agreement with Ref.~\cite{Kitano:2002mt}, to facilitate the calculation of the $\mu\to e$ overlap integrals.}
\beq
\int\dd[3]{r} \big[\psi_\kappa^{\mu, E_e}(\rr)\big]^*\psi_{\kappa'}^{\mu',  E_e'}(\rr)=2\pi\delta_{\kappa\kappa'}\delta_{\mu\mu'}\delta\big(E_e-E_e'\big),
\eeq
and thus
\begin{align}
\label{norm_continuum}
    \int_0^\infty \diff r \Big(g^{E_e}_{\kappa}(r) g^{E'_e}_{\kappa}(r) + f^{E_e}_{\kappa}(r) f^{E'_e}_{\kappa}(r)\Big) =  2 \pi \delta\big(E_e-E'_e\big). 
\end{align}
Continuum solutions can simply be solved for a given energy $E_e$, while for the bound-state solutions we need to scan over energy values starting from $E_e=V_0<0$ going up until a sign flip in the asymptotic behavior of $g_\kappa$ or $f_\kappa$ is found, indicating a bound-state solution in between. Ordering the found bound states by energy makes it possible to associate the corresponding atomic state, which is shown exemplarily for the first few states in Table~\ref{tab:bound_states_names}.

\begin{table}[t]
    \centering
    \renewcommand{\arraystretch}{1.3}
    \begin{tabular}{ccccc}
        \toprule
        \multicolumn{5}{c}{$\kappa=-1$} \\
        \midrule 
        \# & $n$ & $l$ & $j$ & Name \\
        \midrule 
        1. & 1 & 0 & $\frac{1}{2}$ & $1s^{\tfrac{1}{2}}$ \\ 
        2. & 2 & 0 & $\frac{1}{2}$ & $2s^{\tfrac{1}{2}}$ \\ 
        3. & 3 & 0 & $\frac{1}{2}$ & $3s^{\tfrac{1}{2}}$ \\ 
        \bottomrule
    \renewcommand{\arraystretch}{1.0}
    \end{tabular}
    \renewcommand{\arraystretch}{1.3}
    \begin{tabular}{ccccc}
        \toprule
        \multicolumn{5}{c}{$\kappa=+1$} \\
        \midrule 
        \# & $n$ & $l$ & $j$ & Name \\
        \midrule 
        1. & 2 & 1 & $\frac{1}{2}$ & $2p^{\tfrac{1}{2}}$ \\ 
        2. & 3 & 1 & $\frac{1}{2}$ & $3p^{\tfrac{1}{2}}$ \\ 
        3. & 4 & 1 & $\frac{1}{2}$ & $4p^{\tfrac{1}{2}}$ \\ 
        \bottomrule
    \renewcommand{\arraystretch}{1.0}
    \end{tabular}
    \renewcommand{\arraystretch}{1.3}
    \begin{tabular}{ccccc}
        \toprule
        \multicolumn{5}{c}{$\kappa=-2$} \\
        \midrule 
        \# & $n$ & $l$ & $j$ & Name \\
        \midrule 
        1. & 2 & 1 & $\frac{3}{2}$ & $2p^{\tfrac{3}{2}}$ \\ 
        2. & 3 & 1 & $\frac{3}{2}$ & $3p^{\tfrac{3}{2}}$ \\ 
        3. & 4 & 1 & $\frac{3}{2}$ & $4p^{\tfrac{3}{2}}$ \\ 
        \bottomrule
    \renewcommand{\arraystretch}{1.0}
    \end{tabular}
    
    \caption{Associating the found bound states with quantum numbers. \# refers to the ordering by energy for each $\kappa$.}
    \label{tab:bound_states_names}
\end{table}

\subsection{Phase-shift model}

The phase-shift model describes the elastic scattering cross section for a spherically symmetric nucleus in terms of the phase shifts of the different partial waves, which can be deduced numerically using the solutions of the Dirac equation found in the previous section. In the relativistic limit ($m_e\to0$) the cross section takes the form in Eq.~\eqref{eq:crosssection_phaseshift}, with partial-wave expansion~\eqref{eq:phase_shift_amplitude}. We may rewrite this as
\begin{equation}
    A_s(\theta) = \frac{1}{2 i k_e} \sum_{\kappa \ge 0} a_\kappa P_\kappa(\cos\theta), 
\end{equation}
with series coefficients
\beq
    a_\kappa = \kappa e^{2i\delta_\kappa} + (\kappa+1) e^{2i\delta_{-(\kappa+1)}} \equiv a_\kappa^{(0)}. 
\eeq
This series does not converge easily and would generally require an infinite amount of partial waves. However, we may improve convergence by factoring out poles of $(1 - \cos\theta)^{-1}$ responsible for the poor convergence by redefining
\begin{align}
    (1 - \cos\theta)^m 2 i k_e A_\text{s}(\theta) &= \sum_{\kappa\ge0} a^{(m)}_\kappa P_\kappa(\cos\theta),
\end{align}
which implies
\begin{align}
    a_\kappa^{(i+1)} = a_\kappa^{(i)} &- \frac{\kappa+1}{2\kappa+3} a_{\kappa+1}^{(i)} - \frac{\kappa}{2\kappa-1} a_{\kappa-1}^{(i)}.
\end{align}
In the large-$\kappa$ limit $\delta_\kappa$ reduces to the Coulomb phase shift, which behaves according to 
\beq
\delta^C_\kappa \to -\gamma\log|\kappa|,\qquad \gamma=\alpha Z\frac{E_e}{k_e}.
\eeq
Based on this expression,
one can show that 
\beq
\bigg|\frac{a_\kappa^{(1)}}{a_\kappa^{(0)}}\bigg|=\frac{2\gamma^2}{\kappa^2}+\Order\big(\kappa^{-3}\big),
\eeq
and additional steps further improve the convergence, so that after a few reductions the series converges rapidly. We choose $m=3$, in which case it is sufficient to calculate between 10 to 20 partial waves. Note that in the limit of $m_e=0$ it follows that $\delta_\kappa=\delta_{-\kappa}$, which cuts the number of phase extractions that need to be performed in half. 

The phase shifts $\delta_\kappa$ are extracted from the asymptotic behavior of the numerical solutions of the Dirac equation. For a potential that asymptotically approaches the Coulomb potential $V(r)\to-{\alpha Z}/{r} $, we have the following asymptotic plane-wave solutions 
\begin{align}
    g_\kappa^{E_e}(r) & \to - \sgn(\kappa)\sqrt{2\frac{E_e+m_e}{k_e}} \cos\big[k_e r + \tilde\delta_\kappa(r)\big], \nonumber\\
    f_\kappa^{E_e}(r) & \to ~~\sgn(\kappa)\sqrt{2\frac{E_e-m_e}{k_e}} \sin\big[k_e r + \tilde\delta_\kappa(r)\big]. \label{eq:fg_asymp}
\end{align}
The $r$-dependent part of $\tilde\delta_\kappa$ is known analytically and given by 
\begin{align}
    \tilde\delta_\kappa(r) &= \delta^{1/r}_\kappa(r) + \delta_\kappa ,& \delta^{1/r}_\kappa(r) &= \gamma \log(2 k_e r) -  (l+1) \frac{\pi}{2},
\end{align}
which also defines $\delta_\kappa$ as used in Eq.~\eqref{eq:phase_shift_amplitude}. The phase shift can be further separated into several components based on the analytically known solution for a pure Coulomb potential according to
\begin{align}
    \delta_\kappa &= \delta_\kappa^{C,\,r} + \bar{\delta}_\kappa. 
    \label{eq:delta}
\end{align}
The Coulomb phase shifts corresponding to the regular ($r$) and irregular ($i$) solutions are
\begin{align}
\label{Coulomb_phase_shift}
    \delta^{C,\,r/i}_\kappa &= (l+1)\frac{\pi}{2} - \arg \Gamma(\pm \rho_\kappa + i \gamma) + \eta^{r/i}_\kappa \mp \frac{\pi}{2} \rho_\kappa, \nonumber\\
    \eta^r_\kappa & = -\frac{\pi}{2} \tfrac{1+\sgn(\kappa)}{2} - \frac{1}{2}\arg\bigg[\rho_\kappa - \frac{\gamma^2 m_e}{\kappa E_e} + i \gamma \Big(1+\frac{\rho_\kappa m_e}{\kappa E_e}\Big)\bigg], \nonumber\\
     \eta^i_\kappa &= \eta^r_\kappa - \pi - (\eta^r_\kappa + \eta^r_{-\kappa}) = - \eta^r_{-\kappa} - \pi,
\end{align}
where $\Gamma(x)$ is the Euler $\Gamma$ function, 
\beq
    \rho_\kappa = \sqrt{\kappa^2-(\alpha Z)^2},
\eeq
and $r/i$ correspond to upper/lower signs, respectively. The differences between regular and irregular phase shifts take the form
\begin{align}
    \theta_\kappa &= \tilde\delta_\kappa^{C,\,i} - \tilde\delta_\kappa^{C,\,r} = \delta_\kappa^{C,\,i} - \delta_\kappa^{C,\,r} \\
    &=\pi\big(\rho_\kappa-\abs{\kappa}\big)-\arctan\Big[\tan\Big(\pi\big(\abs{\kappa}-\rho_\kappa\big)\Big)\coth(\pi \gamma)\Big] - \frac{\pi}{2} - \arg(\rho_\kappa + i \gamma) - (\eta^r_\kappa + \eta^r_{-\kappa}). \nonumber
\end{align}
The phase-shift difference $\bar{\delta}_\kappa$ tends to zero for large $\kappa$, as with increasing partial waves less of the central region of the potential is probed. Thus it is usually sufficient to extract $\bar{\delta}_\kappa$ for a dozen partial waves. 
This can be done from the numerical solution of the Dirac equation by comparing them to the analytically known solutions for a pure Coulomb potential. In particular, if $V(r \ge a)=-{\alpha Z}/{r}$ exactly, we can match
\begin{align}
    g_\kappa(r \ge a) &= A_\kappa g^{C,\,r}_\kappa(r \ge a) + B_\kappa g^{C,\,i}_\kappa(r \ge a), \nonumber\\ 
    f_\kappa(r \ge a) &= A_\kappa f^{C,\,r}_\kappa(r \ge a) + B_\kappa f^{C,\,i}_\kappa(r \ge a), \label{eq:matchfg}
\end{align} 
where
\begin{align}
    g^{C,\,r/i}_\kappa(r) &= -\sgn(\kappa)\sqrt{2\frac{E_e+m_e}{k_e}} (2 k_e r)^{\pm\rho_\kappa} e^{\frac{\pi}{2} \gamma} \frac{\abs{\Gamma(\pm\rho_\kappa+i\gamma)}}{\Gamma(\pm2\rho_\kappa+1)} \nonumber\\
    & \qquad \times \Re\Big[e^{-i k_e r + i \eta_\kappa^{r/i}} (\pm\rho_\kappa + i \gamma) \, {}_1F_1(\pm\rho_\kappa+1+i\gamma,\pm2\rho_\kappa+1,2ik_e r) \Big], \nonumber \\
    f^{C,\,r/i}_\kappa(r) &= \sgn(\kappa)\sqrt{2\frac{E_e-m_e}{k_e}} (2 k_e r)^{\pm\rho_\kappa}e^{\frac{\pi}{2} \gamma} \frac{\abs{\Gamma(\pm\rho_\kappa+i\gamma)}}{\Gamma(\pm2\rho_\kappa+1)} \nonumber\\
    & \qquad \times \Im\Big[e^{-i k_e r + i \eta_\kappa^{r/i}} (\pm\rho_\kappa + i \gamma) \, {}_1F_1(\pm\rho_\kappa+1+i\gamma,\pm2\rho_\kappa+1,2ik_e r) \Big],
    \label{eq:fg_c} 
\end{align}
are the explicit regular and irregular analytical solutions to the Dirac equation for a pure Coulomb potential. The confluent hypergeometric function 
\beq
{}_1F_1(a,c,z)=\sum_{m=0}^\infty \frac{\Gamma(a+m)\Gamma(c)}{\Gamma(c+m)\Gamma(a)}
\frac{z^m}{m!}
=\frac{\Gamma(c)}{\Gamma(a)\Gamma(c-a)}\int_0^1\diff t\,t^{a-1}(1-t)^{c-a-1}e^{z t}
\eeq
needs to be analytically continued to complex arguments, implementations for which in \texttt{python} can be found for example in the \texttt{mpmath} package \cite{mpmath:2023}. 
From Eq.~\eqref{eq:matchfg} it follows
\begin{align}
    \frac{A_\kappa}{B_\kappa} = -\frac{f^{C, \, i}_\kappa(a) - g^{C, \, i}_\kappa(a) \frac{f_\kappa(a)}{g_\kappa(a)}}{f^{C, \, r}_\kappa(a)-g^{C, \, r}_\kappa(a) \frac{f_\kappa(a)}{g_\kappa(a)}}, 
    \label{eq:AoB}
\end{align}
which is independent of the (global) normalization of $f_\kappa$ and $g_\kappa$.
We may use this knowledge to extract $\bar{\delta}_\kappa$ in the following way
\begin{align}
    \tan(\bar{\delta}_\kappa + \Delta^a_\kappa) &= \frac{\tfrac{A_\kappa}{B_\kappa}\sin(\Delta^a_\kappa) + \sin(\theta_\kappa+\Delta^a_\kappa)}{\tfrac{A_\kappa}{B_\kappa}\cos(\Delta^a_\kappa) + \cos(\theta_\kappa+\Delta^a_\kappa)} , &
    \Delta^a_\kappa &= k_e a + \tilde\delta_\kappa^{C,\,r},
\label{eq:tan(delta_b)}
\end{align}
which follows by inserting Eq.~\eqref{eq:fg_asymp} into Eq.~\eqref{eq:matchfg} and solving for $\bar{\delta}_\kappa$. 
Moreover, while Eq.~\eqref{eq:fg_asymp} is only valid when $f_\kappa$ and $g_\kappa$ really become asymptotic, which for increasing $\kappa$ requires higher and higher values of $r$, we may extract $A_\kappa/B_\kappa$ at a smaller and $\kappa$-independent point ${r=a}$, which denotes the point beyond which the potential can be assumed to be Coulombic. 

\subsection{Recoil effects}
\label{app:recoil}

We can incorporate the leading (kinematic) recoil effects by boosting our reference frame from the Lab frame into the CMS frame, then performing our calculation, and finally transforming our result back into the Lab frame as discussed in Ref.~\cite{Foldy:1959zz}. 
For a nucleus of mass $M$ this means in practice that we transform our energies $E_e$ and angles $\theta$ to the CMS frame via 
\begin{align}
    E_\text{CMS} &= E_e \bigg[1 - \frac{E_e}{M} + \mathcal{O}\bigg(\frac{E^2_e}{M^2}\bigg)\bigg], \notag\\
    \theta_\text{CMS} &= \theta + \frac{E_e}{M} \sin\theta + \mathcal{O}\bigg(\frac{E_e^2}{M^2}\bigg),
\end{align}
before using the phase-shift model. The resulting cross section is then transformed back to the Lab frame via 
\begin{align} 
    \bigg(\dv{\sigma}{\Omega}\bigg)_\text{Lab} = \bigg(\dv{\sigma}{\Omega}\bigg)_\text{CMS} \bigg[ 1 + 2\frac{E_e}{M} \cos\theta + \mathcal{O}\bigg(\frac{E^2_e}{M^2}\bigg) \bigg].
\end{align} 
In the end, the effect of the recoil correction proves rather small, with ${E_e}/{M}$ for the considered nuclei at most of percent order. 
Furthermore, dynamic recoil effects have been shown to be even more suppressed~\cite{Foldy:1959zz}.

\subsection{Distorted-wave Born approximation}

For nuclei with non-zero spin $J$, the elastic scattering cross section also depends on higher (non-spherical) multipole contributions to the charge density as well as magnetic interactions. While these are generally suppressed in comparison to the leading spherically symmetric contributions from the charge density,  such higher-order corrections do become relevant in the vicinity of the minima of the $L=0$ charge form factor. To include these higher multipoles one may use the DWBA, which amounts to a generalization of the phase-shift model to arbitrary spin, and has found applications primarily to inelastic scattering. The main assumption is that Coulomb-distortion effects are restricted to the spherically symmetric charge density, which still defines the potential that enters the solution of the Dirac equation, however, the resulting Coulomb corrections then affect all multipoles, not just the $L=0$ ones.  
The interaction can be characterized by the following Hamiltonian\footnote{Note that the analog equation in Ref.~\cite{Uberall:1971} is given for $\alpha=e^2$ and thus differs by $1/(4\pi)$.} 
\begin{align}
    H &= \frac{e}{4\pi} \int \dd[3]{r} \int \dd[3]{r'} \frac{\rho(\rr)\rho_e(\rr') - \jj(\rr)\cdot\jj_e(\rr')}{|\rr-\rr'|},
\end{align}
where $\rho$ ($\rho_e$) and $\jj$ ($\jj_e$) are the charge density and current of the nucleus (electron). The wave functions of the incoming and outgoing electron $\psi_i$ and $\psi_f$ enter via
\begin{align}
    \rho_e(\rr) = -e \psi_f^\dagger \psi_i, \qquad
    \jj_{e}(\rr) = -e \psi^\dagger_f \gamma^0 \boldsymbol{\gamma} \psi_i.
\end{align}
The expectation value of the Hamiltonian interacting with a nucleus of total spin $J$ and initial (final) spin state $M_i$ ($M_f$) is given by 
\begin{align}
    \bra{J M_i} H \ket{J M_f} = \sum_{LM} \frac{\sqrt{2L+1}}{\sqrt{2J+1}} \braket{JM_i,LM}{JM_f} H_{LM}.
\end{align}
To derive the CMS cross section, we write     
\begin{align}
\dv{\sigma}{\Omega}&=\frac{k'_e}{k_e}\frac{|\overline{\mathcal M}_\text{rel}|^2}{64\pi^2 M^2}=\frac{k'_e}{k_e}\frac{(2m_e2M)^2}{64\pi^2 M^2}\frac{1}{2(2J+1)}\sum_{m,m'}\sum_{M_i,M_f}\big|\bra{J M_i} H \ket{J M_f} \big|^2\notag\\
&=\frac{1}{4 \pi^2} \frac{m_e^2}{2J+1} \frac{k'_e}{k_e} \frac{1}{2} \sum_{m,m'} \sum_{L,M} \abs{H_{LM}}^2, \label{eq:crosssection_H}
\end{align}
where the first identity gives the cross section in the standard relativistic normalization of states, $\langle \pp'|\pp\rangle=(2\pi)^32E_p\delta^{(3)}(\pp-\pp')$, with spin-averaged squared matrix element $|\overline{\mathcal M}_\text{rel}|^2$ and initial (final) CMS momentum $k_e=|\kk|$ ($k'_e=|\kk'|$). We neglected corrections to the CMS squared energy
$s=M^2$, and, by the same reasoning, the second step amounts to a non-relativistic normalization for the nuclear states, $\langle \pp'|\pp\rangle=(2\pi)^3\delta^{(3)}(\pp-\pp')$, while the electron states are normalized to $\langle \pp'|\pp\rangle=(2\pi)^3\delta^{(3)}(\pp-\pp')E_e/m_e$. The spin average extends over the spin projections of the electron $m,m'=\pm 1/2$ and the nuclear spins $-J\leq M_i, M_f\leq J$, where the simplification in the last line follows from the orthogonality of the Clebsch--Gordan coefficients  $\braket{JM_i,LM}{JM_f}$.

Employing for $\psi_i$ and $\psi_f$ the wave functions numerically found by solving the Dirac equation and separating radial and angular components of the integral, the cross section for elastic electron--nucleus scattering can be written as\footnote{The radial wave functions $g_\kappa$  are normalized according to $- \sgn(\kappa) \eval{g_\kappa(r)}_\text{this work} = \sqrt{2 k_e (E_e + m_e)} \eval{r g_\kappa(r)}_{\text{\cite{Uberall:1971}}}$, the charges $\rho_L$  according to $\sqrt{4\pi(2J+1)}\eval{\rho_L}_\text{this work}=\eval{\rho_L}_\text{\cite{Uberall:1971}}$, and likewise for $f_\kappa$ and $j_{LL}$, respectively. \label{footnote:DWBA_norm}}
\begin{align} 
    \dv{\sigma}{\Omega} &= \frac{1}{2} \sum_{m,m'=\pm\frac{1}{2}} \sum^{2J}_{L=0} \sum^L_{M=-L} \bigg|\sum_{\tau=\text{ch},\text{mag}} A^{\tau,\,mm'}_{LM}\bigg|^2, \label{eq:crosssection_DWBA} \\
    A^{\tau,\,mm'}_{LM}
    &= \frac{(4\pi)^{\tfrac{3}{2}} \alpha}{2 i k_e} \sum_{\kappa,\kappa'} (-1)^{l+\abs{\kappa'}} ~ i^{l-l'} ~ e^{i(\delta_\kappa+\delta_{\kappa'})} \times \text{CG}_0^{\kappa\kappa'}  \!\times \text{CG}_\tau^{\kappa\kappa'} \!\times R^{\kappa\kappa'}_{\tau,L} \times Y_{l'm - m' - M}(\theta,\phi), \nonumber
\end{align}
where
\begin{align} 
    \text{CG}_0^{\kappa\kappa'} & = \frac{2l+1}{2L+1} \sqrt{2j+1} \braket{l0,\tfrac{1}{2}m}{jm} \braket{jm,L-M}{j'm-M}\notag\\
    &\qquad\times\braket{l'm-m'-M,\tfrac{1}{2}m'}{j'm-M}, \nonumber\\ 
    \text{CG}_{\text{ch}}^{\kappa\kappa'}&= \braket{l0,L0}{l'0} W(j'l'jl;\tfrac{1}{2}L),\quad  \text{CG}_{\text{mag}}^{\kappa\kappa'}= \braket{l0,L0}{\bar{l'}0} W(j'\bar{l'}jl;\tfrac{1}{2}L). 
\end{align}  
Moreover, the quantities $\text{CG}_{\tau}^{\kappa\kappa'}$  are typically expressed in terms of the Racah coefficients $W(abcd;ef)$~\cite{Tuan:1968,Uberall:1971}, which can be expanded in terms of Clebsch--Gordan coefficients via
\begin{align}
    W(abcd;ef) \delta_{cc'} \delta_{\gamma \gamma'} &= \sum_{\alpha\beta\delta\epsilon\phi} \frac{\braket{a\alpha,b\beta}{e\epsilon} \braket{e\epsilon,d\delta}{c\gamma} \braket{b\beta,d\delta}{f\phi} \braket{a\alpha,f\phi}{c'\gamma'}}{\sqrt{(2e+1)(2f+1)}},
\end{align}
in particular,
\begin{align}
    \braket{l0,L0}{l'0} W(j'l'jl;\tfrac{1}{2}L) = (-1)^l \frac{\braket{j' \tfrac{1}{2},j-\tfrac{1}{2}}{L0}}{\sqrt{(2l+1)(2L+1)}} \frac{1}{2}\Big[1+(-1)^{L+l'+l}\Big].
\end{align}
Finally, one has 
\begin{align}
\label{R_ch_mag}
    R^{\kappa\kappa'}_{\text{ch},\,L} &= \int \diff r \diff r' r'^2  \Big[f_\kappa(r) f_{\kappa'}(r) + g_\kappa(r) g_{\kappa'}(r)\Big]  \frac{r^L_<}{r^{L+1}_>} ~ \rho_L(r'), \nonumber \\
    R^{\kappa\kappa'}_{\text{mag},\,L} &= \tfrac{-i(\kappa+\kappa')}{\sqrt{L(L+1)}} \int \diff r \diff r' r'^2 \Big[f_\kappa(r) g_{\kappa'}(r) + g_\kappa(r) f_{\kappa'}(r)\Big]  \frac{r^L_<}{r^{L+1}_>} ~ j_{LL}(r'),
\end{align} 
with $r_> = \max(r,r')$, $r_< = \min(r,r')$, and $f_\kappa$, $g_\kappa$ are the radial wave functions found by solving the Dirac equation. Here, $\rho_L$ and $j_{LL}$ are the charge density and current contributions to the expectation values of charge density and current with the given angular momentum according to
\begin{align}
    \bra{J M_i} \rho(\rr) \ket{J M_f} &= \sqrt{4\pi} \sum_{LM} \braket{JM_i,LM}{JM_f} \rho_L(r) Y^*_{LM}(\rrhat), \notag\\
    \bra{J M_i} \jj(\rr) \ket{J M_f} &= \sqrt{4\pi} \sum_{LL'M} \braket{JM_i,LM}{JM_f} j_{LL'}(r) \mathbf{Y}^*_{LL'M}(\rrhat),
\end{align}
where $\mathbf{Y}_{LL'M}$ are the vector spherical harmonics defined as 
\beq
    \mathbf{Y}_{LL'M}(\rrhat) = \sum_{M',m} \braket{L'M',1m}{LM} Y_{L'M'}(\rrhat) \mathbf{e}_{m},
\eeq
with spherical basis vectors $\mathbf{e}_{\pm1} = \mp (\mathbf{e}_x \pm i\mathbf{e}_y)/\sqrt{2}$, $\mathbf{e}_{0}=\mathbf{e}_z$.
In the special case $L'=L$ one has
\beq
    \mathbf{Y}_{LLM}(\rrhat) = \frac{\mathbf{L}}{\sqrt{L(L+1)}} Y_{LM}(\rrhat), 
\eeq
with the angular-momentum operator $\mathbf{L}=-i\rr\times\boldsymbol{\nabla}$. 

For improved convergence we may write again
\begin{align}
    (1 - \cos\theta)^{\tilde{m}} 2 i k_e A^{\tau,mm'}_{LM}(\theta) &= \sum_{\kappa'\ge0} a^{(\tilde{m})}_{\kappa'} Y_{\kappa'm-m'-M}(\theta,0),
\end{align}
where we set $\phi=0$ as the absolute value removes any phase from the spherical harmonics. We obtain the series coefficients 
\begin{align}
    a^{(0)}_{\kappa'} &= \sum_{\kappa \neq 0}(-i)^{l-\kappa'}\bigg[e^{i(\delta_\kappa+\delta_{\kappa'})} \times \text{CG}^{\kappa\kappa'}_0  \times \text{CG}^{\kappa\kappa'}_\tau \times R^{\kappa\kappa'}_{\tau,L} \times \big(1-\delta_{\kappa'0}\big)\notag\\
    & \qquad \quad - e^{i(\delta_\kappa+\delta_{-(\kappa'+1)})} \times \text{CG}^{\kappa,-(\kappa'+1)}_0  \times \text{CG}^{\kappa,-(\kappa'+1)}_\tau \times R^{\kappa,-(\kappa'+1)}_{\tau,L}\bigg],
    \label{a_kappap}
\end{align}
with the recursion relation~\cite{Tuan:1968}
\begin{align}
    a^{(\tilde{m})}_{\kappa'} &= a^{(\tilde{m}-1)}_{\kappa'} - \sqrt{\frac{(\kappa'+m-m'-M+1)(\kappa'-m+m'+M+1)}{(2\kappa'+1)(2\kappa'+3)}} a^{(\tilde{m}-1)}_{\kappa'+1} \nonumber\\
    & \qquad - \sqrt{\frac{(\kappa'+m-m'-M)(\kappa'-m+m'+M)}{(2\kappa'-1)(2\kappa'+1)}} a^{(\tilde{m}-1)}_{\kappa'-1}. 
\end{align}
Due to the coefficients $\braket{l0,L0}{l'0}$, $\braket{l0,L0}{\bar{l'}0}$ in $\text{CG}_\tau^{\kappa\kappa'}$, the summation in Eq.~\eqref{a_kappap} only extends over finitely many values of $\kappa$, and half the terms vanish since $\braket{l0,L0}{l'0}=0$ for $l+L+l'$ odd. In addition to the convergence in $\kappa$ also the integrals over Coulombic solutions in Eq.~\eqref{R_ch_mag} become delicate, and methods have been developed to perform such integrals analytically~\cite{Reynolds:1964,Alder:1966,Gargaro:1970}. In this work, we have not implemented Coulomb-distortion effects beyond $L=0$, but such corrections could be taken into account along the lines described above if improved data sensitive to the $L>1$ multipoles became available. 

\subsection{Limiting cases}

It is instructive to study some limiting cases of the general expression~\eqref{eq:crosssection_H}.
If $J=0$, the sums collapse to $L=M=0$, and the expression simplifies to
\begin{align}
\label{cross_section_I_kappa}
\dv{\sigma}{\Omega}&=(4\pi\alpha)^2\Bigg\{\bigg|\frac{1}{2ik_e}\sum_\kappa e^{2i\delta_\kappa}|\kappa|P_l(\cos\theta)I_\kappa\bigg|^2
+\bigg|-\frac{\sin\theta}{2ik_e}\sum_\kappa e^{2i\delta_\kappa}\sgn(\kappa)P_l'(\cos\theta)I_\kappa\bigg|^2\Bigg\}\notag\\
&=(4\pi\alpha)^2\Big(1 + \tan^2\frac{\theta}{2}\Big)\bigg|\frac{1}{2ik_e}\sum_{\kappa>0} e^{2i\delta_\kappa}\kappa \Big(P_\kappa(\cos\theta)+P_{\kappa-1}(\cos\theta)\Big)I_\kappa\bigg|^2,
\end{align}
where the second identity holds for $\delta_\kappa=\delta_{-\kappa}$, $I_\kappa=I_{-\kappa}$, with 
\beq
\label{I_kappa}
I_\kappa\equiv R^{\kappa\kappa}_{\text{ch},\,0}
=\int \diff r \diff r' r'^2  \Big(f_\kappa^{E_e}(r) f^{E_e'}_{\kappa}(r) + g_\kappa^{E_e}(r) g_{\kappa}^{E_e'}(r)\Big)  \frac{\rho_0(r')}{r_>}.
\eeq
The resulting expression~\eqref{cross_section_I_kappa} already looks similar to Eqs.~\eqref{eq:crosssection_phaseshift} and~\eqref{eq:phase_shift_amplitude}. For further comparisons, we note that $I_\kappa$ can be written as 
\begin{align}
\label{I_kappa_2}
I_\kappa&=\frac{1}{4\pi}\int_0^\infty \diff r   \Big(f_\kappa^{E_e}(r) f^{E_e'}_{\kappa}(r) + g_\kappa^{E_e}(r) g_{\kappa}^{E_e'}(r)\Big) \int\diff^3r' \frac{\rho_0(r')}{|\rr'-\rr|}\notag\\
&=\int\frac{\diff^3\tilde q}{(2\pi)^3}\frac{Z F_\text{ch}(\tilde q)}{\tilde q^2} \int_0^\infty \diff r\Big(f_\kappa^{E_e}(r) f^{E_e'}_{\kappa}(r) + g_\kappa^{E_e}(r) g_{\kappa}^{E_e'}(r)\Big) j_0(\tilde q r),
\end{align}
after transforming to momentum space and using Eq.~\eqref{eq:Fch<->rho0}. In our normalization, the free solutions are given by 
\begin{align}
 g_\kappa^{E_e}(r)&=-\sgn(\kappa)\sqrt{2k_e(E_e+m_e)}\,rj_l(k_e r),\notag\\
 f_\kappa^{E_e}(r)&=-\sqrt{2k_e(E_e-m_e)}\,rj_{\bar l}(k_e r),
\end{align}
which indeed fulfill~\eqref{norm_continuum}
\beq
 \int_0^\infty \diff r \Big(g^{E_e}_{\kappa}(r) g^{E'_e}_{\kappa}(r) + f^{E_e}_{\kappa}(r) f^{E'_e}_{\kappa}(r)\Big) =
 4k_eE_e\frac{\pi}{2k_e^2}\delta(k_e-k_e')
 =2\pi\delta(E_e-E_e'),
\eeq
due to the relation
\beq
\int_0^\infty \diff r\, r^2 j_\kappa(\alpha r)j_\kappa(\beta r)=\frac{\pi}{2\alpha^2}\delta(\alpha-\beta).
\eeq
In this case, one also has $\delta_{\kappa}=0$ and the sum in Eq.~\eqref{cross_section_I_kappa} can be performed via
\beq
j_0(q r)=\sum_{\kappa=0}^\infty (2\kappa+1)P_\kappa(\cos\theta)j_\kappa(k_e r)j_\kappa(k_e' r),
\eeq
while the $r$ integration 
\beq
\int_0^\infty\diff r\,r^2 j_0(\tilde q r) j_0( q r)=\frac{\pi}{2q^2}\delta(\tilde q - q)
\eeq
gives a $\delta$ function that puts $\tilde q$ to its physical value. Collecting all factors, including a factor $1+\cos\theta$ that emerges from the Legendre polynomials via
\beq
(1+\cos\theta)(2\kappa+1)P_\kappa(\cos\theta)=\kappa \big[P_{\kappa-1}(\cos\theta)+P_\kappa(\cos\theta)\big]+(\kappa+1)\big[P_{\kappa+1}(\cos\theta)+P_\kappa(\cos\theta)\big],
\eeq
this limit reproduces precisely the $L=0$ term in Eq.~\eqref{elastic_cross_section_PWBA}, thereby providing a strong consistency check for the normalization in Eq.~\eqref{eq:crosssection_DWBA}.

As second special case, we consider a pure Coulomb potential, evaluated using the phase-shift model~\eqref{eq:crosssection_phaseshift} and working directly in the $m_e=0$ limit. Inserting the Coulomb phase shifts~\eqref{Coulomb_phase_shift}, one finds
\beq
e^{2i\delta_\kappa}=\kappa\frac{\Gamma(\rho_\kappa-i\gamma)}{\Gamma(\rho_\kappa+i\gamma+1)}e^{-i\pi(\rho_\kappa-l)}\equiv \kappa C_\kappa (-1)^l,
\eeq
leading to 
\beq
A_s(\theta)=\frac{1}{2ik_e}\sum_{\kappa>0}\kappa^2(-1)^\kappa C_\kappa\big[P_\kappa(\cos\theta)+P_{\kappa-1}(\cos\theta)\big].
\eeq
Expanding $\rho_\kappa=\kappa+\Order(\alpha Z)$, the coefficients become 
\beq
\label{C_kappa}
C_\kappa=(-1)^\kappa \frac{\Gamma(\kappa-i\gamma)}{\Gamma(\kappa+i\gamma+1)},
\eeq
and the sum can be performed in closed form~\cite{Mott:1928,Mott:1929,McKinley:1948zz,Curr:1955}
\beq
\label{A_s_Mott}
A_s(\theta)=\frac{\gamma}{2k_e}\exp\Big\{i\gamma \log\sin^2\frac{\theta}{2}\Big\}\frac{\Gamma(1-i\gamma)}{\Gamma(1+i\gamma)}\cot^2\frac{\theta}{2}.
\eeq
Accordingly, Eq.~\eqref{eq:crosssection_phaseshift} reproduces the Mott cross section~\eqref{Mott_cross_section} at leading order in $\gamma=\alpha Z$. 

In the general case, the evaluation of the integrals in Eqs.~\eqref{I_kappa} or~\eqref{I_kappa_2}  has to proceed numerically, which again requires a careful treatment of slowly converging Coulomb integrals and the forward limit. That is, even for the free solutions and a point-like charge distribution, 
\beq
I_\kappa=\frac{Zk_e^2}{2\pi}\int_0^\infty dr\, r\Big(j_\kappa(k_e r)j_\kappa(k_e'r)+j_{\kappa-1}(k_e r)j_{\kappa-1}(k_e'r)\Big),
\eeq
one cannot directly take the limit $k_e=k_e'$. Instead, one encounters a logarithmic divergence according to 
\beq
\label{I_kappa_free}
\lim_{\alpha\to\beta}\int_0^\infty dr\,r\Big(j_\kappa(\alpha r)j_\kappa(\beta r)+
j_{\kappa-1}(\alpha r)j_{\kappa-1}(\beta r)\Big)=\frac{1}{\alpha^2}\bigg[\log\frac{2\alpha}{|\alpha-\beta|}-\gamma_E-\frac{1+2\kappa\psi^{(0)}(\kappa)}{2\kappa}\bigg],
\eeq
with digamma function $\psi^{(0)}(z)=\Gamma'(z)/\Gamma(z)$ and Euler--Mascheroni constant $\gamma_E$. This divergence disappears in the sum over $\kappa$, because of~\cite{Yennie:1954zz}
\beq
\label{P_kappa_sum_1}
\sum_{\kappa>0}\kappa\big[P_\kappa(\cos\theta)+P_{\kappa-1}(\cos\theta)\big]=0,
\eeq
which holds away from the forward direction $\theta=0$. In fact, this feature is related to the phase-shift model in the Coulomb case, where one has 
\beq
\kappa^2(-1)^\kappa C_\kappa =\kappa -i\gamma \big[1+2\kappa \psi^{(0)}(\kappa)\big]+\Order(\gamma^2), 
\eeq
so that the first term vanishes due to Eq.~\eqref{P_kappa_sum_1}, while the second one matches onto the $\kappa$-dependent term in Eq.~\eqref{I_kappa_free}. The corresponding sum 
can then be inferred from Eq.~\eqref{A_s_Mott}, which ultimately reproduces again the Mott cross section~\eqref{Mott_cross_section}. Similar subtleties arise in the evaluation for Coulomb wave functions, which in the $\gamma\to 0$ limit reduce to the free solutions due to 
\beq
{}_1F_1(\kappa+1,2\kappa+1,2iz)=\frac{\Gamma(2\kappa+1)}{\Gamma(\kappa+1)}z^{1-\kappa}2^{-\kappa}e^{iz}\big[j_{\kappa-1}(z)+ij_\kappa(z)\big].
\eeq
For the practical application, one should thus try to perform the Coulomb part of the resulting integrals analytically, using the methods developed in Refs.~\cite{Reynolds:1964,Alder:1966,Gargaro:1970}. For the $L=0$ case, however, it is much simpler to directly work with the phase-shift model~\eqref{eq:crosssection_phaseshift}, which is the avenue we pursue in this work.

\subsection{Analytical expressions for quantities deduced from  FB charge densities}
\label{app:FBparam}

To use a charge density characterized by a FB series as given in Eq.~\eqref{eq:FB} in the phase-shift model in an efficient manner, analytical expressions for derived quantities need to be provided. 
First,  if the charge density is normalized to $Z$, it follows 
\begin{align}
Z = 4 \pi R  \sum^N_{n=1} (-1)^{n+1} \frac{a_n}{q_n^2}.
\label{eq:norm}
\end{align}
Second, in accordance with Eq.~\eqref{eq:Fch<->rho0}, the charge form factor is given by
\begin{align}
    F^\text{ch}_0(q) &= \frac{4 \pi}{Z} R ~ j_0(q R) \sum^N_{n=1} (-1)^{n} \frac{a_n}{q^2 -q_n^2}. 
    \label{eq:F(q)}
\end{align}
In particular, one has $F^\text{ch}_0(q \to q_n) \to 2 \pi R {a_n}/(Z q_n^2)$, which implies that the form factor in the vicinity of ${q = q_n}$ is mainly determined by $a_n$. 
Next, a nucleus with the electric charge density of Eq.~\eqref{eq:FB} possesses the following electric field and potential
\begin{align}
    E(r) &= \frac{\sqrt{4\pi \alpha}}{r^2} \int_0^r \diff r' r'^2 \rho(r') 
    = \begin{cases} \sqrt{4\pi \alpha} \sum^N_{n=1} \frac{a_n}{q_n} j_1(q_n r) & ,\quad r\leq R\\ \sqrt{\frac{\alpha}{4 \pi}} \frac{Z}{r^2}  &,\quad r > R  \end{cases}, \notag\\
    V(r) &= -\sqrt{4\pi \alpha} \int_r^\infty \diff r' E(r') 
    = \begin{cases} - \frac{\alpha Z}{R} - 4\pi\alpha\sum^N_{n=1} \frac{a_n}{q_n^2} j_0(q_n r) & ,\quad r\leq R\\ - \frac{\alpha Z}{r}  &,\quad r > R  \end{cases}, 
    \label{eq:V_FB_E_FB}
\end{align}
which serves as input for solving the Dirac equation. As one can see, due to the finite extent of the nucleus, the potential behaves asymptotically exactly like a Coulomb potential. Finally, the charge radius and Barrett moments can also be calculated analytically, they are given by
\begin{align}
    \expval{r^2} &= \frac{4 \pi}{Z} \int_0^\infty \diff r' r'^4 \rho(r') = \frac{4 \pi}{Z} R^2 \sum^N_{n=1} (-1)^{n+1} \Big(\pi n - \frac{6}{\pi n}\Big) \frac{a_n}{q_n^3}, \\ 
    \expval{r^k e^{-\alpha r}} &= \frac{4 \pi}{Z} \int_0^\infty \diff r' r'^{2+k} e^{-\alpha r} \rho(r') = \frac{4 \pi}{Z} \sum^N_{n=1} \frac{a_n}{q_n} \Im\bigg[\frac{\Gamma\big(2+k,0,R(\alpha-i q_n)\big)}{(\alpha-i q_n)^{k+2}}\bigg],\notag
\end{align}
where the generalized incomplete $\Gamma$ function is defined by 
\beq
\Gamma(z,a,b)=\Gamma(z,a)-\Gamma(z,b),\qquad \Gamma(z,a)=\int_a^\infty\diff t\, t^{z-1}e^{-t},
\eeq
as well as the appropriate analytic continuation for complex arguments. In the limit $\alpha\to 0$, $k\to 2$ one indeed recovers the expression for $\expval{r^2}$.

\section{Fit details and constraints}
\label{app:fit_details}

We fit the charge-density parameterizations directly to the experimental cross-section data. As we need to solve the Dirac equation numerically for all partial waves in every iteration, calculation speed is of crucial importance for a working fit routine. We set the precision goals of the underlying numerical routines in such a way that further improvement in precision only marginally changes the resulting $\chi^2$, verifying the latter for a representative  set of cases. 

As the data sets are generally not dominated by statistical uncertainties, systematic uncertainties need to be included in a robust manner, for which we follow the procedure established for fits to $e^+e^-\to\text{hadrons}$ cross-section data~\cite{Aoyama:2020ynm} in Refs.~\cite{Colangelo:2018mtw,Hoferichter:2019mqg,Hoid:2020xjs,Stamen:2022uqh,Colangelo:2022prz,Hoferichter:2023bjm}. Accordingly, our $\chi^2$ function is defined by 
\begin{align}
    \chi^2 = \sum_{i,j} (f(x_i)-y_i) X_{ij}^{-1} (f(x_j)-y_j),
\end{align}
with covariance matrix
\begin{align}
    X_{ij} = X^\text{stat}_{ij} + \frac{f(x_i) f(x_j)}{y_i y_j} X^\text{syst}_{ij}.
\end{align}
For the data sets used in this work, information on correlations is minimal, but most systematic effects concern normalizations, suggesting fully correlated errors, while for the statistical errors no relevant correlations have been reported. Therefore, we assume  
\begin{align}
    X^\text{stat}_{ij}=\delta_{ij} \sigma_i^2 , \qquad X^\text{syst}_{ij}=\sigma_i \sigma_j.
\end{align}
Especially for correlations of the normalization type, care needs to be taken to avoid 
a D'Agostini-bias~\cite{DAgostini:1993arp,Ball:2009qv}. To this end, we calculate $X_{ij}$ for our initial parameters once and only adjust it once the fit converged for that covariance matrix. Then we calculate a new covariance and restart the fit, iterating the procedure until the fit parameters do not change anymore. 

In the cases for which an additional systematic uncertainty arises from the output of a previous fit of ours, we propagate the known correlations. This is particularly relevant for $^{46,50}$Ti, as the available data were taken relative to $^{48}$Ti, whose cross-section measurements involve sizable uncertainties themselves. 
Regarding fit routines, due to the high correlation among the different FB parameters a standard gradient-descent approach proved insufficient. After benchmarking over different fit routines, we settled on the \texttt{'Powell'} method~\cite{Powell:1964rxf} as implemented in \texttt{scipy.optimize.minimize}~\cite{Virtanen:2019joe} with interface \texttt{lmfit}~\cite{lmfit:2015}, which consistently gave the best performance.  

\subsection{Normalization constraint}
\label{subsec:norm}

To improve convergence and not scan unnecessarily over unphysical parameter space, we implement the charge-conservation constraint of Eq.~\eqref{eq:norm} explicitly. 
Implementing such a condition for two parameters is trivial, but already with three parameters one may run into the scenario in which the first two parameters are chosen within reasonable bounds, while the third is forced to a highly unphysical value. To solve this problem, we employ the following strategy:
by setting some conservative bounds for the FB parameters $a_n$, we can reparameterize them in such  a way that they can only adopt values which (i) are still allowed by total charge conservation and (ii) no parameter needs to be chosen outside these bounds. 
In essence, instead of parameterizing $a_n$ directly, we parameterize the position within the remaining parameter space that is left to still fulfill the normalization, after the $a_{n'}$ with ${n'<n}$ are chosen. To avoid introducing a fit bias,
we choose a very generous limit of
\begin{align}
    \abs{F_0^\text{ch}(q_n)} \leq \pi\Big(\frac{q_1}{q_n}\Big)^3 = \frac{\pi}{n^3}\qquad
    \Leftrightarrow\qquad \abs{a_n} \leq \frac{1}{n} \frac{\pi^2 Z}{2 R^3} \equiv \tilde{a}_{n,\text{lim}},
\end{align}
which loosely reflects the expected asymptotic suppression of the form factor with $q$ (one power less than the perturbative-QCD scaling~\cite{Lepage:1979zb,Lepage:1980fj}). In particular, this is the minimal assumption to implement a suppression of $|a_n|$ for increasing values of $n$, while previous FB fits, e.g.,  in the compilation~\cite{DeVries:1987atn}, typically imposed constraints on the asymptotic behavior of the FB coefficients that were much more stringent. 

Next, to employ our reparameterization we define a version of the parameters without alternating signs via
\begin{align}
    \tilde{a}_n=-(-1)^n a_n,
\end{align}
for which the normalization condition of Eq.~\eqref{eq:norm} becomes  
\begin{align}
    \tilde{a}_n =  \frac{Z}{4 \pi R} q_n^2 - \sum^{N}_{\substack{n' \neq n \\ n'=1}} \tilde{a}_{n'} \frac{q_n^2}{q_{n'}^2}.
    \label{eq:anu_norm}
\end{align}
Initially, the parameter space for each parameter $\tilde{a}_n$ is given by $\qty[-\tilde{a}_{n, \text{lim}},\tilde{a}_{n, \text{lim}}]$. However, once  $\tilde{a}_{n}$ have been chosen for some $n$, not the entire parameter space $\qty[-\tilde{a}_{n, \text{lim}},\tilde{a}_{n, \text{lim}}]$ may be available anymore, due to the normalization condition of Eq.~\eqref{eq:anu_norm}. Hence, one can deduce which maximal and minimal values within $\qty[-\tilde{a}_{n, \text{lim}},\tilde{a}_{n, \text{lim}}]$ are still possible. Consequently, choosing the $\tilde{a}_{n}$ in ascending order starting at ${n=1}$, we may assume that for any $\tilde{a}_{n}$ all the $\tilde{a}_{n'}$ with ${n'<n}$ are already set, while the ones with ${n'>n}$ are still to be determined. Hence the remaining parameter space for $\tilde{a}_{n}$ denoted by $\tilde{a}_n \in \qty[\tilde{a}_{n, \overline{\text{min}}},\tilde{a}_{n, \overline{\text{max}}}]$ is limited by
\begin{align}
    \tilde{a}_{n, \overline{\text{max}}} &= \min\qty(\tilde{a}_{n, \text{lim}},\frac{Z}{4 \pi R} q_n^2 + \sum^{N}_{\substack{n' > n \\ n'=1}} \tilde{a}_{n',\text{lim}} \frac{q_n^2}{q_{n'}^2} - \sum^{N}_{\substack{n' < n \\ n'=1}} \tilde{a}_{n'} \frac{q_n^2}{q_{n'}^2}),\notag\\
    \tilde{a}_{n, \overline{\text{min}}} &= \max\qty(-\tilde{a}_{n, \text{lim}},\frac{Z}{4 \pi R} q_n^2 - \sum^{N}_{\substack{n' > n \\ n'=1}} \tilde{a}_{n',\text{lim}} \frac{q_n^2}{q_{n'}^2} - \sum^{N}_{\substack{n' < n \\ n'=1}} \tilde{a}_{n'} \frac{q_n^2}{q_{n'}^2}).
\end{align}
Once the remaining parameter space has been constrained in this way, we can reparameterize the $a_n$ in terms of $x_n \in \qty[0,1]$ via
\begin{align}
    a_n(x_n) = -(-1)^n \Big[ x_n (\tilde{a}_{n, \overline{\text{max}}} - \tilde{a}_{n, \overline{\text{min}}}) + \tilde{a}_{n, \overline{\text{min}}} \Big].
\end{align}
By fitting these $x_n$ instead of the $a_n$ parameters, we always make sure that the normalization constraint is exactly fulfilled and no unphysical parameter space is probed during the fit. Furthermore, in this way the fit is encouraged to fix the parameters in ascending order, which projects the relative importance of the factors for the shape of the charge density and is in line with ascending momentum transfers in the form factor.

\subsection{Constraints from Barrett moments}
\label{subsec:barrett}

Measurements of $2p\to1s$ transition energies of muonic atoms  can be used as an independent constraint on the charge radius, and thus can be included as additional data points into our fits. These constraints are usually quoted in terms of the so-called Barrett moments~\cite{Barrett:1970wpc}, which can be extracted relatively model-independently from the measured transition energies, and can be easily calculated from a given charge density via
\begin{align}
    \expval{r^k e^{-\alpha r}} = \frac{4\pi}{Z} \int^\infty_0 \diff r\,r^{k+2}\rho(r)  e^{-\alpha r}, \label{eq:Barrett}
\end{align}
where $k$ and $\alpha$ are chosen in a way that allows for the best extraction of the relevant information from the transition energies. We use the values from Refs.~\cite{Fricke:1995zz,Fricke:1992zza,Wohlfahrt:1981zz} as reproduced in Table~\ref{tab:barrett}, included in the $\chi^2$ via 
\begin{align}
    \chi^2 \to \chi^2 + \qty(\frac{\expval{r^k e^{-\alpha r}} - \expval{r^k e^{-\alpha r}}_\text{ref}}{\Delta\expval{r^k e^{-\alpha r}}_\text{ref}})^{2}.
\end{align}

\begin{table}[tp]
    \centering 
    \renewcommand{\arraystretch}{1.3}
    \begin{tabular}{ccclc} 
        \toprule
         Nucleus & $k$ & $\alpha$ [fm${}^{-1}$] & $\expval{r^k e^{-\alpha r}}_\text{ref}$  [fm${}^{k}$]  & Reference\\
         \midrule
         $^{27}$Al & 2.0573 & 0.0419 & 8.6616(17)(101)& \cite{Fricke:1992zza}\\
         $^{40}$Ca & 2.0911 & 0.0596 & 10.7759(50)(81)& \cite{Wohlfahrt:1981zz}\\
         $^{48}$Ca & 2.0912 & 0.0596 & 10.7809(50)(86)& \cite{Wohlfahrt:1981zz}\\
         $^{46}$Ti & 2.1009 & 0.0640 & 11.4796(37)(110)& \cite{Wohlfahrt:1981zz}\\
         $^{48}$Ti & 2.1007 & 0.0641 & 11.3877(37)(101)& \cite{Wohlfahrt:1981zz}\\
         $^{50}$Ti & 2.1003 & 0.0642 & 11.2525(36)(91)& \cite{Wohlfahrt:1981zz}\\
         \bottomrule
    \renewcommand{\arraystretch}{1.0}
    \end{tabular}
    \caption{Barrett moments as quoted in Ref.~\cite{Fricke:1995zz} via Barrett radii. The first error is statistical, the second one refers to a $30\%$ uncertainty on the nuclear-polarization correction from Ref.~\cite{Rinker:1978kh}.}
    \label{tab:barrett}
\end{table}

\subsection{Constraints to prevent oscillations}
\label{subsec:oscillations}

To assess the systematic uncertainties of the fit, it is crucial to consider fit variants with a multitude of values for $N$ and $R$. However, with increasing $R$ and $N$, at some point energies are reached that are not constrained by the data anymore. Accordingly, the fit is enticed to add large contributions there, to balance minor improvements of the fit in the low-energy region and/or to accommodate additional constraints such as the Barrett moment. This results in unphysical oscillations in the shape of the charge density as well as unphysically large values for the form factor in the high-energy region. To counteract this behavior, we adopted the following strategy. For sufficiently large $r$ (and relatively low $Z$), one expects a monotonically decreasing charge density to ensure a smooth transition to its asymptotic value zero, so that demanding the derivative to be negative, ${\rho'(r) \leq 0}$, suppresses oscillations. We implemented this idea by first performing an integration by parts in the sum rule for the charge radius, yielding 
\begin{align}
    \expval{r^2} &= \frac{4\pi}{Z} \int^\infty_0  \diff r\,r^4 \rho(r) = \frac{4\pi}{Z} \int^\infty_0\diff r\, \frac{r^5}{5} ~ \qty(-\rho'(r))  \nonumber\\
    & = \underbrace{\frac{4\pi}{Z} \int^\infty_0\diff r\, \frac{r^5}{5} ~ \qty(-\rho'(r)) ~ \theta(-\rho'(r)) }_{\equiv \expval{r^2}_+ }  + \underbrace{\frac{4\pi}{Z} \int^\infty_0\diff r\, \frac{r^5}{5} ~ \qty(-\rho'(r)) ~ \theta(\rho'(r))}_{\equiv\expval{r^2}_-}.
\end{align}
If one strictly demanded ${\rho'(r) \leq 0}$, one would have ${\expval{r^2}_-=0}$ exactly, but of course this would constitute too strong a constraint. 
In practice, we impose $\big|\expval{r^2}_-| \lesssim \Delta\expval{r^2}$, where we choose $\Delta\expval{r^2}=0.04\,\text{fm}^2$, as a rough upper limit on the typical statistical uncertainty of the charge radius, see Sec.~\ref{sec:charge}. In particular, due to the $r^5$ weighting, this implementation only suppresses oscillations for large $r$, while the low-energy region remains essentially unaffected. The additional term in the $\chi^2$ is
\begin{align}
    \chi^2 \to \chi^2 + \qty(\frac{\expval{r^2}_-}{\Delta\expval{r^2}})^2.
\end{align}

\subsection{Veto on asymptotics}
\label{subsec:asymp}

Imposing the monotony constraint from App.~\ref{subsec:oscillations} removes most of the unphysical oscillations, but some solutions in which the fit clearly exploits the lack of data in certain regions to produce implausible charge distributions remain, especially for fit variants in which large values of $N$ are admitted. 
One could prevent such cases by simply restricting $R$ and $N$ to the range in which data are available, but we also considered another variant in which we impose a more stringent high-energy behavior. To this end, we estimate an upper limit on a reasonable asymptotic behavior of the form factor in the experimentally constrained region and exclude solutions that exceed this limit for higher energies. 
Assuming  the form factor  to decay asymptotically, we used the ansatz
\begin{align}
    |F_0^\text{ch}(q)| < A q^{-m},    
\end{align}
with $A>0$, $m>0$. Arguments from perturbative QCD~\cite{Lepage:1979zb,Lepage:1980fj} suggest $m\geq 4$, while determining  $A$ and $m$ by mapping this form onto the data-constrained part of the form factor tends to suggest
even larger values of $m$. 
To arrive at such estimates, we took
the maximal values between the last and second-to-last as well as second-to-last and third-to-last unambiguous minimum of the form factor to determine $A$ and $m$. In fact, as the explicit fits in App.~\ref{app:fit_strat} show, the real envelope tends to be even steeper. 
 As we have a variety of $R$ and $N$ pairs for each scenario at our disposal, another way to relax the constraint is to use only the worst limit from all fits within that scenario. In the end, such considerations become most relevant for fit variants that already show signs of overparameterization, but with reasonable constraints on the asymptotic behavior it is relatively straightforward to veto such cases, while minimizing the bias towards the selection of acceptable fits that form the basis for our uncertainty quantification.

\begin{figure}[t]
    
    \hfil
    \includegraphics[width=0.49\linewidth]{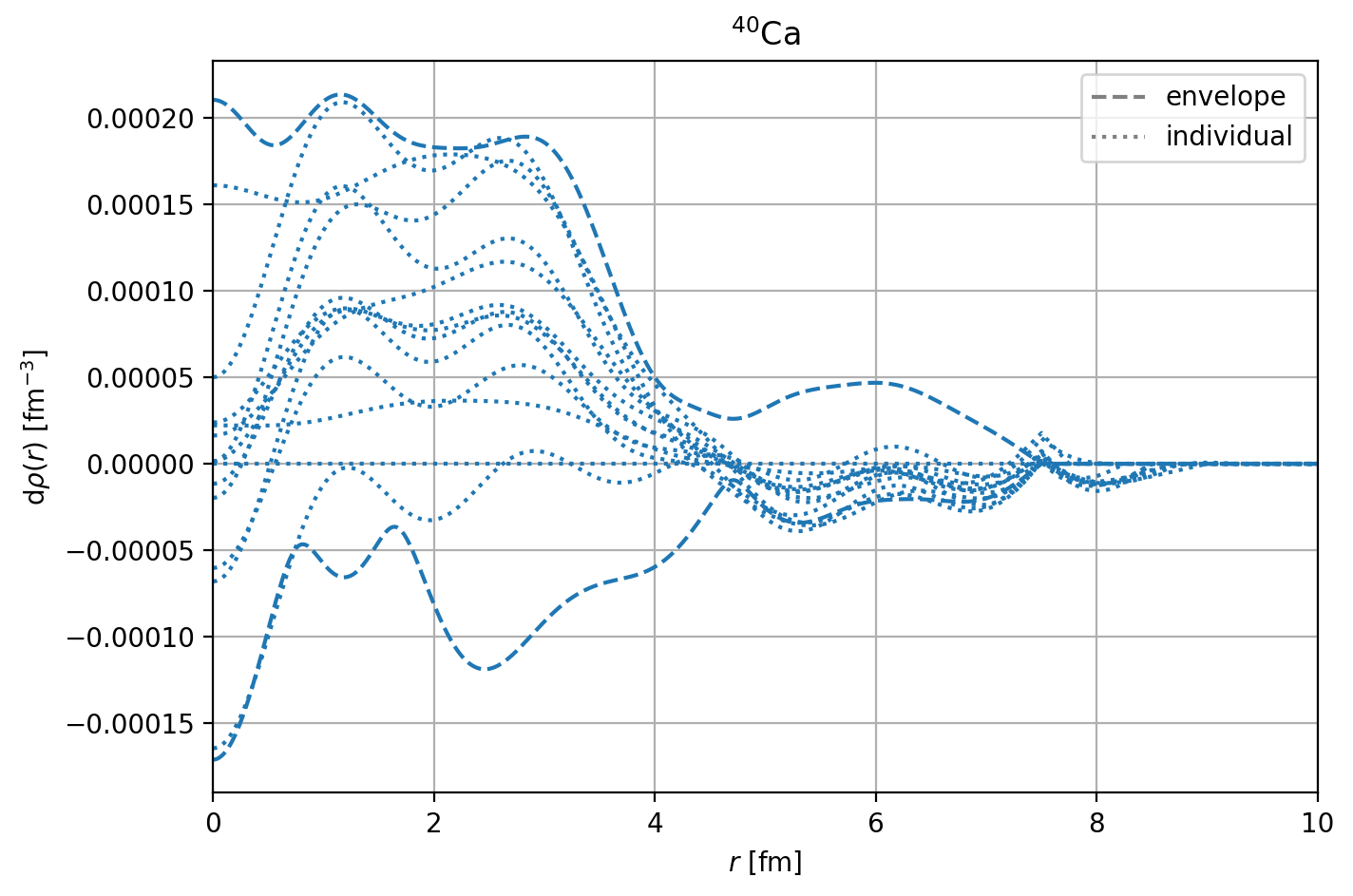}
    \hfil
    \hfil
    \includegraphics[width=0.49\linewidth]{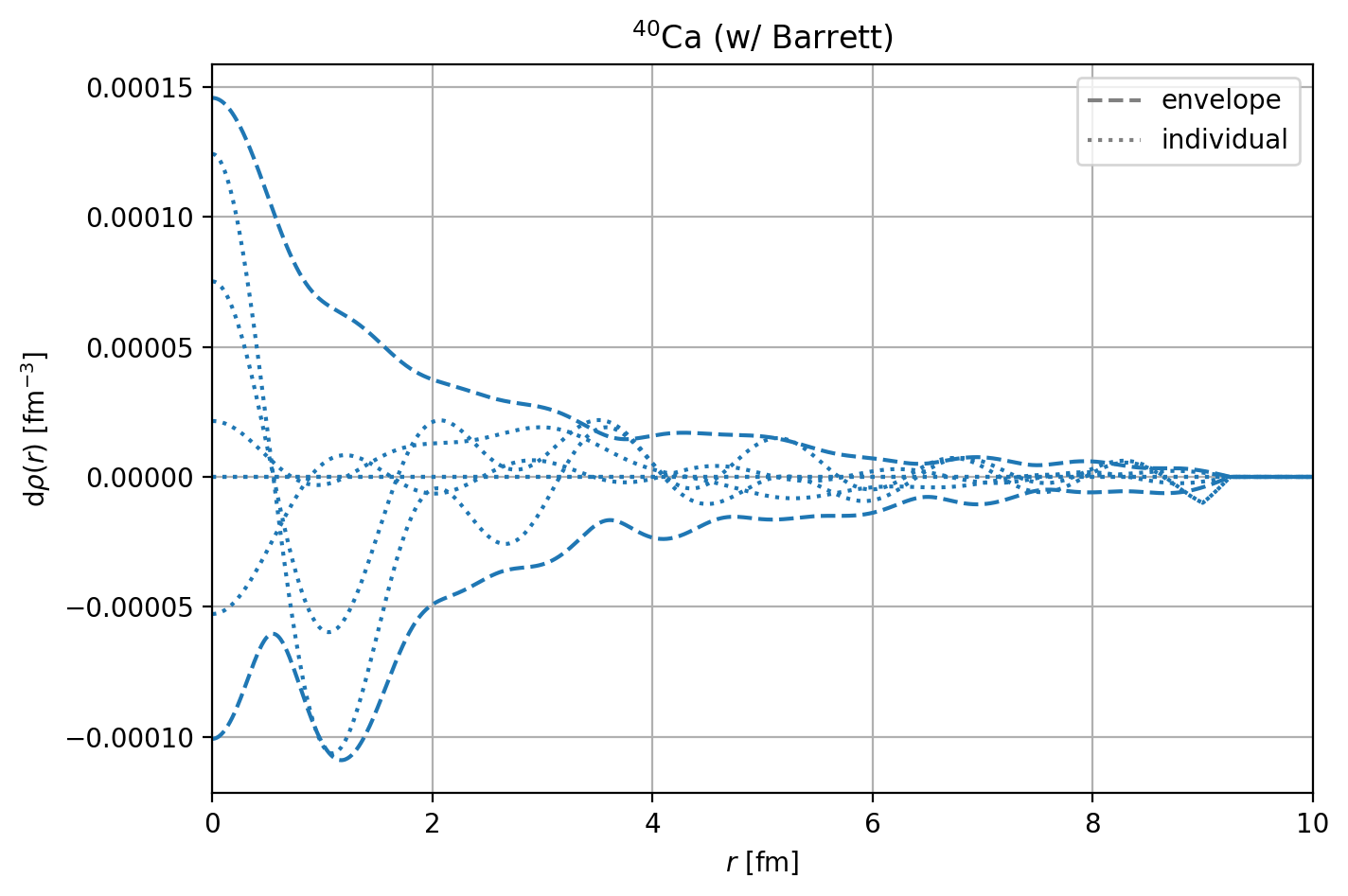}
    \hfil

    \hfil
    \includegraphics[width=0.49\linewidth]{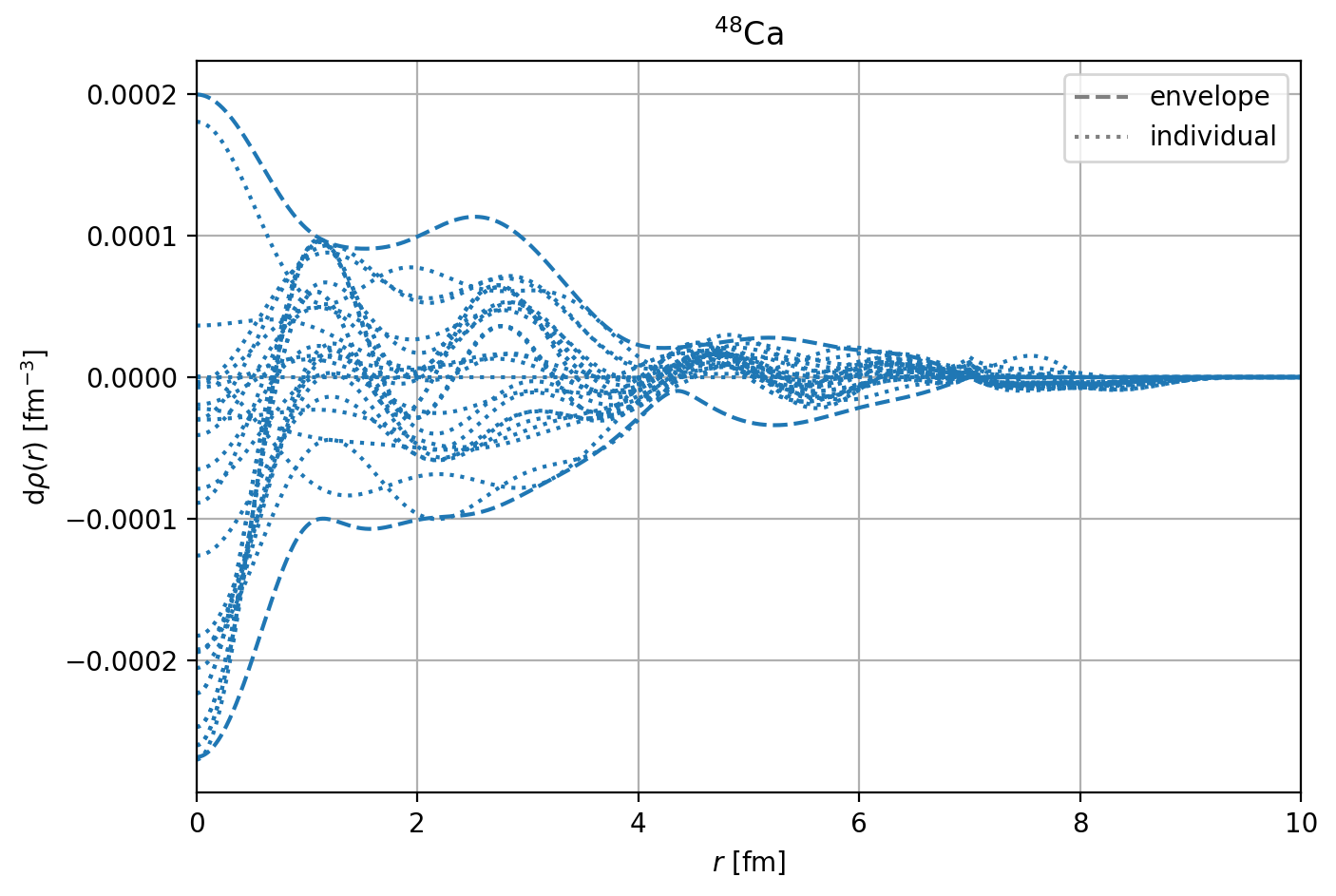}
    \hfil
    \hfil
    \includegraphics[width=0.49\linewidth]{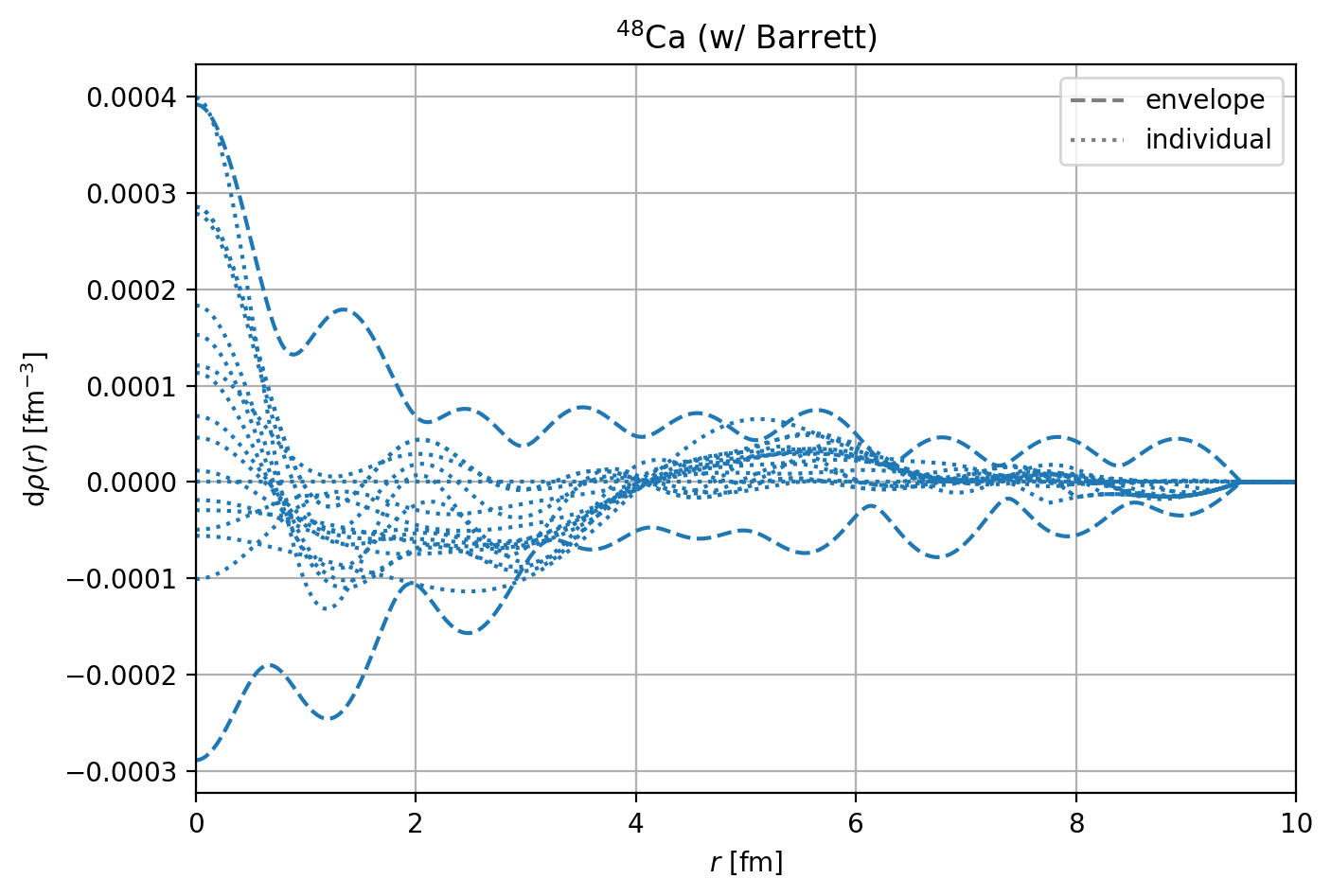}
    \hfil
    
    \caption{Deviations from central solution and fit uncertainty envelope for $^{40,48}$Ca.}
    \label{fig:Ca_drho_syst}
\end{figure}

\section{Fitting strategy and intermediate selection}
\label{app:fit_strat}

In this appendix we expand on the fitting strategy described in Sec.~\ref{sec:FB_fit}.
In addition to the procedure explained in the main text, with every set of fits we performed, we identified clear outliers (for example, cases in which $\chi^2(R,N+1)>\chi^2(R,N)$) and redid these fits with starting values based on adjacent solutions, to counteract the possibility of hitting local minima.
Moreover, since a lot of the data are likely to be affected by systematic uncertainties not fully reflected by the quoted uncertainties, we follow the standard procedure to inflate the fit errors by scale factors $S=\sqrt{\chi^2/\text{dof}}$. In practice, these scale factors mainly affect the fits to titanium and aluminum. In the documentation of the various fits, we provide the reduced $\chi^2$ and corresponding $p$-values, from which the cases that require an error inflation can be inferred.  
 
To assess the systematic uncertainties in our fits, we employ two main strategies. On the one hand, we parameterize the envelope of all acceptable fits again in terms of a FB series, and then propagate uncertainties for the quantities of interest such as the charge radii or overlap integrals. On the other hand, we can calculate these quantities of interest for all individual fits and then analyze the spread directly for the observable in question. The second strategy probably  represents the  systematic spread within this particular quantity more accurately, but requires one to keep track of a large set of parameterizations, necessitating the calculation of the observable for each one. The first strategy, therefore, is a lot more practical, requiring only a single parameterization. Furthermore, the implicit interpolation might represent the systematic uncertainty more accurately if the number of underlying fits is small. For the quantities considered in this work we employed both strategies, allowing one to compare the outcome to be able to assess the stability of the uncertainty quantification. For the calcium and aluminum fits we observe reasonable agreement between the two strategies, while for titanium in some cases the envelope strategy tends to overestimate the uncertainties.    

\begin{figure}[t]
    
    \hfil
    \includegraphics[width=0.45\linewidth]{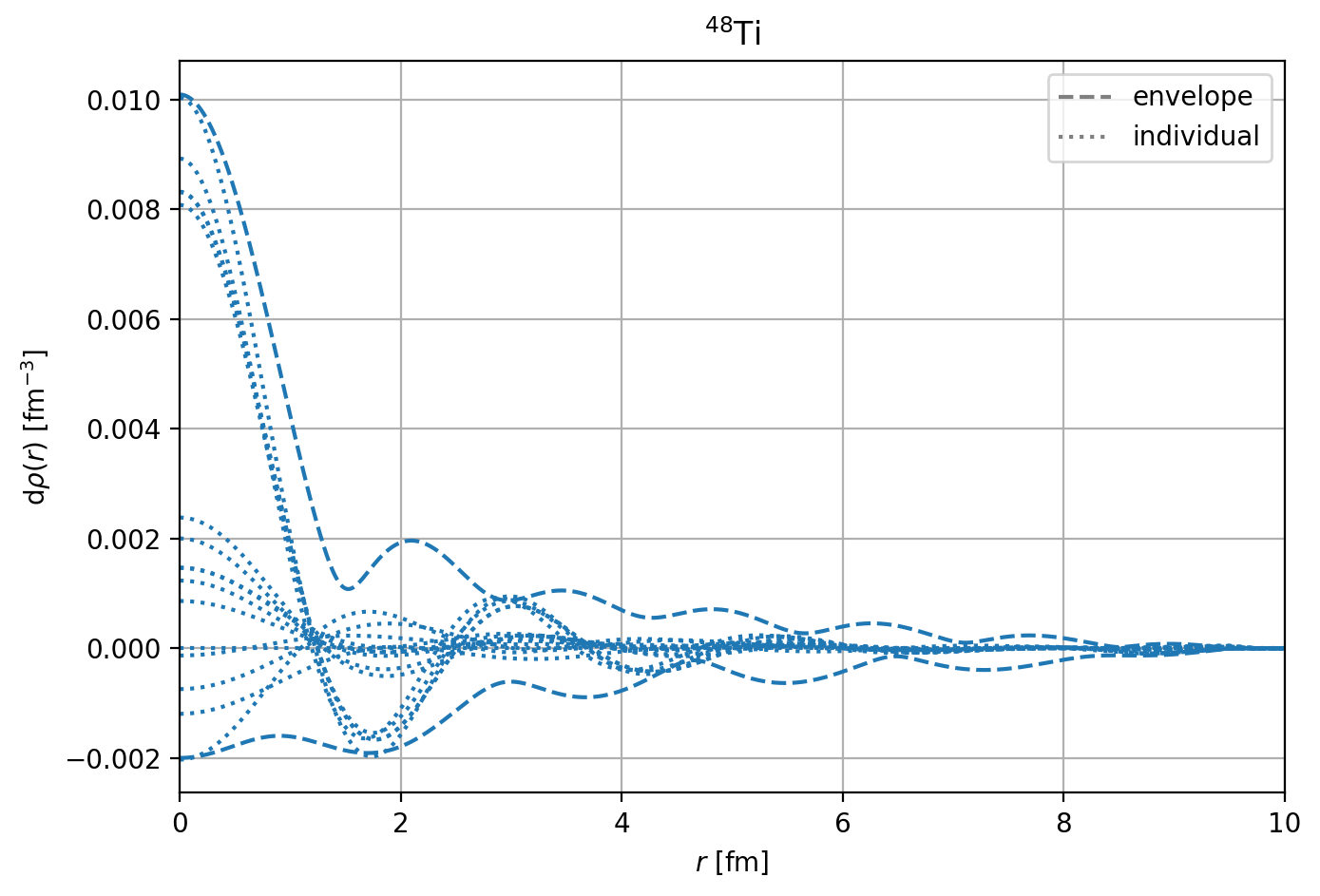}
    \hfil
    \hfil
    \includegraphics[width=0.45\linewidth]{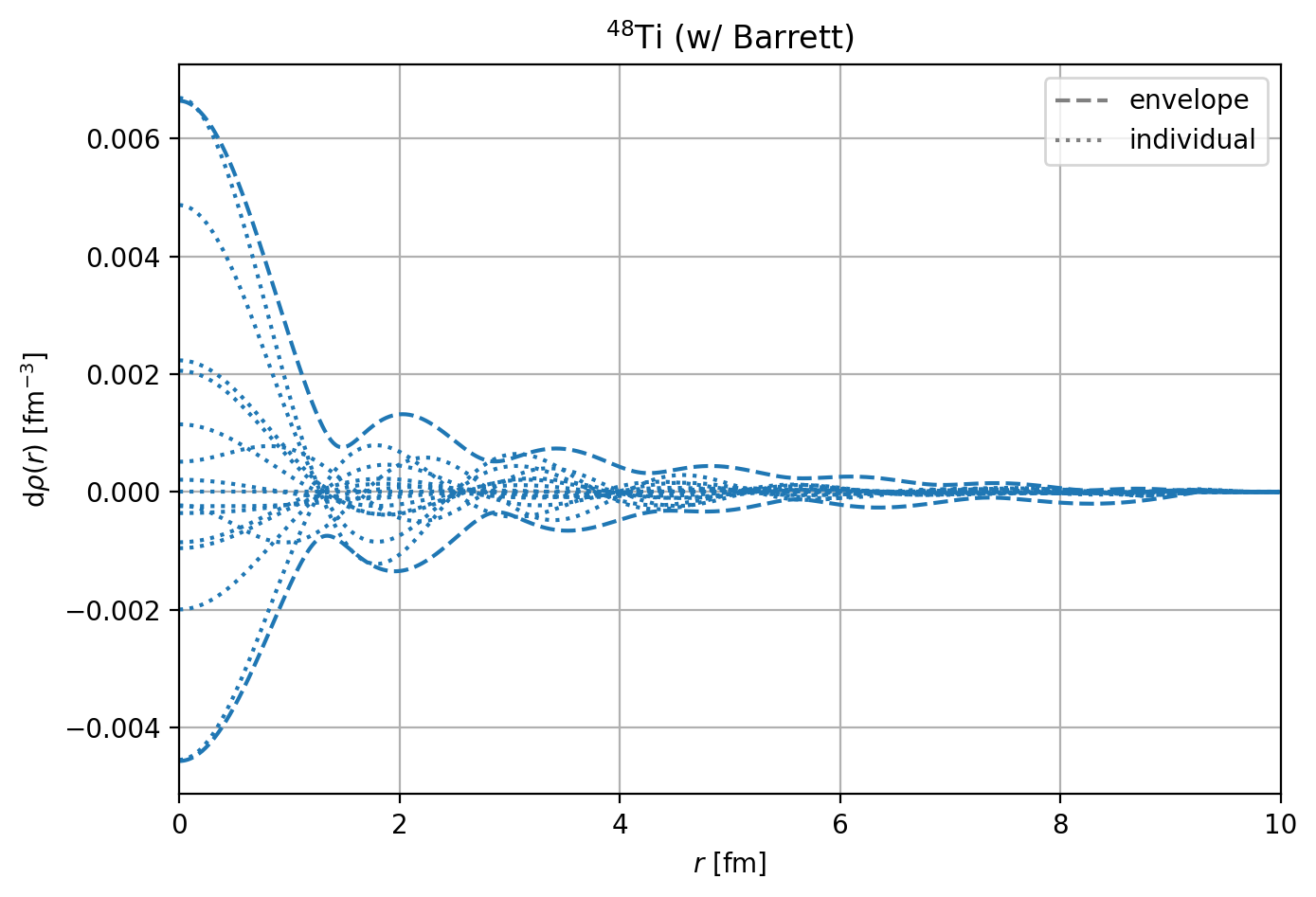}
    \hfil

    \hfil
    \includegraphics[width=0.45\linewidth]{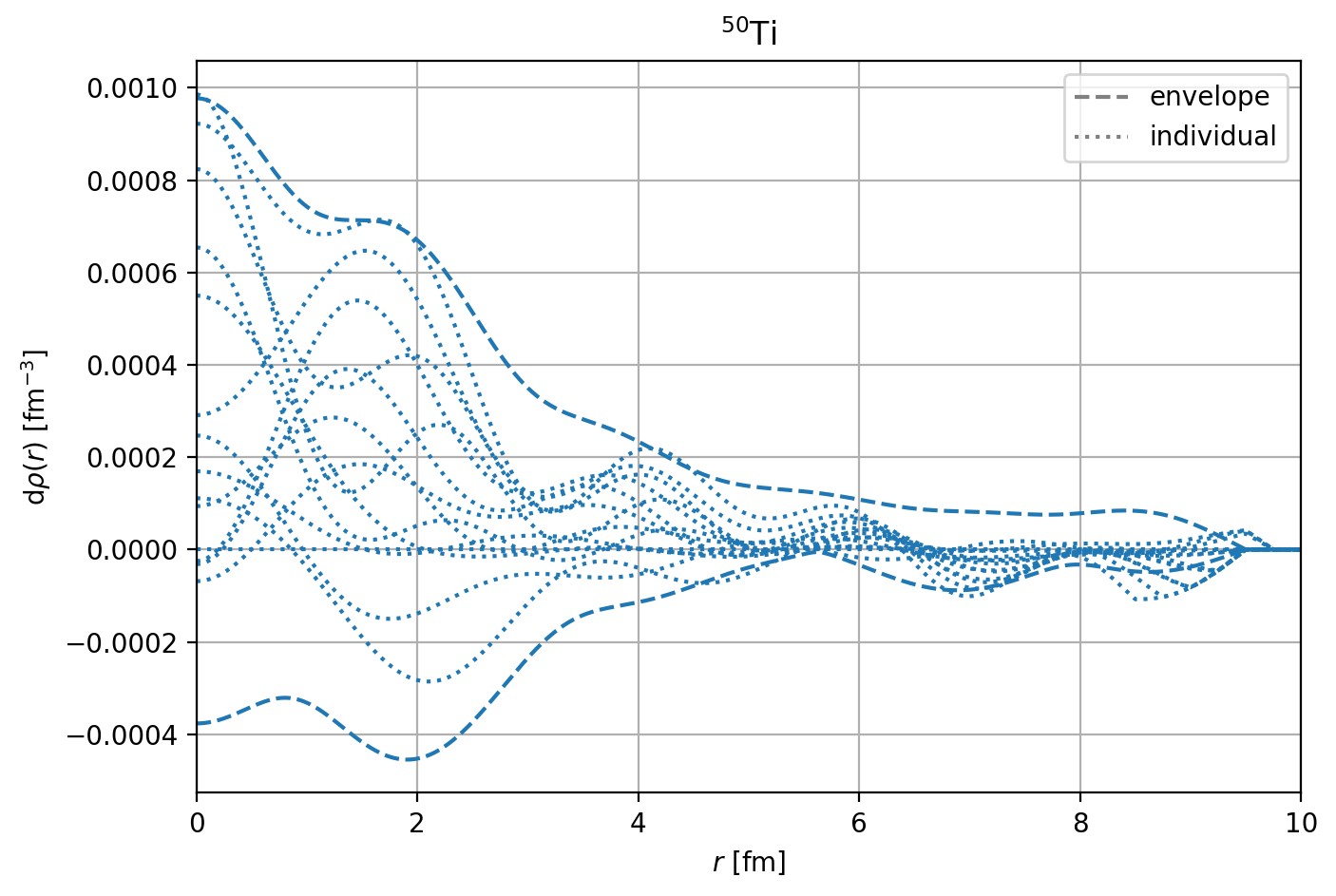}
    \hfil
    \hfil
    \includegraphics[width=0.45\linewidth]{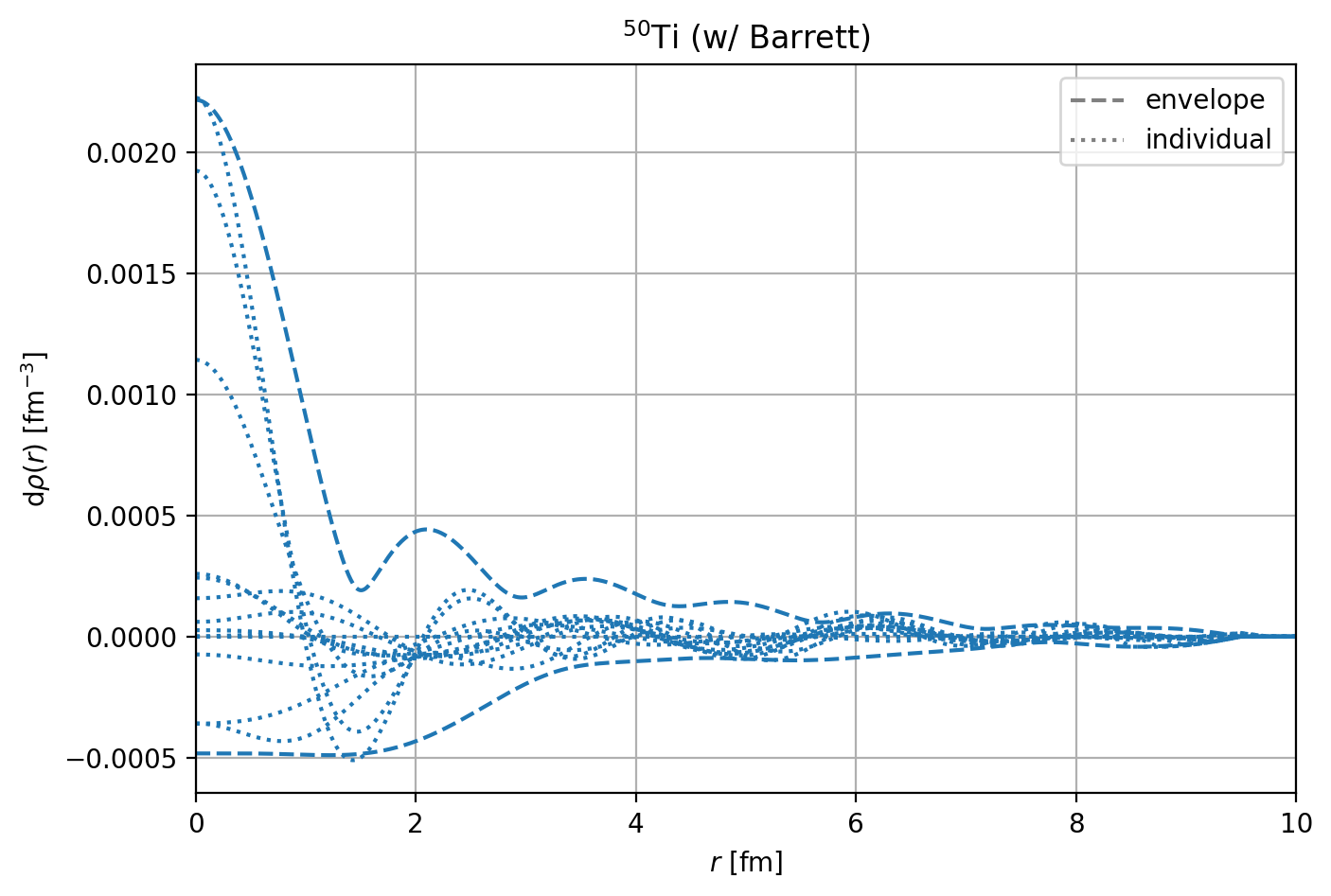}
    \hfil
    
    \caption{Deviations from central solution and fit uncertainty envelope for $^{48,50}$Ti.}
    \label{fig:Ti_drho_syst}
\end{figure}

\subsection{Fitting systematic uncertainty bands}
\label{subsec:bands}

\begin{figure}[t]
    
    \hfil
    \includegraphics[width=0.45\linewidth]{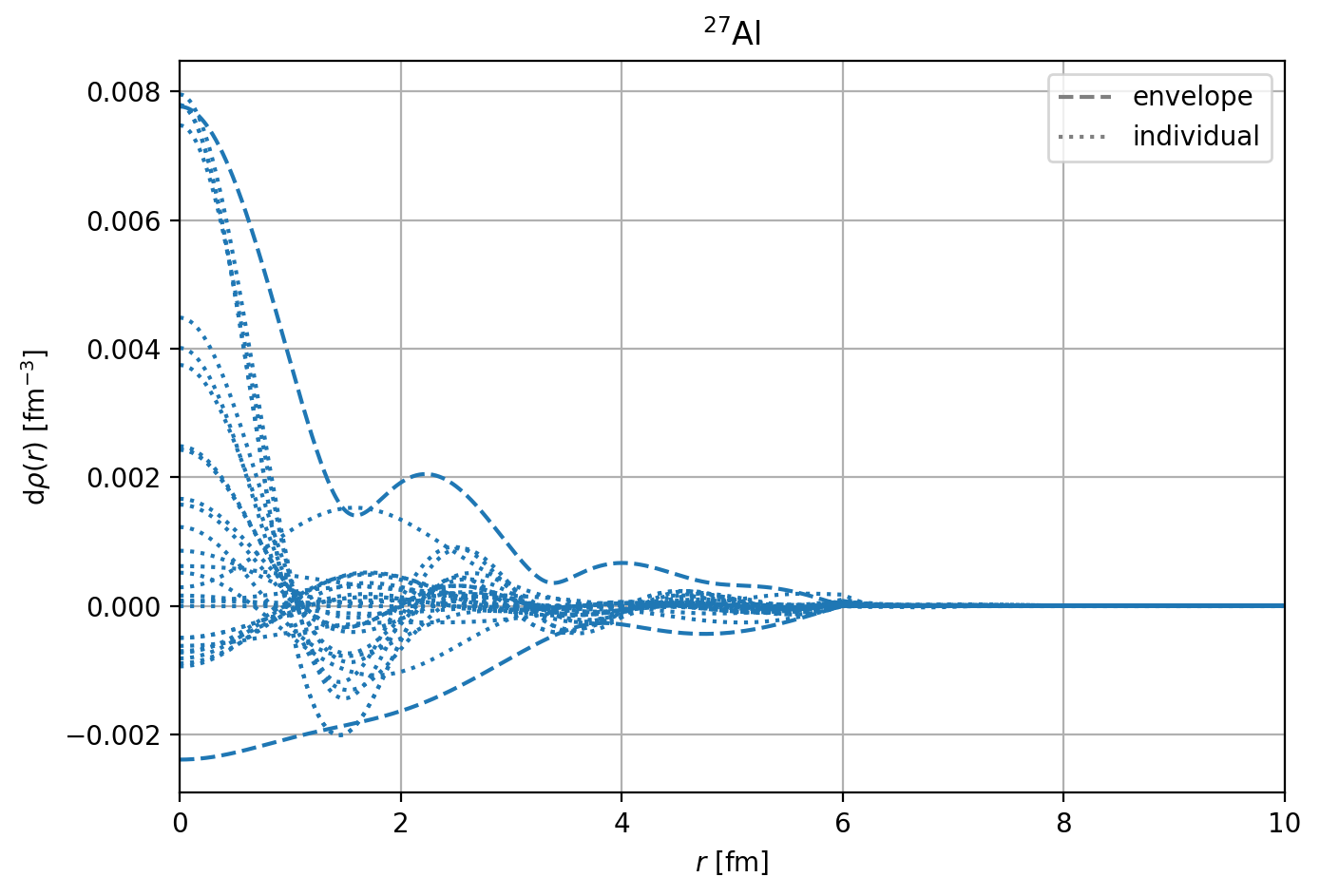}
    \hfil
    \hfil
    \includegraphics[width=0.45\linewidth]{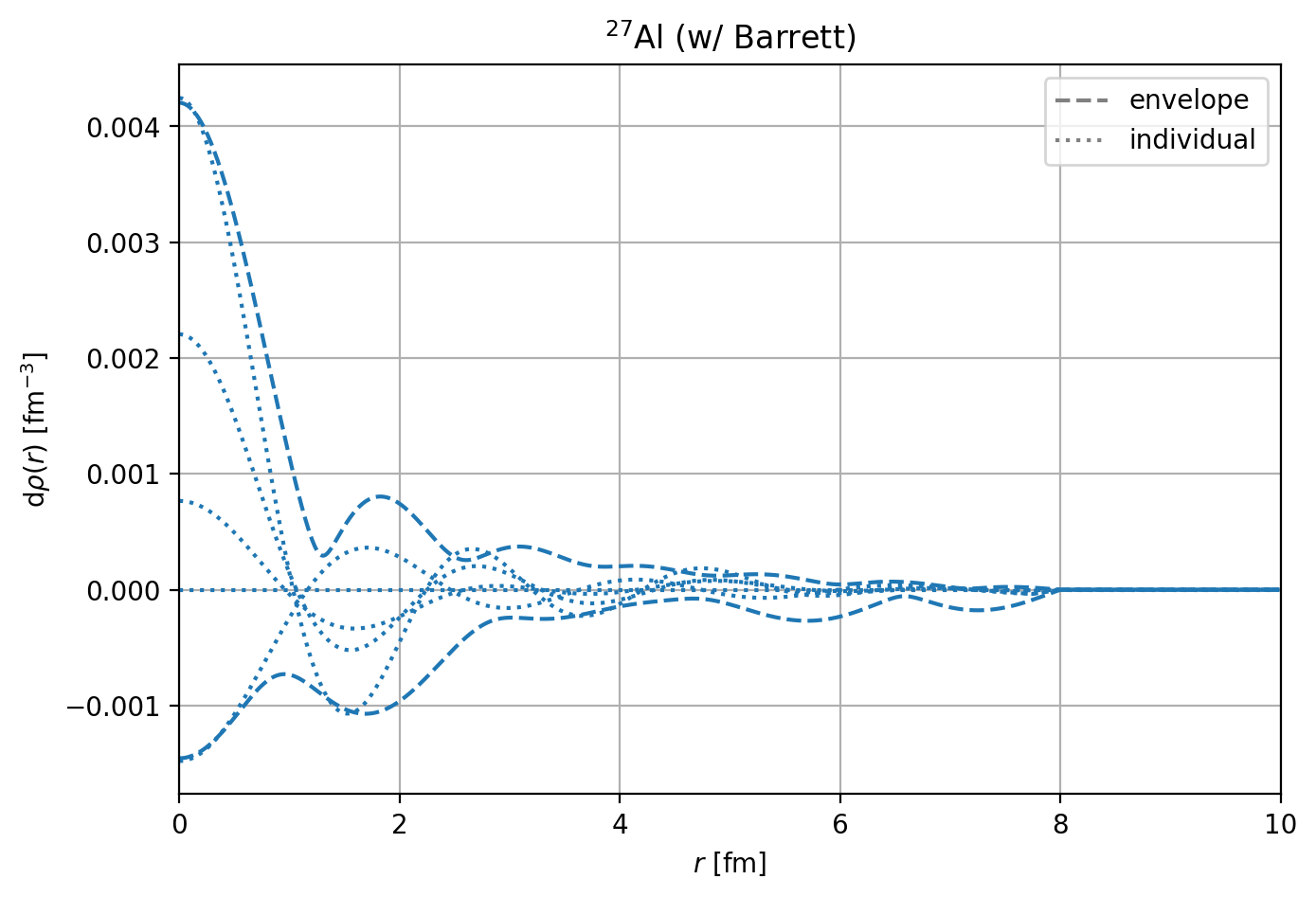}
    \hfil
    
    \caption{Deviations from central solution and fit uncertainty envelope for $^{27}$Al.}
    \label{fig:Al_drho_syst}
\end{figure}

To fit a systematic uncertainty band, we take a set of solutions for the charge density $\rho(r)$ and calculate the maximum and minimum at 100 values between $r=0$ and $r=10\,\text{fm}$. We call the upper/lower distance to the central value $\diff\rho^{\text{syst},\pm}_\text{data}$. To define a conservative systematic uncertainty error band, we aim at constructing an envelope around these limits, which is complicated by the fact  
that charge conservation should be fulfilled exactly. Accordingly, we choose to fit values of $\diff x^{\text{syst},\pm}_i$ that behave in the same way as the statistical uncertainties with the same correlations, and thus by construction fulfill charge conservation. Therefore, our model $\diff\rho^{\text{syst},\pm}_\text{model}$ amounts to the Gaussian uncertainty propagation of $\diff x^{\text{syst},\pm}_i$ using the correlations of the statistical uncertainties. 
Since we want to fit an outer envelope, we penalize $\diff\rho^{\text{syst},\pm}_\text{model}<\diff\rho^{\text{syst},\pm}_\text{data}$ and write our loss function as
\begin{align}
    \chi^2 = \qty(\diff\rho^{\text{syst},\pm}_\text{model}-\diff\rho^{\text{syst},\pm}_\text{data})^2 \times \qty(1 + \omega^2 \times \theta\qty(\diff\rho^{\text{syst},\pm}_\text{data}-\diff\rho^{\text{syst},\pm}_\text{model})),
\end{align}
which penalizes fits lying inside the uncertainty band a lot stronger than those outside. We find that ${\omega=10}$ produces satisfactory fit results. Note that we are not able to describe any systematic effects for ${r>R}$, as by construction the charge density and also any propagated uncertainty goes to zero at ${r=R}$. On the other hand, we tend to overestimate systematic effects for $r\leq R$, as the envelope constructed in this way encompasses more parameter space than spanned by the individual fits. Finally, we round the values of $\diff x^{\text{syst},\pm}_i$ to their given uncertainty as extracted from the fit. We show the resulting envelope parameterizations of $\diff\rho^{\text{syst},\pm}_\text{data}$ for all nuclei in Figs.~\ref{fig:Ca_drho_syst}--\ref{fig:Al_drho_syst} with all the individual deviations enclosed. In particular for $^{48}$Ca, without inclusion of the Barrett moment, the envelope sits very tight around the contributing solutions, while with inclusion of the Barrett moment the fit is forced to (strongly) overestimate the uncertainties. In general, we find that this description of the systematic errors gives a fair representation, with a tendency to overestimate the uncertainty in some cases, see also the discussion in Sec.~\ref{sec:charge_radii}.

\begin{figure}[t]

    \hfil
    \includegraphics[width=0.49\textwidth]{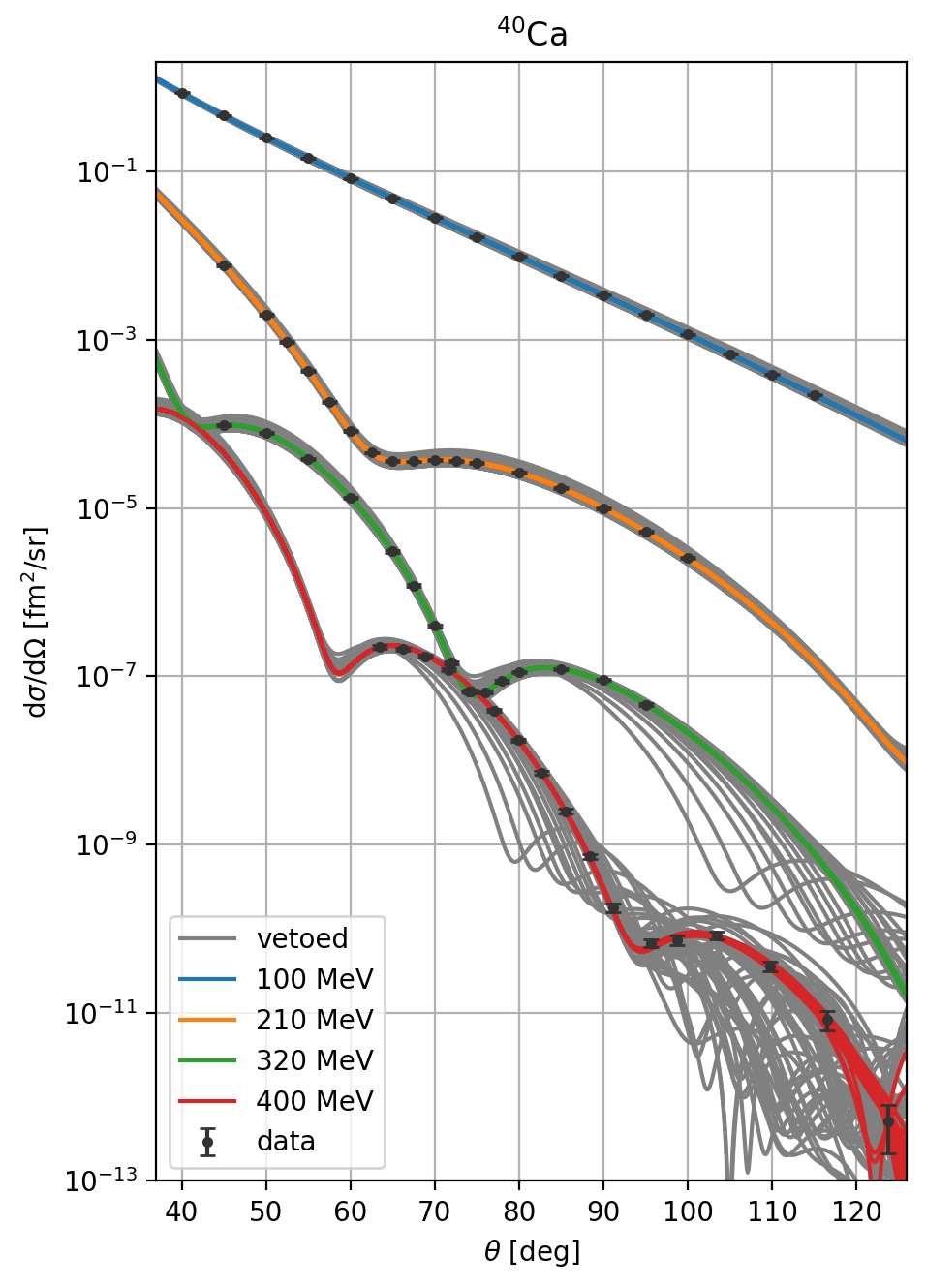}
    \hfil
    \includegraphics[width=0.49\textwidth]{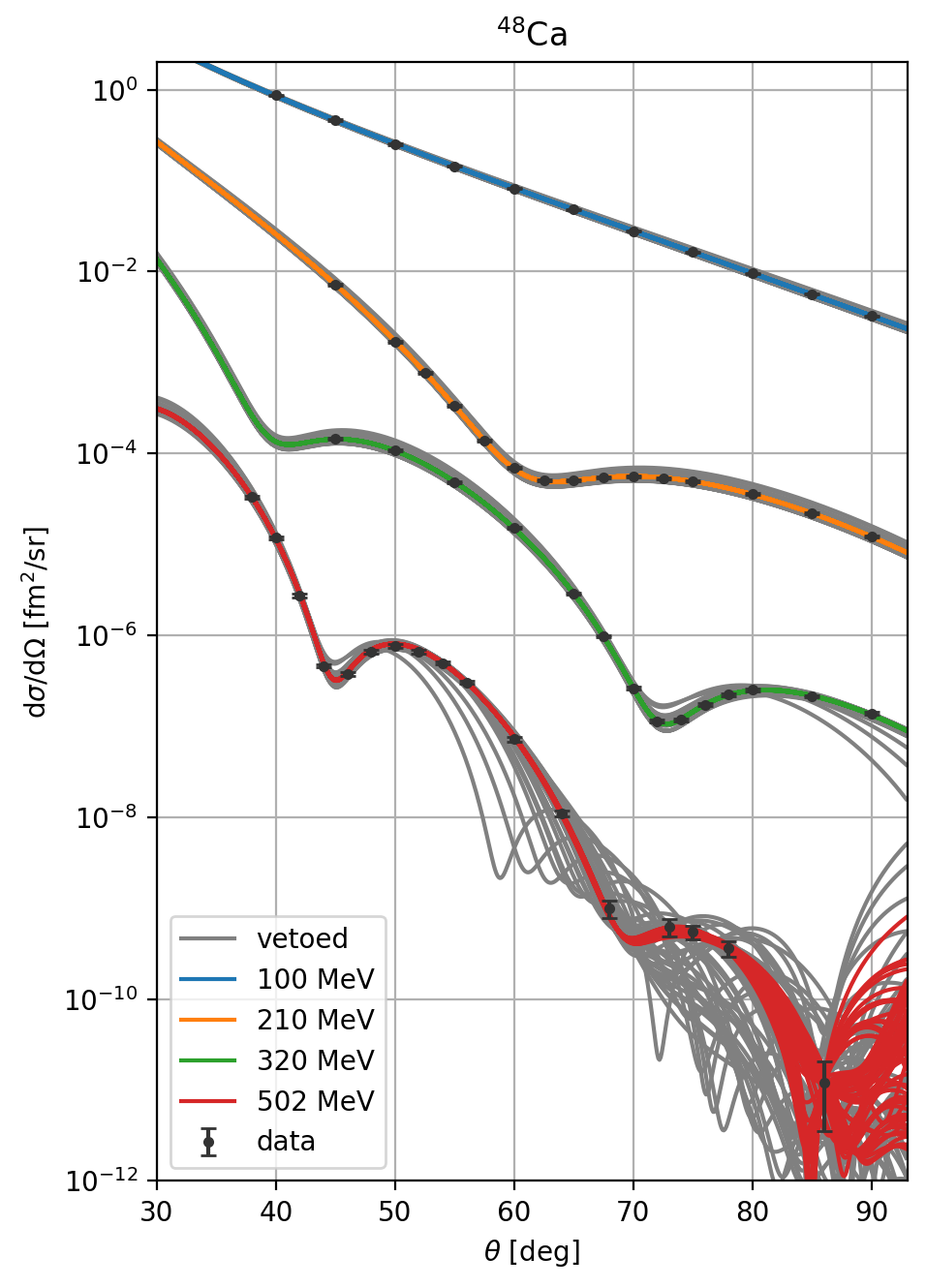}
    \hfil
    
    \hfil
    \resizebox{!}{0.145\textheight}{%
    \renewcommand{\arraystretch}{1.3}
    \begin{tabular}{ccccccccccccc}
        \toprule
        \multicolumn{13}{c}{$^{40}$Ca}\\ \midrule
        $R$\textbackslash{{}}$N_x$ & 3 &  4 &  5 &  6 &  7 &  8 &  9 & 10 & 11 & 12 & 13 & 14 \\ \midrule
        6.00 & {\color{gray} 125.7 } & {\color{gray} 27.4 } & {\color{gray} 25.8 } & {\color{gray} 20.8 } & {\color{gray} 8.3 } & {\color{gray} 6.1 } & {\color{gray} 5.1 } &  &  &  &  &  \\
        6.25 &  & {\color{gray} 12.0 } & {\color{gray} 9.9 } & {\color{gray} 9.5 } & {\color{gray} 5.1 } & {\color{gray} 2.1 } & {\color{gray} 1.8 } & {\color{gray} 1.7 } &  &  &  &  \\
        6.50 &  & {\color{gray} 39.6 } & {\color{gray} 6.8 } & {\color{gray} 4.5 } & {\color{gray} 3.3 } & {\color{gray} 1.8 } & {\color{gray} 1.8 } & 1.2 &  &  &  &  \\
        6.75 &  & {\color{gray} 62.0 } & {\color{gray} 4.0 } & {\color{gray} 2.7 } & {\color{gray} 2.3 } & 1.2 & 1.1 & 1.1 &  &  &  &  \\
        7.00 &  & {\color{gray} 72.0 } & {\color{gray} 4.1 } & {\color{gray} 2.3 } & {\color{gray} 1.6 } & 1.1 & 1.1 & 1.1 &  &  &  &  \\
        7.25 &  &  & {\color{gray} 2.7 } & {\color{gray} 2.7 } & {\color{gray} 1.3 } & 1.1 & 1.1 & 1.1 & 1.1 &  &  &  \\
        7.50 &  &  & {\color{gray} 13.1 } & {\color{gray} 3.6 } & 1.1 & 1.1 & 1.1 & 1.1 & 1.1 &  &  &  \\
        7.75 &  &  & {\color{gray} 29.8 } & {\color{gray} 4.5 } & {\color{gray} 1.2 } & 1.1 & 1.1 & 1.1 & 1.1 &  &  &  \\
        8.00 &  &  &  & {\color{gray} 4.1 } & {\color{gray} 1.8 } & {\color{gray} 1.2 } & 1.1 & 1.1 & 1.1 & 1.1 &  &  \\
        8.25 &  &  &  & {\color{gray} 2.8 } & {\color{gray} 2.8 } & 1.2 & 1.1 & 1.1 & 1.1 & 1.1 &  &  \\
        8.50 &  &  &  & {\color{gray} 8.7 } & {\color{gray} 3.7 } & 1.1 & 1.1 & 1.1 & 1.1 & 1.1 &  &  \\
        8.75 &  &  &  & {\color{gray} 19.6 } & {\color{gray} 4.5 } & {\color{gray} 1.3 } & 1.1 & 1.1 & 1.1 & 1.1 &  &  \\
        9.00 &  &  &  &  & {\color{gray} 4.1 } & {\color{gray} 1.9 } & 1.1 & 1.1 & 1.1 & 1.1 & 1.2 &  \\
        9.25 &  &  &  &  & {\color{gray} 2.8 } & {\color{gray} 2.7 } & 1.1 & 1.1 & 1.2 & 1.1 & 1.1 &  \\
        9.50 &  &  &  &  & {\color{gray} 6.6 } & {\color{gray} 3.7 } & {\color{gray} 1.2 } & {\color{gray} 1.2 } & {\color{gray} 1.2 } & 1.1 & 1.2 &  \\
        9.75 &  &  &  &  &  & {\color{gray} 11.8 } & {\color{gray} 8.0 } & {\color{gray} 8.1 } & {\color{gray} 8.0 } & {\color{gray} 7.6 } & {\color{gray} 7.6 } & {\color{gray} 7.7 } \\%
        \bottomrule
    \end{tabular}%
    \renewcommand{\arraystretch}{1.0}}%
    \hfil
    \resizebox{!}{0.145\textheight}{%
    \renewcommand{\arraystretch}{1.3}
    \begin{tabular}{ccccccccccccc}
        \toprule
        \multicolumn{12}{c}{$^{48}$Ca}\\ \midrule
        $R$\textbackslash{{}}$N_x$ & 3 &  4 &  5 &  6 &  7 &  8 &  9 & 10 & 11 & 12 & 13 \\ \midrule
        6.00 & {\color{gray} 74.9 } & {\color{gray} 13.0 } & {\color{gray} 12.1 } & {\color{gray} 5.5 } & {\color{gray} 2.2 } & {\color{gray} 1.8 } & {\color{gray} 1.6 } &  &  &  &  \\
        6.25 & {\color{gray} 118.2 } & {\color{gray} 3.6 } & {\color{gray} 2.8 } & {\color{gray} 2.2 } & {\color{gray} 1.4 } & {\color{gray} 1.3 } & {\color{gray} 1.3 } &  &  &  &  \\
        6.50 &  & {\color{gray} 1.4 } & {\color{gray} 1.3 } & {\color{gray} 1.3 } & {\color{gray} 1.2 } & 1.2 & 1.2 & 1.2 &  &  &  \\
        6.75 &  & {\color{gray} 6.9 } & {\color{gray} 2.0 } & 1.2 & 1.2 & 1.2 & 1.2 & 1.2 &  &  &  \\
        7.00 &  & {\color{gray} 21.0 } & {\color{gray} 2.3 } & 1.1 & 1.2 & 1.2 & 1.2 & 1.2 &  &  &  \\
        7.25 &  &  & {\color{gray} 1.9 } & {\color{gray} 1.2 } & 1.2 & 1.2 & 1.2 & 1.2 & 1.2 &  &  \\
        7.50 &  &  & {\color{gray} 1.5 } & {\color{gray} 1.5 } & 1.2 & 1.2 & 1.2 & 1.2 & 1.2 &  &  \\
        7.75 &  &  & {\color{gray} 3.4 } & {\color{gray} 2.1 } & 1.2 & 1.2 & 1.2 & 1.2 & 1.2 &  &  \\
        8.00 &  &  & {\color{gray} 8.7 } & {\color{gray} 2.2 } & 1.2 & 1.2 & 1.2 & 1.1 & 1.2 &  &  \\
        8.25 &  &  &  & {\color{gray} 2.0 } & {\color{gray} 1.3 } & 1.2 & 1.1 & 1.1 & 1.1 & 1.2 &  \\
        8.50 &  &  &  & {\color{gray} 1.9 } & {\color{gray} 1.8 } & 1.1 & 1.1 & 1.1 & 1.1 & 1.2 &  \\
        8.75 &  &  &  & {\color{gray} 2.9 } & {\color{gray} 2.2 } & 1.1 & 1.1 & 1.1 & 1.1 & 1.2 &  \\
        9.00 &  &  &  & {\color{gray} 5.6 } & {\color{gray} 2.3 } & 1.1 & 1.1 & 1.1 & 1.1 & 1.2 &  \\
        9.25 &  &  &  &  & {\color{gray} 2.0 } & {\color{gray} 1.3 } & 1.2 & 1.2 & 1.1 & 1.2 & 1.2 \\
        9.50 &  &  &  &  & {\color{gray} 1.9 } & {\color{gray} 1.8 } & 1.1 & 1.2 & 1.1 & 1.2 & 1.2 \\
        9.75 &  &  &  &  & {\color{gray} 7.3 } & {\color{gray} 6.8 } & {\color{gray} 5.6 } & {\color{gray} 5.6 } & {\color{gray} 5.5 } & {\color{gray} 5.5 } & {\color{gray} 5.6 } \\
        \bottomrule
    \end{tabular}%
    \renewcommand{\arraystretch}{1.0}}%
    \hfil
    \caption{Steps \ref{item:base} and \ref{item:sen} of the fits for ${}^{40}$Ca and ${}^{48}$Ca. The plots show the cross-section data and the fit solutions, while the accompanying tables show the $\chi^2/\text{dof}$ ($\text{dof}=65-N_x$). Solutions in gray are excluded for ${p_\text{val}<15\%}$ for ${}^{40}$Ca and ${p_\text{val}<12\%}$ for ${}^{48}$Ca.}
    \label{tab:Ca4048BASE}
\end{figure}


\begin{figure}[t]

    \hfil
    \includegraphics[width=0.49\textwidth]{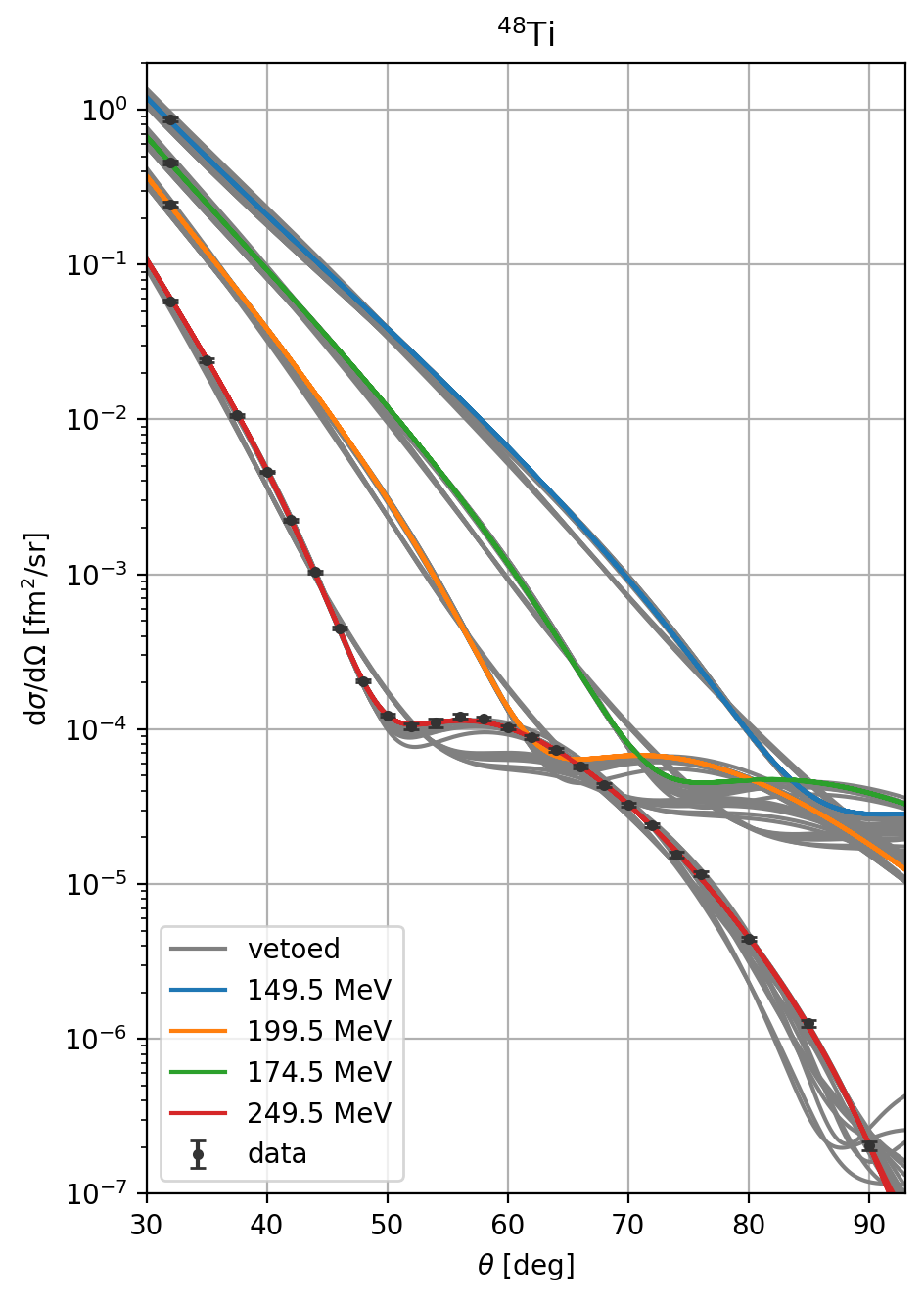}
    \hfil
    \includegraphics[width=0.49\textwidth]{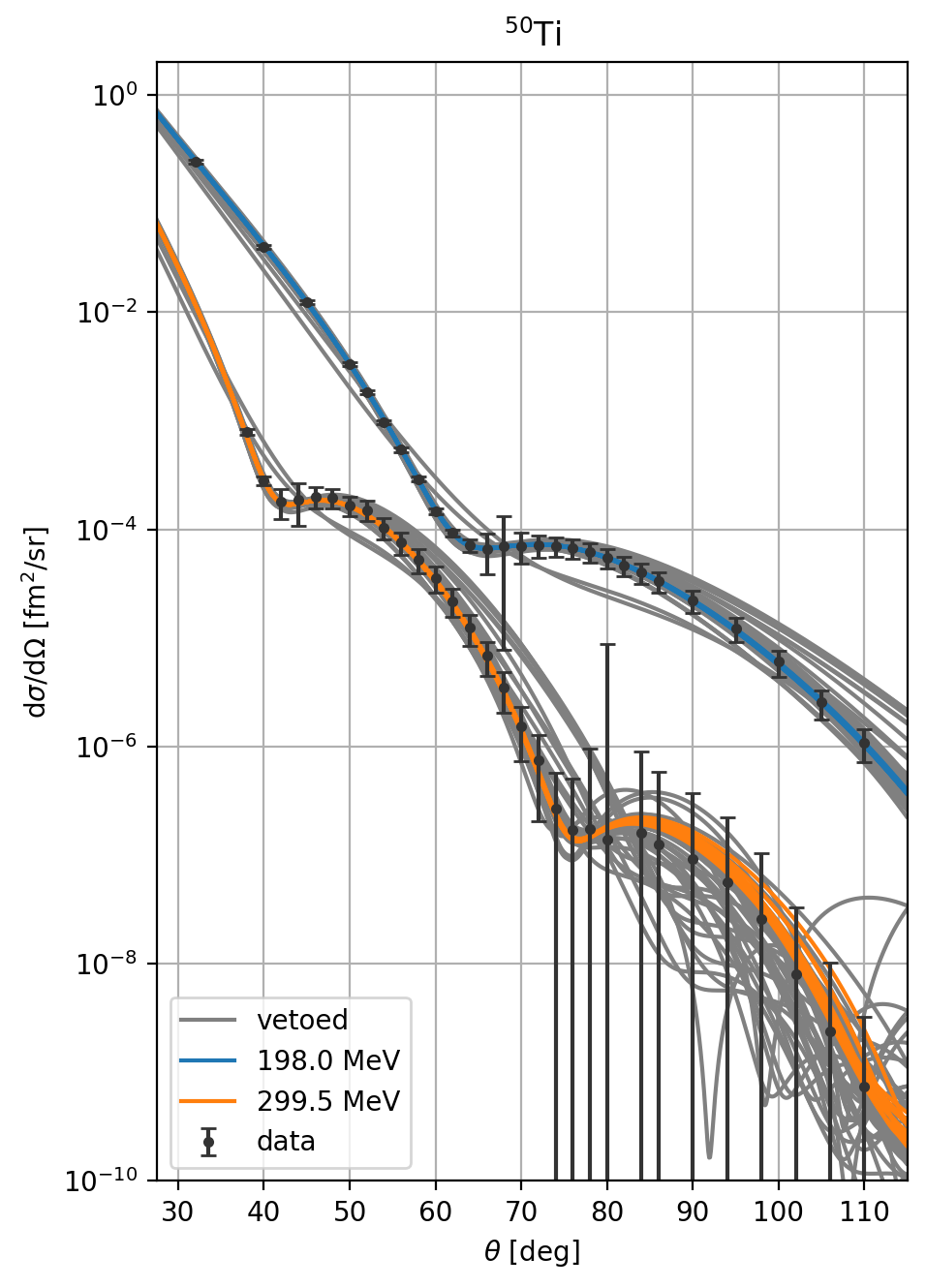}
    \hfil
    
    \hfil
    \resizebox{!}{0.145\textheight}{%
    \renewcommand{\arraystretch}{1.3}
    \begin{tabular}{cccccccc}
        \toprule
        \multicolumn{8}{c}{$^{48}$Ti}\\ \midrule
        $R$\textbackslash{{}}$N_x$ & 3 &  4 &  5 &  6 &  7 &  8 &  9 \\ \midrule
        8.50 & {\color{gray} 119.7 } & {\color{gray} 117.5 } & {\color{gray} 102.2 } & {\color{gray} 99.8 } & {\color{gray} 101.8 } & {\color{gray} 105.9 } &  \\
        8.75 & {\color{gray} 32.2 } & {\color{gray} 1.8 } & {\color{gray} 1.7 } & 1.4 & 1.4 & 1.5 &  \\
        9.00 &  & {\color{gray} 1.9 } & {\color{gray} 1.7 } & 1.4 & 1.5 & 1.5 & 1.6 \\
        9.25 &  & {\color{gray} 1.8 } & {\color{gray} 1.7 } & 1.4 & 1.5 & 1.5 & 1.6 \\
        9.50 &  & {\color{gray} 1.7 } & {\color{gray} 1.7 } & 1.4 & 1.5 & 1.5 & 1.6 \\
        9.75 &  & {\color{gray} 2.6 } & {\color{gray} 1.8 } & 1.5 & 1.5 & 1.5 & 1.6 \\
        10.00 &  & {\color{gray} 7.9 } & {\color{gray} 2.6 } & {\color{gray} 2.0 } & {\color{gray} 2.0 } & {\color{gray} 2.1 } & {\color{gray} 2.2 } \\
        10.25 &  & {\color{gray} 37.4 } & {\color{gray} 18.0 } & {\color{gray} 16.6 } & {\color{gray} 17.2 } & {\color{gray} 18.0 } & {\color{gray} 18.8 } \\
        \bottomrule 
    \end{tabular}%
    \renewcommand{\arraystretch}{1.0}}%
    \hfil
    \resizebox{!}{0.145\textheight}{%
    \renewcommand{\arraystretch}{1.3}
    \begin{tabular}{ccccccccc}
        \toprule
        \multicolumn{9}{c}{$^{50}$Ti}\\ \midrule
        $R$\textbackslash{{}}$N_x$ & 2 &  3 &  4 &  5 &  6 &  7 &  8 &  9 \\ \midrule
        6.25 & {\color{gray} 4.9 } & {\color{gray} 3.6 } & {\color{gray} 2.6 } & {\color{gray} 1.9 } & {\color{gray} 1.7 } &  &  &  \\
        6.50 &  &  & {\color{gray} 110.7 } & {\color{gray} 56.6 } &  &  &  &  \\
        6.75 &  & {\color{gray} 9.7 } & {\color{gray} 1.8 } & {\color{gray} 1.8 } & {\color{gray} 1.7 } & {\color{gray} 1.7 } &  &  \\
        7.00 &  & {\color{gray} 6.7 } & {\color{gray} 1.7 } & {\color{gray} 1.7 } & {\color{gray} 1.7 } & {\color{gray} 1.7 } &  &  \\
        7.25 &  & {\color{gray} 4.9 } & {\color{gray} 1.6 } & {\color{gray} 1.7 } & {\color{gray} 1.7 } & {\color{gray} 1.7 } &  &  \\
        7.50 &  & {\color{gray} 4.2 } & {\color{gray} 2.2 } & {\color{gray} 1.6 } & {\color{gray} 1.6 } & {\color{gray} 1.7 } &  &  \\
        7.75 &  &  & {\color{gray} 5.4 } & {\color{gray} 1.6 } & {\color{gray} 1.6 } & {\color{gray} 1.7 } & {\color{gray} 1.7 } &  \\
        8.00 &  &  & {\color{gray} 7.7 } & {\color{gray} 1.6 } & {\color{gray} 1.6 } & {\color{gray} 1.7 } & {\color{gray} 1.6 } &  \\
        8.25 &  &  & {\color{gray} 8.9 } & {\color{gray} 1.6 } & {\color{gray} 4.2 } & {\color{gray} 1.6 } & {\color{gray} 1.6 } &  \\
        8.50 &  &  & {\color{gray} 6.1 } & {\color{gray} 1.7 } & {\color{gray} 1.6 } & {\color{gray} 1.6 } & 1.6 &  \\
        8.75 &  &  & {\color{gray} 5.7 } & {\color{gray} 2.4 } & {\color{gray} 1.6 } & {\color{gray} 1.7 } & 1.5 &  \\
        9.00 &  &  &  & {\color{gray} 4.7 } & {\color{gray} 1.6 } &  & 1.5 & 1.5 \\
        9.25 &  &  &  & {\color{gray} 7.0 } & {\color{gray} 1.6 } & 1.5 & 1.5 & 1.5 \\
        9.50 &  &  &  & {\color{gray} 8.3 } & 1.5 & 1.5 & 1.5 & 1.5 \\
        9.75 &  &  &  & {\color{gray} 11.3 } & {\color{gray} 6.4 } & 1.5 & 1.5 & 1.5 \\
        10.00 &  &  &  & {\color{gray} 11.0 } & {\color{gray} 5.4 } & {\color{gray} 1.8 } &  &  \\
        10.25 &  &  &  &  & {\color{gray} 8.2 } & {\color{gray} 6.6 } & {\color{gray} 5.0 } &  \\
        \bottomrule 
    \end{tabular}%
    \renewcommand{\arraystretch}{1.0}}%
    \hfil
    
    \caption{Steps \ref{item:base} and \ref{item:sen} of the fits for ${}^{48}$Ti and ${}^{50}$Ti. The plots show the cross-section data and the fit solutions, while the accompanying tables show the $\chi^2/\text{dof}$ ($\text{dof}=28-N_x$ for ${}^{48}$Ti, $\text{dof}=57-N_x$ for ${}^{50}$Ti). Solutions in gray are excluded for $p_\text{val}<3\%$ for ${}^{48}$Ti and $p_\text{val}<0.5\%$ for ${}^{50}$Ti.}
    \label{tab:Ti4850BASE}
\end{figure}


\begin{figure}[t]

    \hfil
    \includegraphics[width=0.49\textwidth]{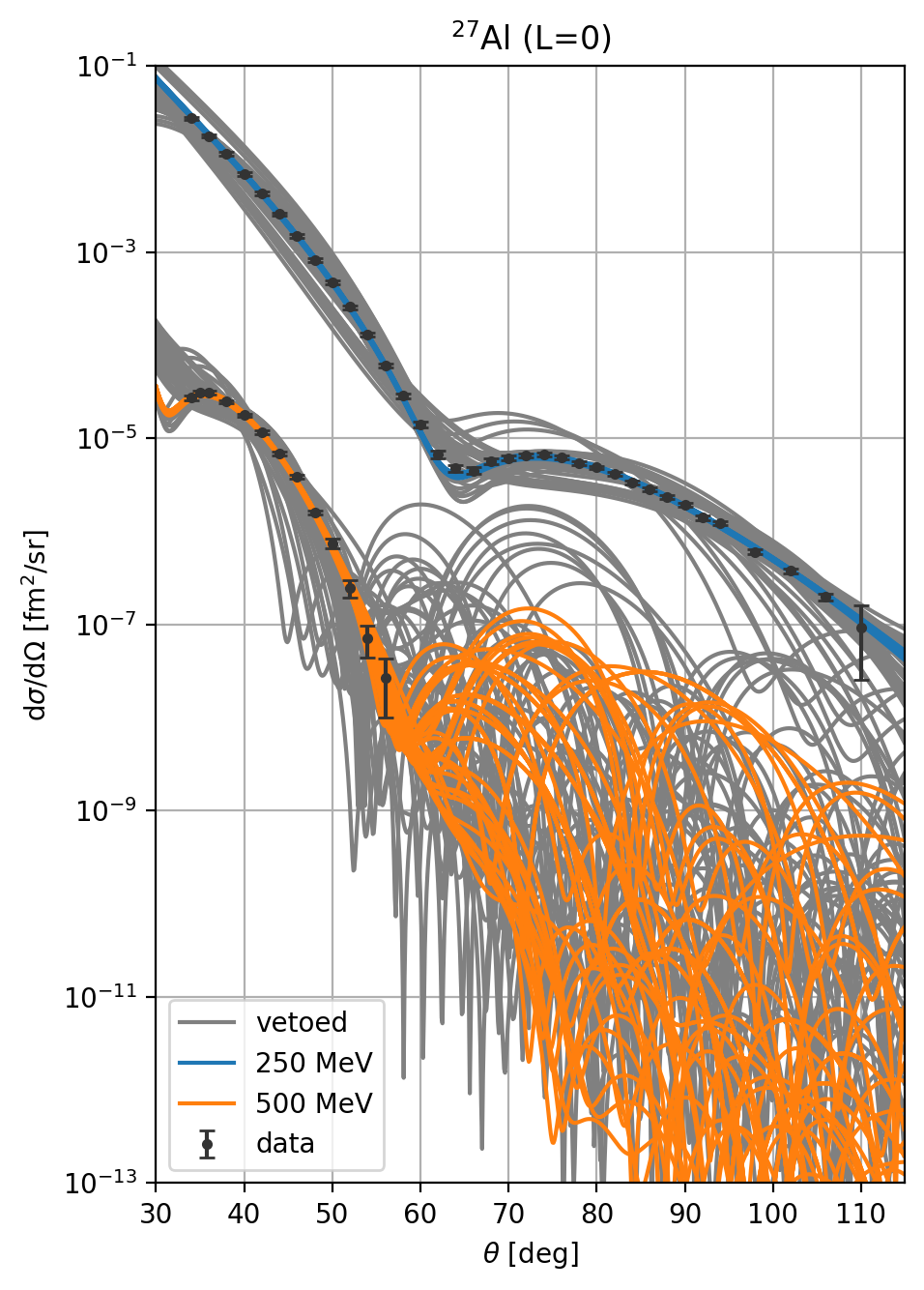}
    \hfil
    \hfil
    \raisebox{4cm}{\resizebox{!}{0.145\textheight}{%
    \renewcommand{\arraystretch}{1.3}
    \begin{tabular}{cccccccccc}
        \toprule
        \multicolumn{10}{c}{$^{27}$Al}\\ \midrule
        $R$\textbackslash{{}}$N_x$ & 2 &  3 &  4 &  5 &  6 &  7 &  8 &  9 & 10 \\ \midrule
        5.50 & {\color{gray} 74.4 } & {\color{gray} 76.0 } & {\color{gray} 73.4 } & {\color{gray} 71.7 } & {\color{gray} 70.8 } & {\color{gray} 70.8 } & {\color{gray} 72.8 } &  &  \\
        5.75 & {\color{gray} 122.2 } & {\color{gray} 74.9 } & {\color{gray} 72.3 } & {\color{gray} 70.8 } & {\color{gray} 70.1 } & {\color{gray} 69.8 } & {\color{gray} 71.3 } &  &  \\
        6.00 & {\color{gray} 93.4 } & 2.0 & 2.0 & 2.0 & 2.0 & 2.1 & 2.1 &  &  \\
        6.25 & {\color{gray} 149.2 } & 2.1 & 2.1 & 2.0 & 2.1 & 2.1 & 2.1 &  &  \\
        6.50 & {\color{gray} 201.5 } & {\color{gray} 3.1 } & 2.1 & 2.0 & 2.0 & 2.1 & 2.1 &  &  \\
        6.75 &  & {\color{gray} 6.6 } & 2.1 & 2.0 & 2.0 & 2.1 & 2.1 & 2.2 &  \\
        7.00 &  & {\color{gray} 17.0 } & 2.0 & 2.0 & 2.0 & 2.1 & 2.1 & 2.2 &  \\
        7.25 &  & {\color{gray} 41.3 } & 2.0 & 2.0 & 2.0 & 2.1 & 2.1 & 2.2 &  \\
        7.50 &  & {\color{gray} 82.2 } & 2.1 & 2.0 & 2.0 & 2.1 & 2.1 & 2.2 &  \\
        7.75 &  & {\color{gray} 132.8 } & {\color{gray} 2.8 } & 2.1 & 2.0 & 1.9 & 1.9 & 1.9 &  \\
        8.00 &  &  & {\color{gray} 5.1 } & 2.1 & 1.9 & 1.8 & 1.9 & 1.9 & 2.0 \\
        8.25 &  &  & {\color{gray} 42.5 } & {\color{gray} 29.1 } & {\color{gray} 21.2 } & {\color{gray} 16.6 } & {\color{gray} 14.3 } & {\color{gray} 13.6 } & {\color{gray} 13.0 } \\
        8.50 &  &  & {\color{gray} 43.1 } & {\color{gray} 16.3 } & {\color{gray} 11.1 } & {\color{gray} 10.4 } & {\color{gray} 9.5 } & {\color{gray} 9.0 } & {\color{gray} 8.9 } \\
        8.75 &  &  & {\color{gray} 61.2 } & {\color{gray} 11.9 } & {\color{gray} 10.9 } & {\color{gray} 10.0 } & {\color{gray} 9.9 } & {\color{gray} 9.9 } & {\color{gray} 10.0 } \\
        \bottomrule
    \end{tabular}%
    \renewcommand{\arraystretch}{1.0}}%
    }
    \hfil
    \caption{Steps \ref{item:base} and \ref{item:sen} of the fits for ${}^{27}$Al. The plot shows the cross-section data and the fit solutions, while the table shows the $\chi^2/\text{dof}$ ($\text{dof}=48-N_x$). Solutions in gray are excluded for $p_\text{val}<10^{-5}$.}
    \label{tab:Al27BASE}
\end{figure}


\subsection{Documentation of the intermediate steps of the fitting strategy}

For each nucleus we show the reduced $\chi^2$ and electron scattering cross sections after steps \ref{item:base} and \ref{item:sen} in Sec.~\ref{sec:FB_fit}, excluding fits based on the $p$-value ($p_\text{val}$), see Figs.~\ref{tab:Ca4048BASE}--\ref{tab:Al27BASE}. One sees how with the chosen limits all remaining fits describe the data points reasonably well, while beyond the data points the shape of the cross section becomes largely unconstrained. In particular for $^{40,48}$Ca a quite large set of reasonable fits can be selected with a wide range of $R$ and $N$ combinations, reflecting the  high quality of the data without any major inconsistencies. For the other nuclei the data situation is markedly worse, which is reflected by the smaller set of reasonable fits that could be found. 


\begin{figure}[t]

    \hfil
    \includegraphics[width=0.49\textwidth]{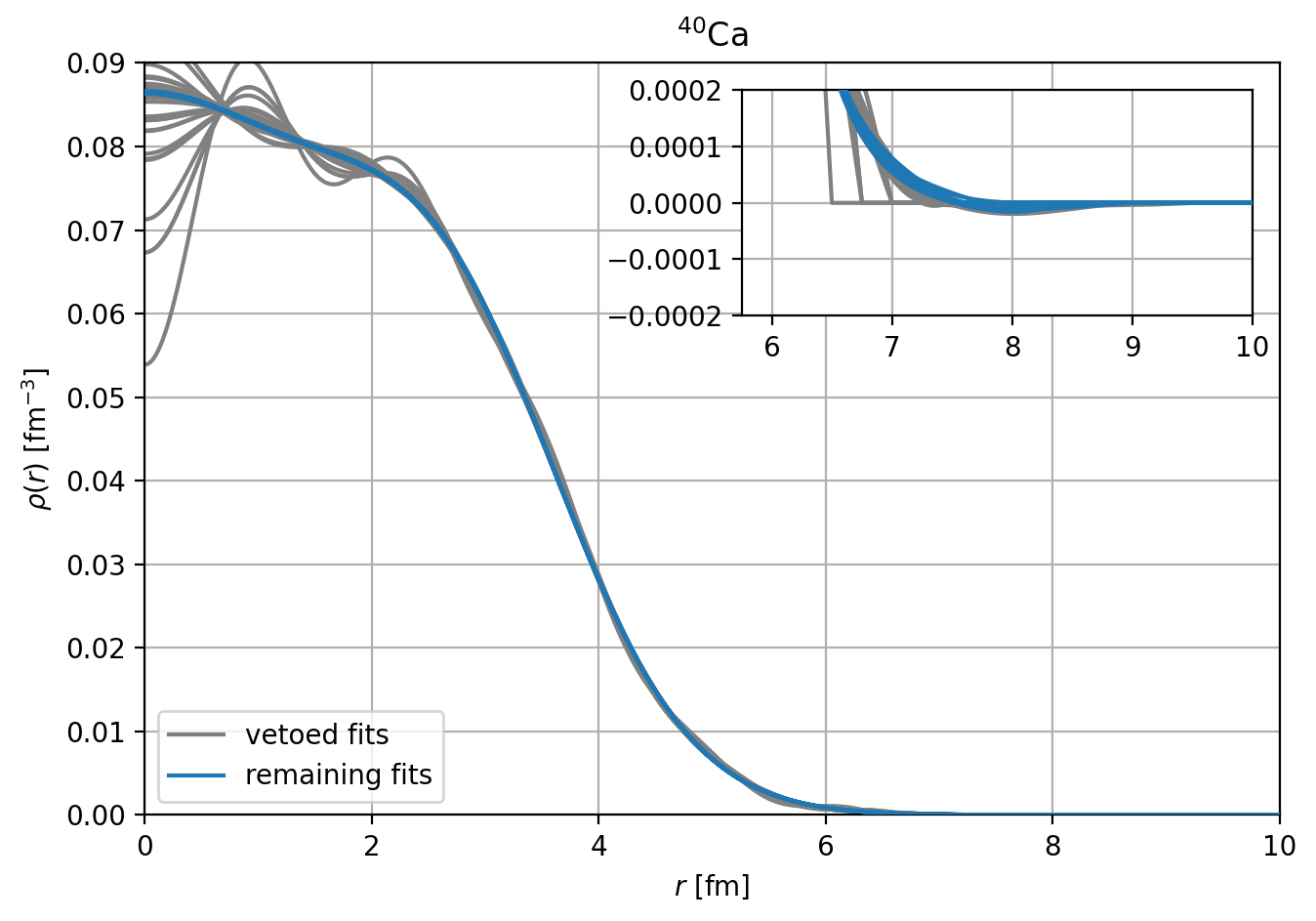}
    \hfil
    \includegraphics[width=0.49\textwidth]{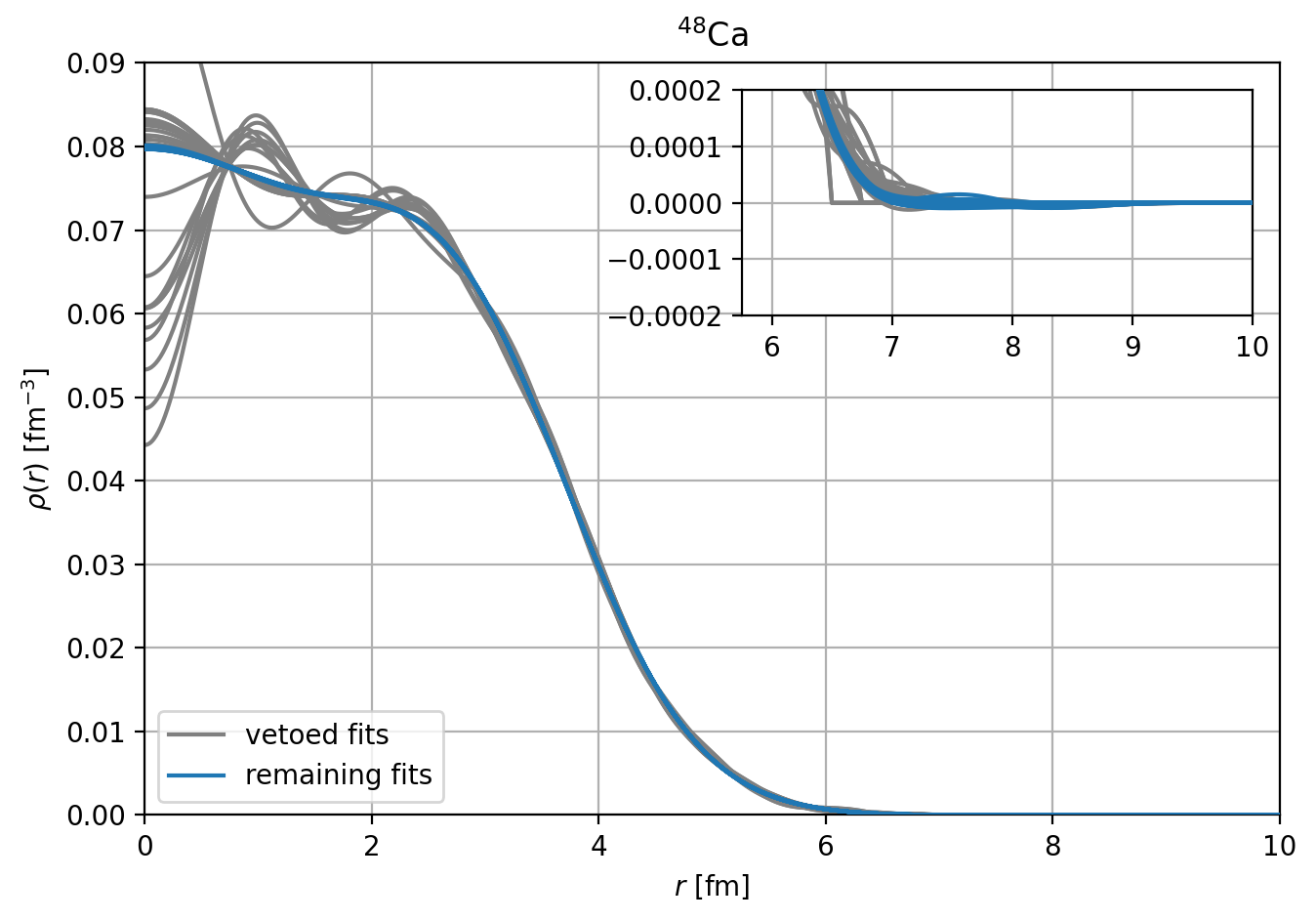}
    \hfil
    
    \hfil
    \includegraphics[width=0.49\textwidth]{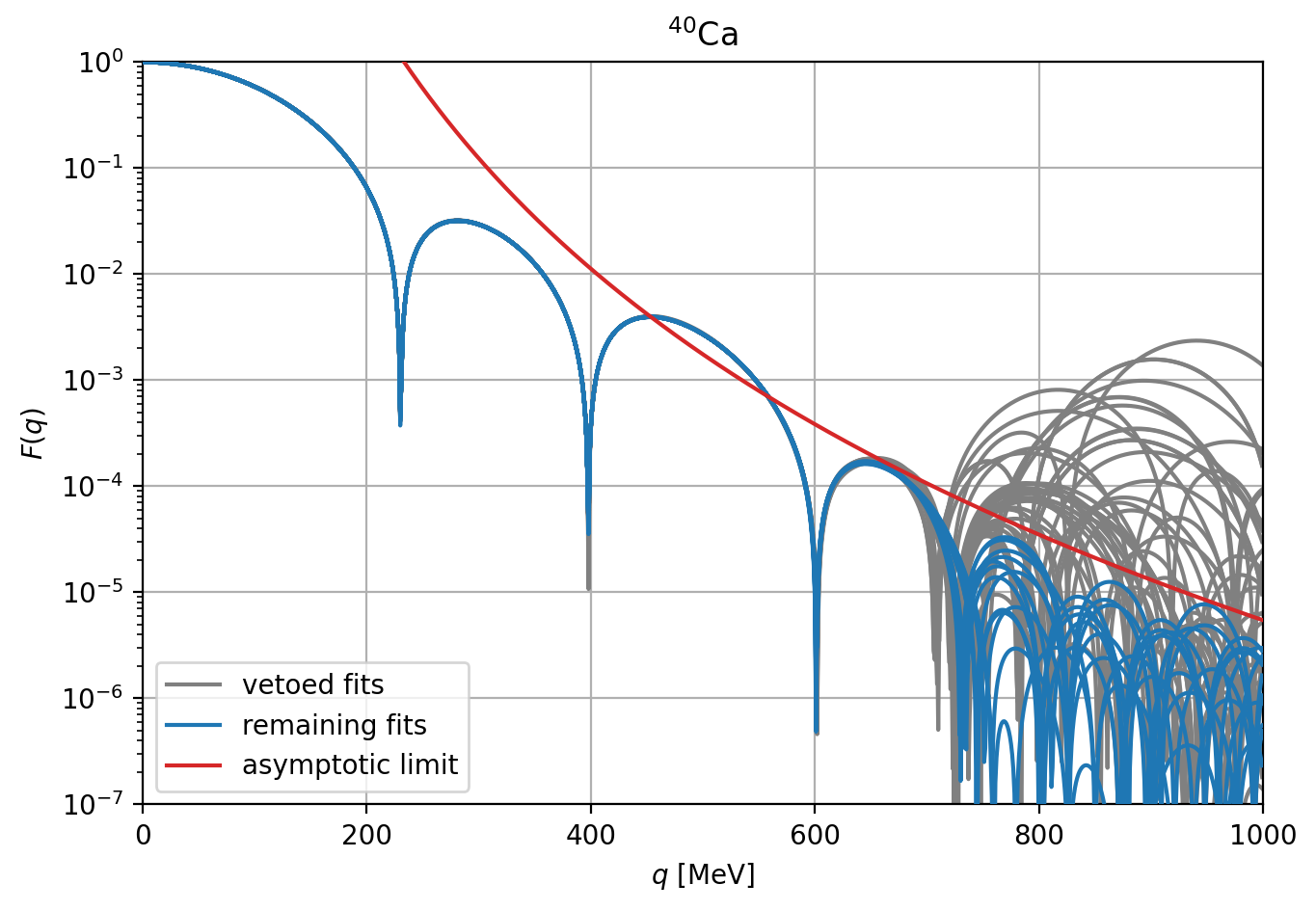}
    \hfil
    \includegraphics[width=0.49\textwidth]{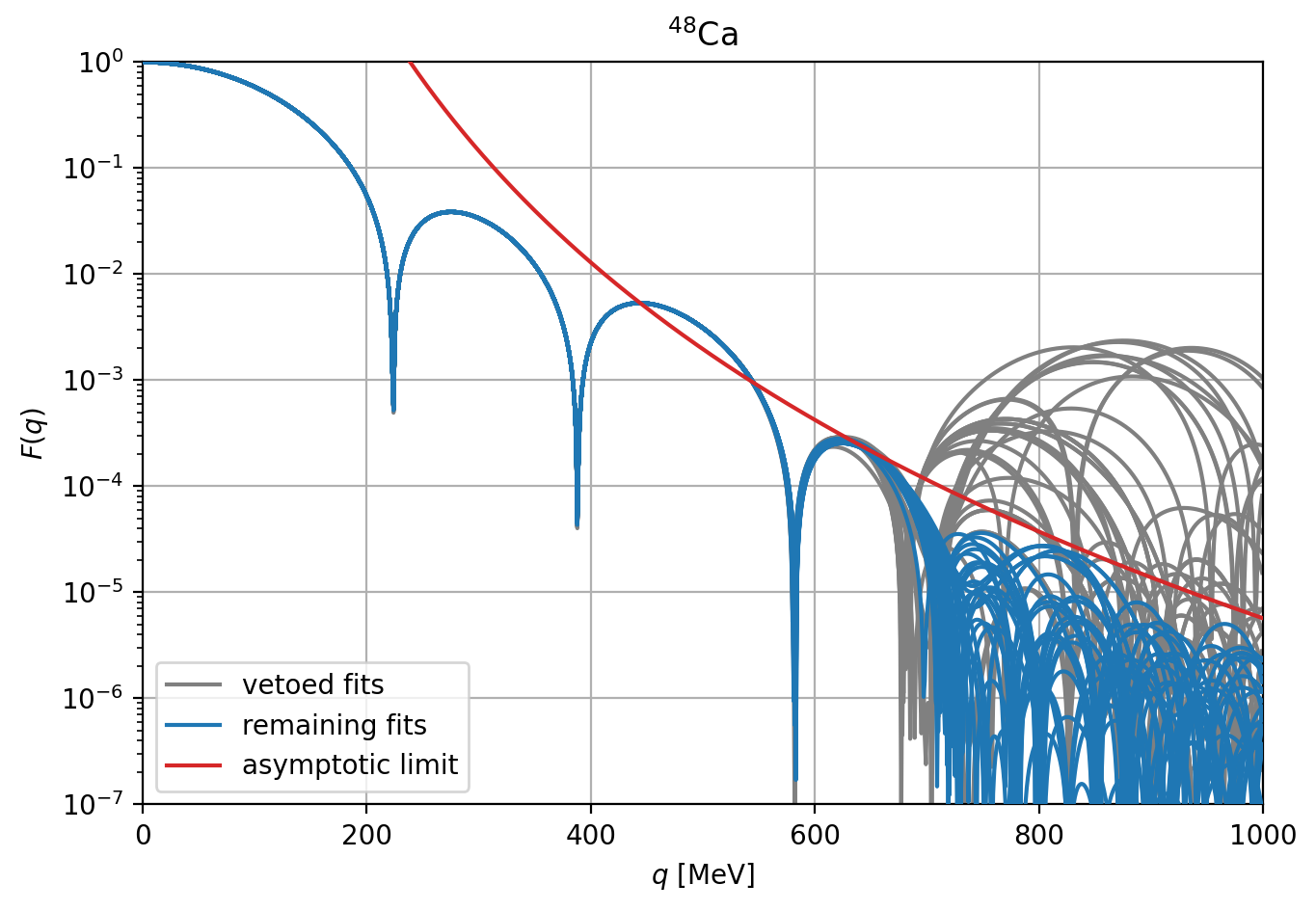}
    \hfil
    
    \hfil
    \resizebox{!}{0.145\textheight}{%
    \renewcommand{\arraystretch}{1.3}
    \begin{tabular}{cccccccc}
        \toprule
        \multicolumn{8}{c}{$^{40}$Ca}\\ \midrule
        $R$\textbackslash{{}}$N_x$ & 7 &  8 &  9 & 10 & 11 & 12 & 13 \\ \midrule
        6.50 &  &  &  & {\color{gray} 1.416 } &  &  &  \\
        6.75 &  & {\color{gray} 1.153 } & {\color{gray} 1.109 } & {\color{gray} 1.129 } &  &  &  \\
        7.00 &  & {\color{gray} 1.101 } & {\color{gray} 1.077 } & {\color{gray} 1.096 } &  &  &  \\
        7.25 &  & {\color{gray} 1.071 } & {\color{gray} 1.082 } & {\color{gray} 1.099 } & {\color{gray} 1.119 } &  &  \\
        7.50 & 1.107 & 1.085 & {\color{gray} 1.087 } & {\color{gray} 1.101 } & {\color{gray} 1.121 } &  &  \\
        7.75 &  & 1.123 & {\color{gray} 1.085 } & {\color{gray} 1.103 } & {\color{gray} 1.117 } &  &  \\
        8.00 &  &  & {\color{gray} 1.087 } & {\color{gray} 1.104 } & {\color{gray} 1.120 } & {\color{gray} 1.141 } &  \\
        8.25 &  & 1.140 & 1.096 & {\color{gray} 1.105 } & {\color{gray} 1.121 } & {\color{gray} 1.139 } &  \\
        8.50 &  & 1.107 & 1.111 & 1.128 & {\color{gray} 1.119 } & {\color{gray} 1.137 } &  \\
        8.75 &  &  & 1.135 & 1.113 & {\color{gray} 1.113 } & {\color{gray} 1.131 } &  \\
        9.00 &  &  & 1.139 & 1.122 & {\color{gray} 1.107 } & {\color{gray} 1.127 } & {\color{gray} 1.148 } \\
        9.25 &  &  & {\color{gray} 1.208 } & 1.150 & {\color{gray} 1.135 } & {\color{gray} 1.149 } & {\color{gray} 1.171 } \\
        9.50 &  &  &  &  &  & {\color{gray} 1.196 } & {\color{gray} 1.186 } \\
        \bottomrule
    \end{tabular}%
    \renewcommand{\arraystretch}{1.0}}%
    \hfil
    \resizebox{!}{0.145\textheight}{%
    \renewcommand{\arraystretch}{1.3}
    \begin{tabular}{ccccccccc}
        \toprule
        \multicolumn{9}{c}{$^{48}$Ca}\\ \midrule
        $R$\textbackslash{{}}$N_x$ & 6 &  7 &  8 &  9 & 10 & 11 & 12 & 13 \\ \midrule
        6.50 &  &  & {\color{gray} 1.168 } & {\color{gray} 1.177 } & {\color{gray} 1.197 } &  &  &  \\
        6.75 & {\color{gray} 1.135 } & {\color{gray} 1.154 } & {\color{gray} 1.162 } & {\color{gray} 1.167 } & {\color{gray} 1.186 } &  &  &  \\
        7.00 & 1.124 & 1.140 & {\color{gray} 1.155 } & {\color{gray} 1.163 } & {\color{gray} 1.183 } &  &  &  \\
        7.25 &  & 1.145 & {\color{gray} 1.154 } & {\color{gray} 1.168 } & {\color{gray} 1.186 } & {\color{gray} 1.206 } &  &  \\
        7.50 &  & 1.155 & {\color{gray} 1.152 } & {\color{gray} 1.164 } & {\color{gray} 1.185 } & {\color{gray} 1.206 } &  &  \\
        7.75 &  & 1.140 & 1.155 & {\color{gray} 1.164 } & {\color{gray} 1.185 } & {\color{gray} 1.206 } &  &  \\
        8.00 &  & 1.171 & 1.167 & {\color{gray} 1.167 } & {\color{gray} 1.184 } & {\color{gray} 1.205 } &  &  \\
        8.25 &  &  & 1.180 & {\color{gray} 1.172 } & {\color{gray} 1.181 } & {\color{gray} 1.203 } & {\color{gray} 1.224 } &  \\
        8.50 &  &  & 1.137 & 1.152 & 1.168 & 1.189 & 1.211 &  \\
        8.75 &  &  & 1.162 & 1.149 & 1.169 & 1.187 & 1.209 &  \\
        9.00 &  &  & {\color{gray} 1.244 } & 1.151 & 1.170 & 1.188 & 1.210 &  \\
        9.25 &  &  &  & 1.189 & 1.190 & {\color{gray} 1.206 } & {\color{gray} 1.228 } & {\color{gray} 1.252 } \\
        9.50 &  &  &  & 1.192 & 1.212 & 1.219 & {\color{gray} 1.241 } & {\color{gray} 1.264 } \\
        \bottomrule
    \end{tabular}%
    \renewcommand{\arraystretch}{1.0}}%
    \hfil
    
    \caption{Resulting charge densities and form factors after steps \ref{item:pen} and \ref{item:sel} for ${}^{40}$Ca and ${}^{48}$Ca. The tables show $\chi^2/\text{dof}$ ($\text{dof}=65+1-N_x$). Solutions in gray are excluded for $p_\text{val}<17\%$ for ${}^{40}$Ca and $p_\text{val}<12\%$ for ${}^{48}$Ca or when violating the asymptotic limit (indicated in red).}
    \label{tab:Ca4048PEN}
\end{figure}

\begin{figure}[t]

    \hfil
    \includegraphics[width=0.49\textwidth]{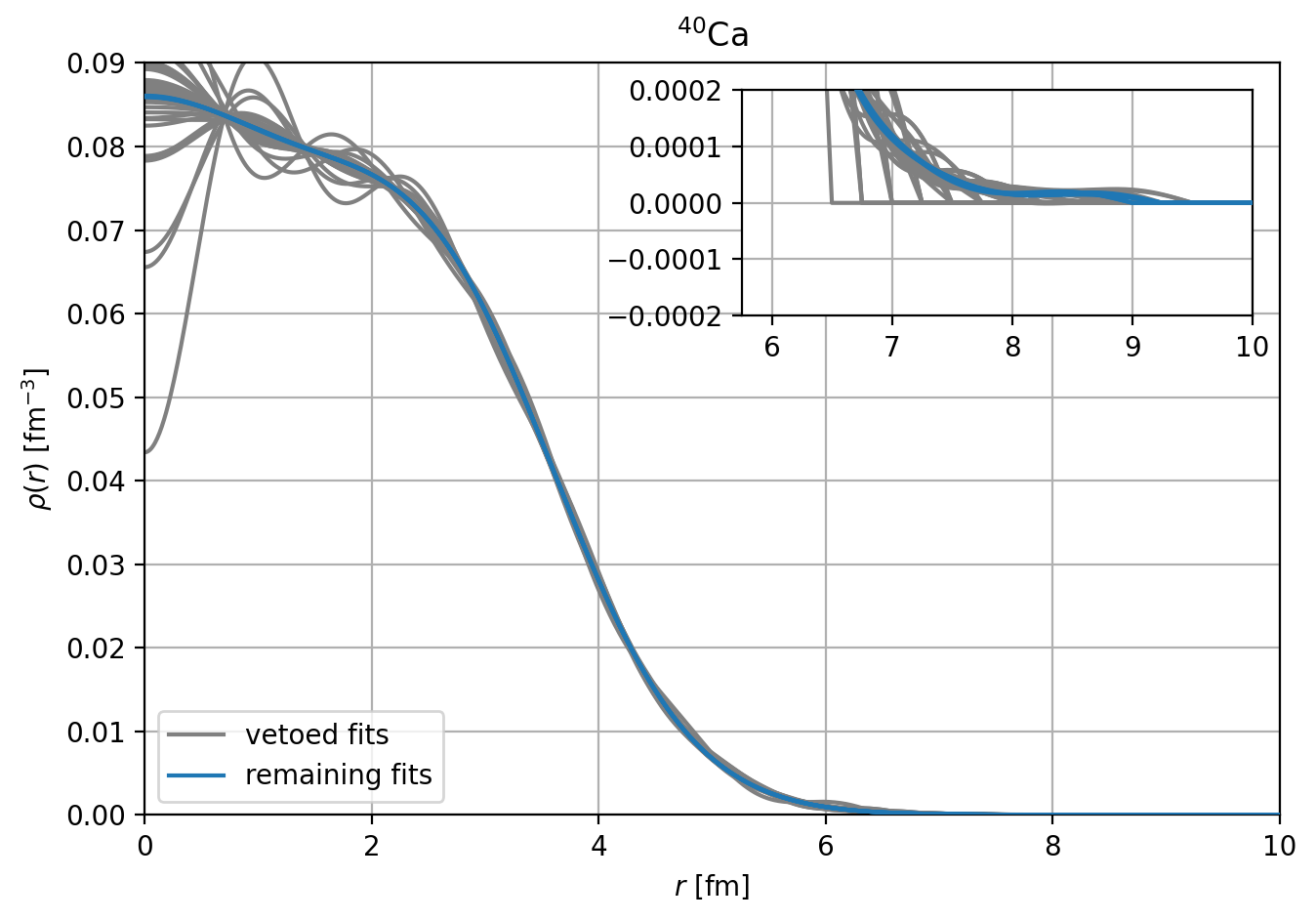}
    \hfil
    \includegraphics[width=0.49\textwidth]{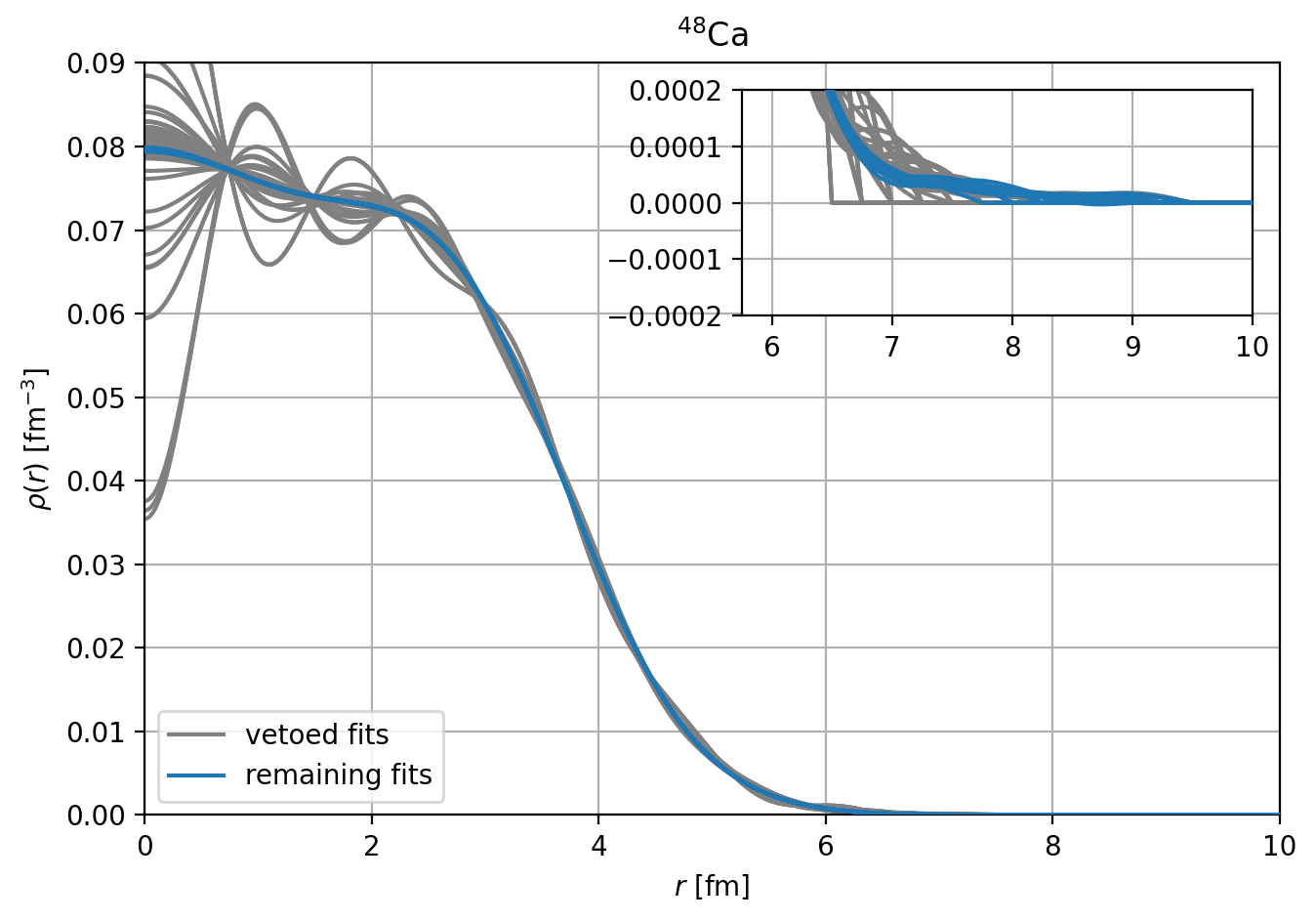}
    \hfil
    
    \hfil
    \includegraphics[width=0.49\textwidth]{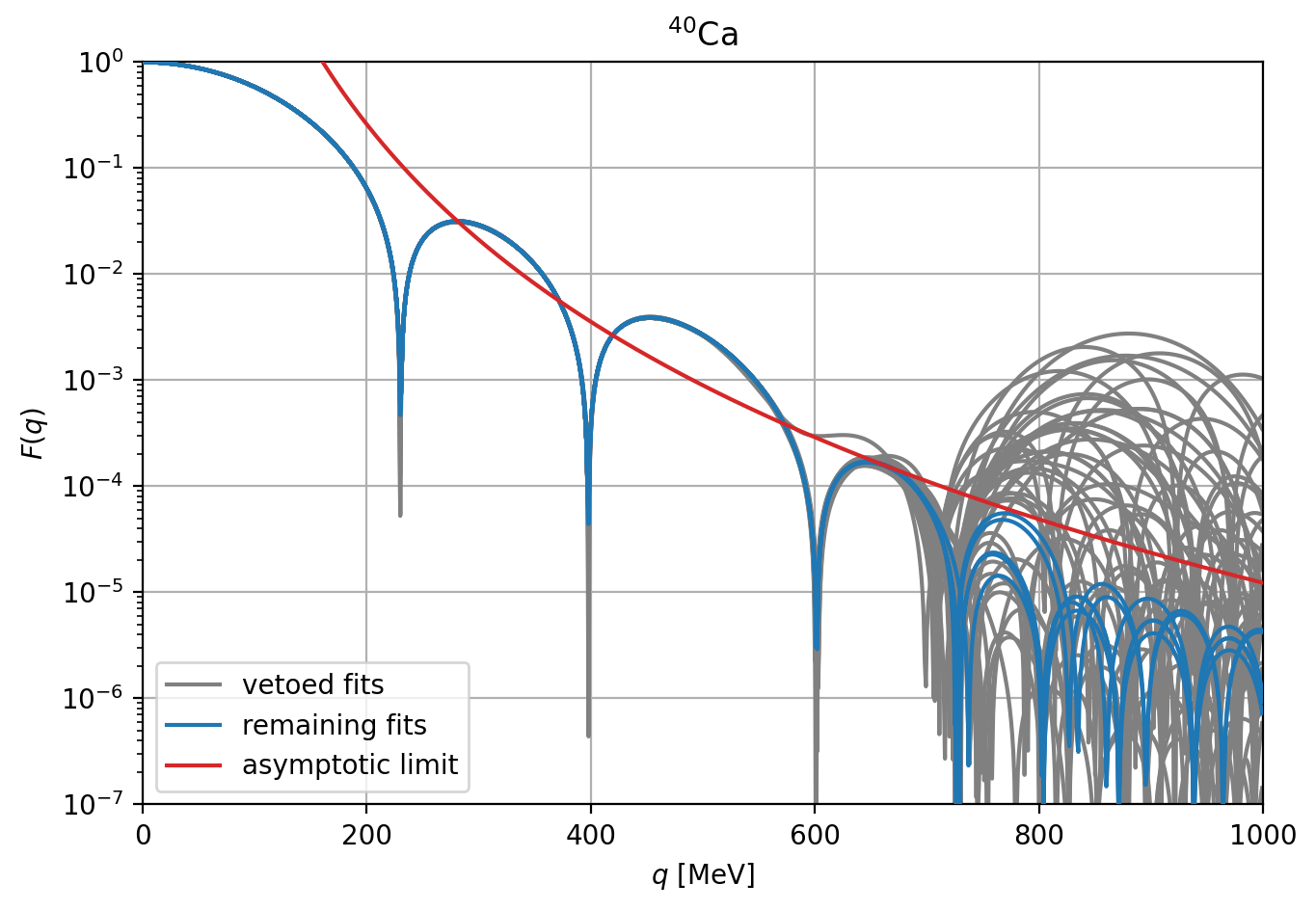}
    \hfil
    \includegraphics[width=0.49\textwidth]{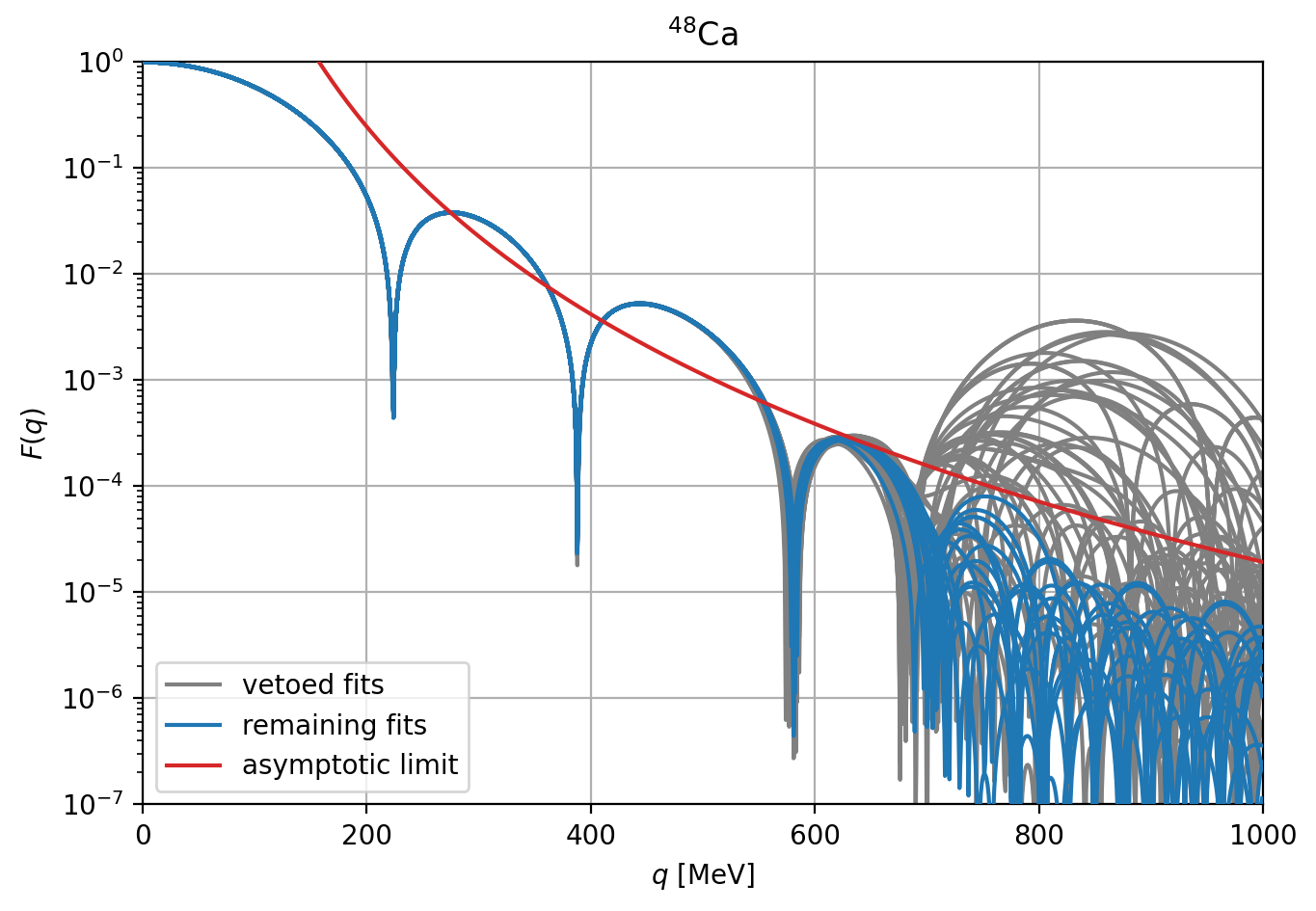}
    \hfil
    
    \hfil
    \resizebox{!}{0.145\textheight}{%
    \renewcommand{\arraystretch}{1.3}
    \begin{tabular}{cccccccc}
        \toprule
        \multicolumn{8}{c}{$^{40}$Ca}\\ \midrule
        $R$\textbackslash{{}}$N_x$ & 7 &  8 &  9 & 10 & 11 & 12 & 13 \\ \midrule
        6.50 &  &  &  & {\color{gray} 3.950 } &  &  &  \\
        6.75 &  & {\color{gray} 3.190 } & {\color{gray} 2.720 } & {\color{gray} 2.764 } &  &  &  \\
        7.00 &  & {\color{gray} 2.300 } & {\color{gray} 1.948 } & {\color{gray} 1.953 } &  &  &  \\
        7.25 &  & {\color{gray} 1.875 } & {\color{gray} 1.740 } & {\color{gray} 1.769 } & {\color{gray} 1.784 } &  &  \\
        7.50 & {\color{gray} 1.992 } & {\color{gray} 1.936 } & {\color{gray} 1.672 } & {\color{gray} 1.671 } & {\color{gray} 1.701 } &  &  \\
        7.75 &  & {\color{gray} 1.611 } & {\color{gray} 1.632 } & {\color{gray} 1.623 } & {\color{gray} 1.648 } &  &  \\
        8.00 &  &  & {\color{gray} 1.612 } & {\color{gray} 1.594 } & {\color{gray} 1.622 } & {\color{gray} 1.651 } &  \\
        8.25 &  & {\color{gray} 1.580 } & {\color{gray} 1.599 } & {\color{gray} 1.571 } & {\color{gray} 1.599 } & {\color{gray} 1.628 } &  \\
        8.50 &  & {\color{gray} 1.608 } & {\color{gray} 1.601 } & {\color{gray} 1.550 } & {\color{gray} 1.576 } & {\color{gray} 1.604 } &  \\
        8.75 &  &  & {\color{gray} 1.867 } & {\color{gray} 1.530 } & {\color{gray} 1.553 } & {\color{gray} 1.579 } &  \\
        9.00 &  &  & {\color{gray} 1.862 } & 1.505 & 1.531 & {\color{gray} 1.551 } & {\color{gray} 1.577 } \\
        9.25 &  &  & 1.453 & 1.459 & 1.485 & 1.510 & {\color{gray} 1.528 } \\
        9.50 &  &  &  &  &  & {\color{gray} 1.533 } & {\color{gray} 1.488 } \\
    \bottomrule
    \end{tabular}
    \renewcommand{\arraystretch}{1.0}}%
    \hfil
    \resizebox{!}{0.145\textheight}{%
    \renewcommand{\arraystretch}{1.3}
    \begin{tabular}{ccccccccc}
        \toprule
        \multicolumn{9}{c}{$^{48}$Ca}\\ \midrule
        $R$\textbackslash{{}}$N_x$ & 6 &  7 &  8 &  9 & 10 & 11 & 12 & 13 \\ \midrule
        6.50 &  &  & {\color{gray} 1.985 } & {\color{gray} 2.019 } & {\color{gray} 2.055 } &  &  &  \\
        6.75 & {\color{gray} 2.256 } & {\color{gray} 2.164 } & {\color{gray} 1.721 } & {\color{gray} 1.731 } & {\color{gray} 1.761 } &  &  &  \\
        7.00 & {\color{gray} 1.929 } & {\color{gray} 1.886 } & {\color{gray} 1.658 } & {\color{gray} 1.626 } & {\color{gray} 1.644 } &  &  &  \\
        7.25 &  & {\color{gray} 1.762 } & {\color{gray} 1.592 } & {\color{gray} 1.563 } & {\color{gray} 1.573 } & {\color{gray} 1.598 } &  &  \\
        7.50 &  & {\color{gray} 1.679 } & {\color{gray} 1.502 } & {\color{gray} 1.516 } & {\color{gray} 1.537 } & {\color{gray} 1.565 } &  &  \\
        7.75 &  & 1.535 & {\color{gray} 1.472 } & {\color{gray} 1.497 } & {\color{gray} 1.521 } & {\color{gray} 1.547 } &  &  \\
        8.00 &  & 1.453 & 1.476 & {\color{gray} 1.489 } & {\color{gray} 1.514 } & {\color{gray} 1.541 } &  &  \\
        8.25 &  &  & 1.463 & {\color{gray} 1.486 } & {\color{gray} 1.508 } & {\color{gray} 1.535 } & {\color{gray} 1.563 } &  \\
        8.50 &  &  & 1.460 & 1.485 & {\color{gray} 1.500 } & {\color{gray} 1.527 } & {\color{gray} 1.554 } &  \\
        8.75 &  &  & 1.459 & 1.481 & {\color{gray} 1.498 } & {\color{gray} 1.522 } & {\color{gray} 1.544 } &  \\
        9.00 &  &  & 1.540 & 1.481 & 1.481 & {\color{gray} 1.505 } & {\color{gray} 1.528 } &  \\
        9.25 &  &  &  & 1.474 & 1.454 & {\color{gray} 1.472 } & {\color{gray} 1.492 } & {\color{gray} 1.511 } \\
        9.50 &  &  &  & 1.407 & 1.422 & 1.446 & 1.472 & {\color{gray} 1.485 } \\ 
        \bottomrule
    \end{tabular}
    \renewcommand{\arraystretch}{1.0}}%
    \hfil
    
    \caption{Same as Fig.~\ref{tab:Ca4048PEN}, but including the Barrett moment as an additional constraint. The tables show $\chi^2/\text{dof}$ ($\text{dof}=65+1+1-N_x$). Solutions in gray are excluded for $p_\text{val}<0.3\%$ for ${}^{40}$Ca and $p_\text{val}<0.25\%$ for ${}^{48}$Ca or when violating the asymptotic limit (indicated in red).}
    \label{tab:Ca4048BAR}
\end{figure}


\begin{figure}[t]

    \hfil
    \includegraphics[width=0.49\textwidth]{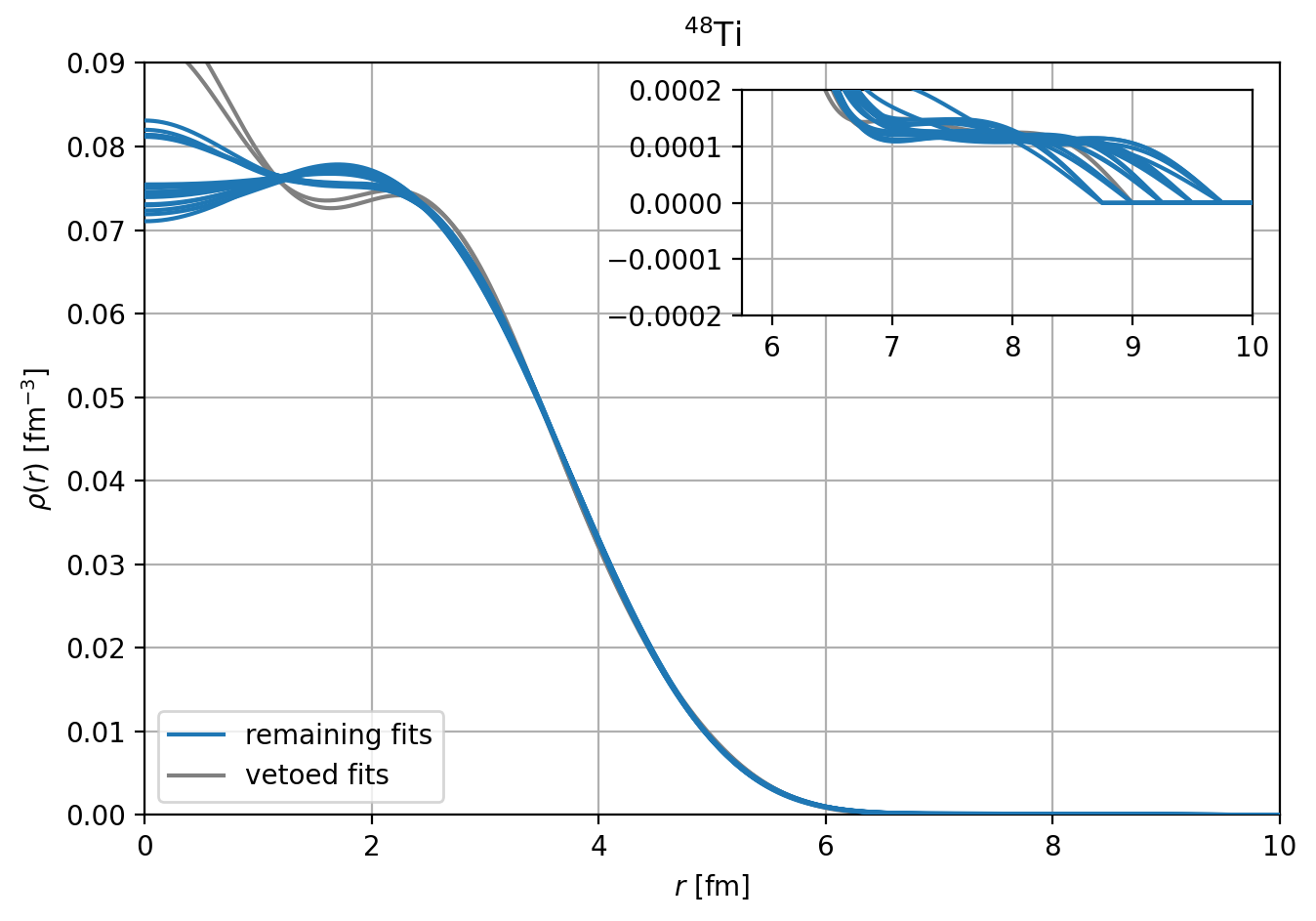}
    \hfil
    \includegraphics[width=0.49\textwidth]{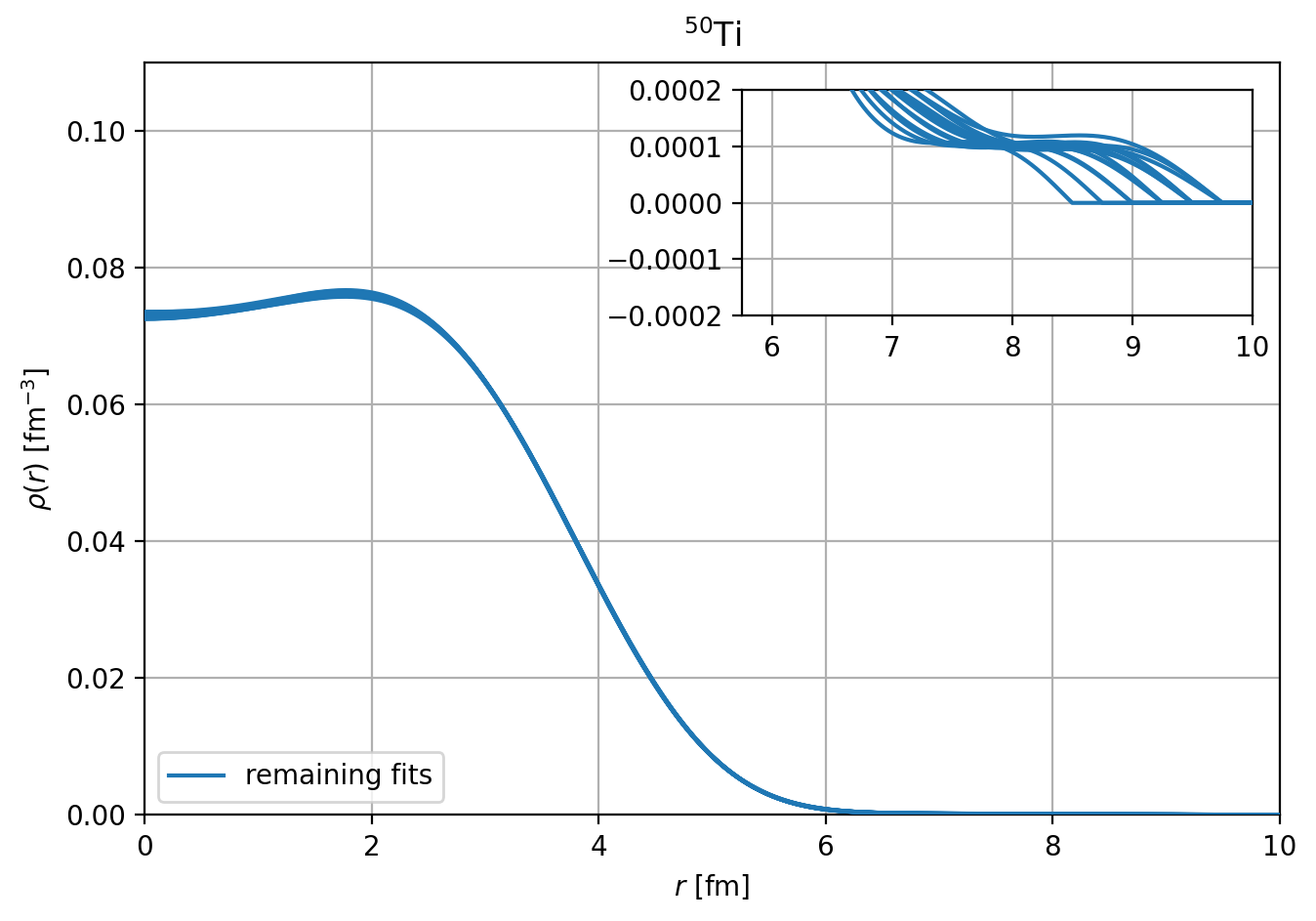}
    \hfil
    
    \hfil
    \includegraphics[width=0.49\textwidth]{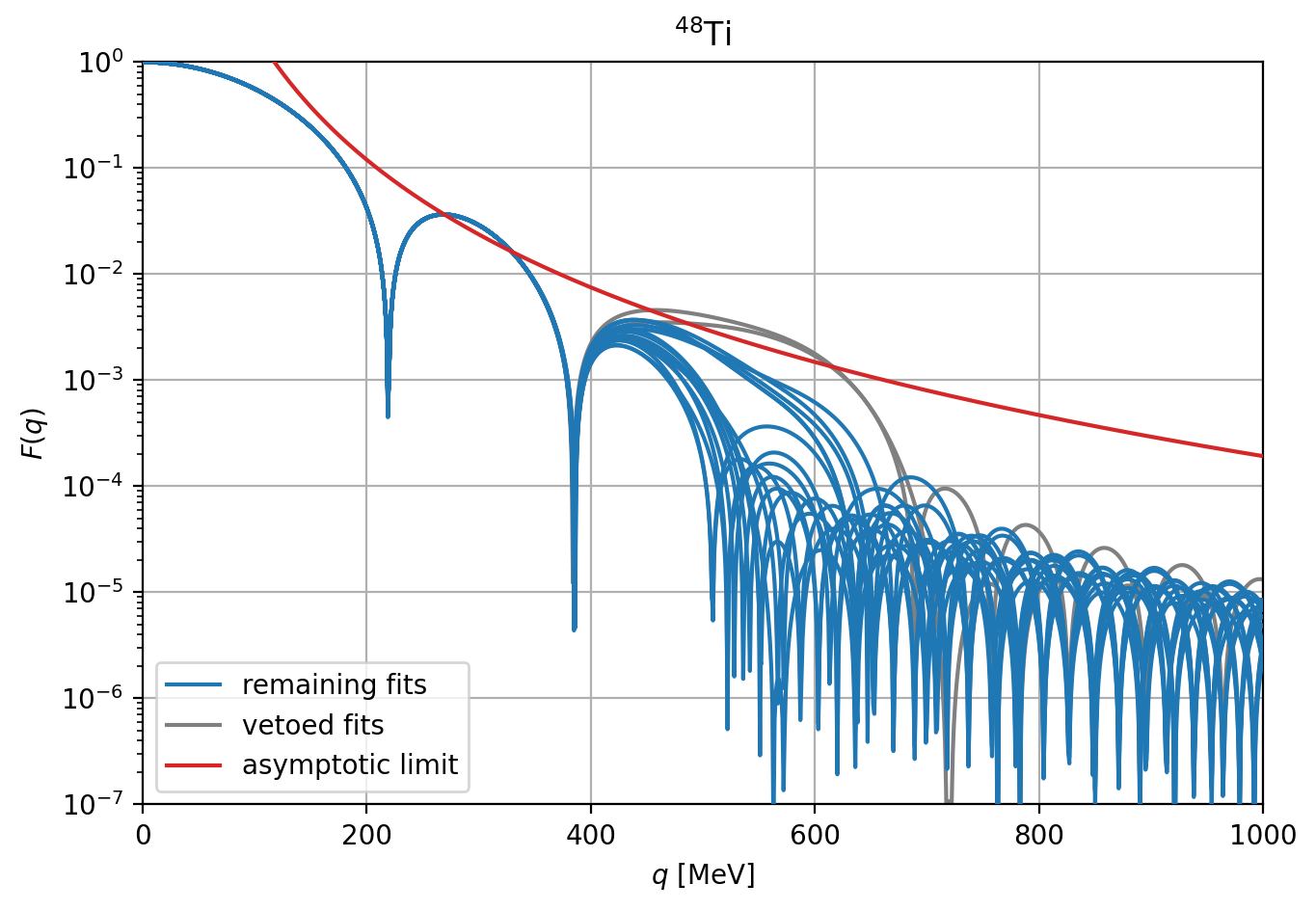}
    \hfil
    \includegraphics[width=0.49\textwidth]{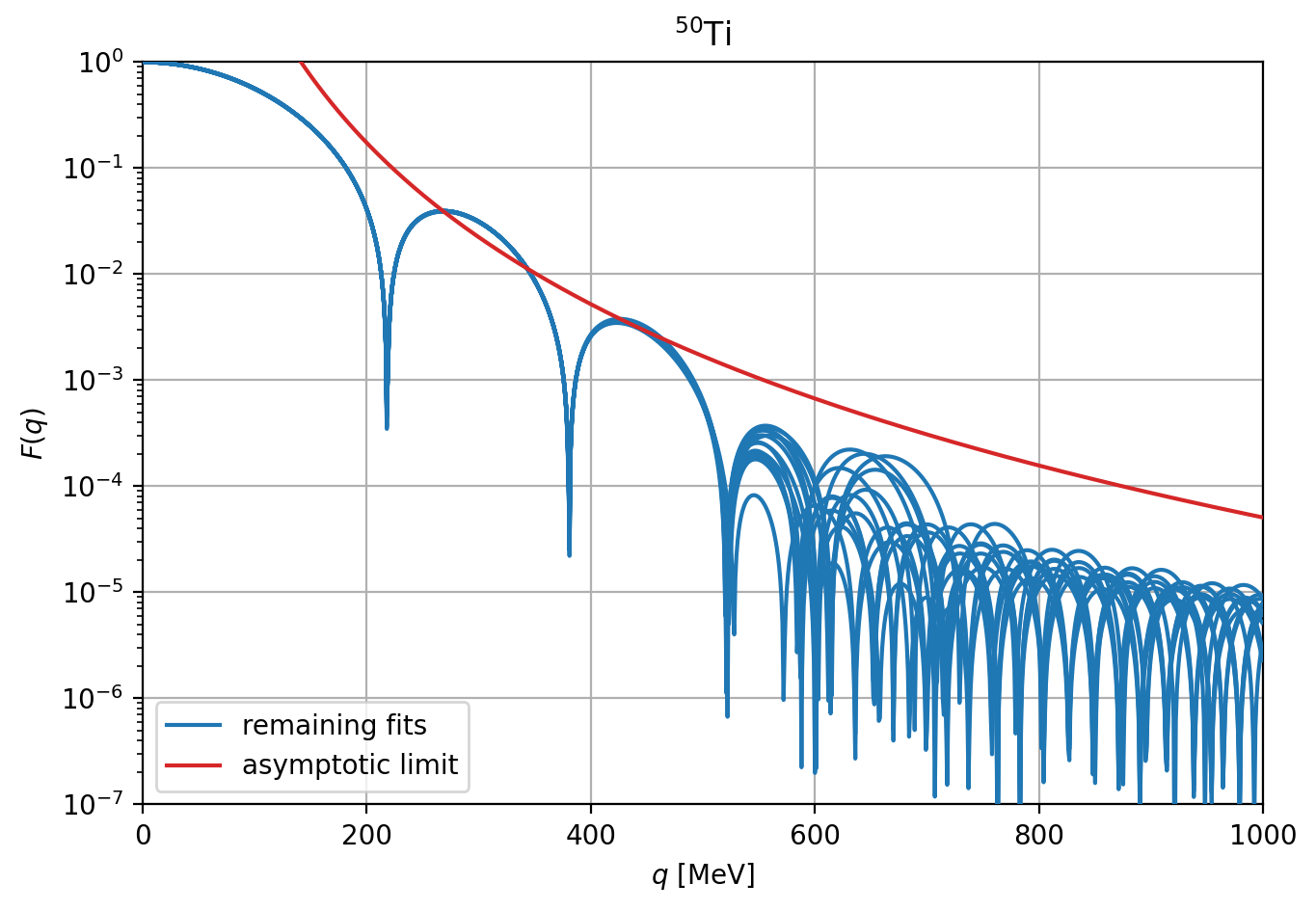}
    \hfil
    
    \hfil
    \resizebox{!}{0.115\textheight}{%
    \renewcommand{\arraystretch}{1.3}
    \begin{tabular}{ccccc}
        \toprule
        \multicolumn{5}{c}{$^{48}$Ti}\\ \midrule
        $R$\textbackslash{{}}$N_x$ & 6 &  7 &  8 &  9 \\ \midrule
        8.75 & 1.659 & 1.729 & 1.809 &  \\
        9.00 & 1.623 & 1.694 & {\color{gray} 1.737 } &  \\
        9.25 & 1.587 & 1.659 & 1.716 & 1.792 \\
        9.50 & 1.565 & 1.625 & 1.669 & {\color{gray} 1.747 } \\
        9.75 & 1.568 & 1.598 & 1.639 & 1.721 \\
        \bottomrule 
    \end{tabular}%
    \renewcommand{\arraystretch}{1.0}}%
    \hfil
    \resizebox{!}{0.115\textheight}{%
    \renewcommand{\arraystretch}{1.3}
    \begin{tabular}{cccccc}
        \toprule
        \multicolumn{6}{c}{$^{50}$Ti}\\ \midrule
        $R$\textbackslash{{}}$N_x$ & 6 &  7 &  8 &  9 \\ \midrule
        8.50 &  &  & 1.627 &  \\
        8.75 &  &  & 1.606 &  \\
        9.00 &  &  & 1.584 & 1.617 \\
        9.25 &  & 1.532 & 1.562 & 1.594 \\
        9.50 & 1.486 & 1.514 & 1.543 & 1.574 \\
        9.75 &  & 1.500 & 1.528 & 1.557 \\
        \bottomrule
    \end{tabular}%
    \renewcommand{\arraystretch}{1.0}}%
    \hfil
    
    \caption{Resulting charge densities and form factors after steps \ref{item:pen} and \ref{item:sel} for ${}^{48}$Ti and ${}^{50}$Ti. The tables show $\chi^2/\text{dof}$ ($\text{dof}=28+1-N_x$ for ${}^{48}$Ti, $\text{dof}=57+1-N_x$ for ${}^{50}$Ti). Solutions in gray are excluded for $p_\text{val}<1\%$ for ${}^{48}$Ti and $p_\text{val}<0.1\%$ for ${}^{50}$Ti or when violating the asymptotic limit (indicated in red).}
    \label{tab:Ti4850PEN}
\end{figure}

\begin{figure}[t]

    \hfil
    \includegraphics[width=0.49\textwidth]{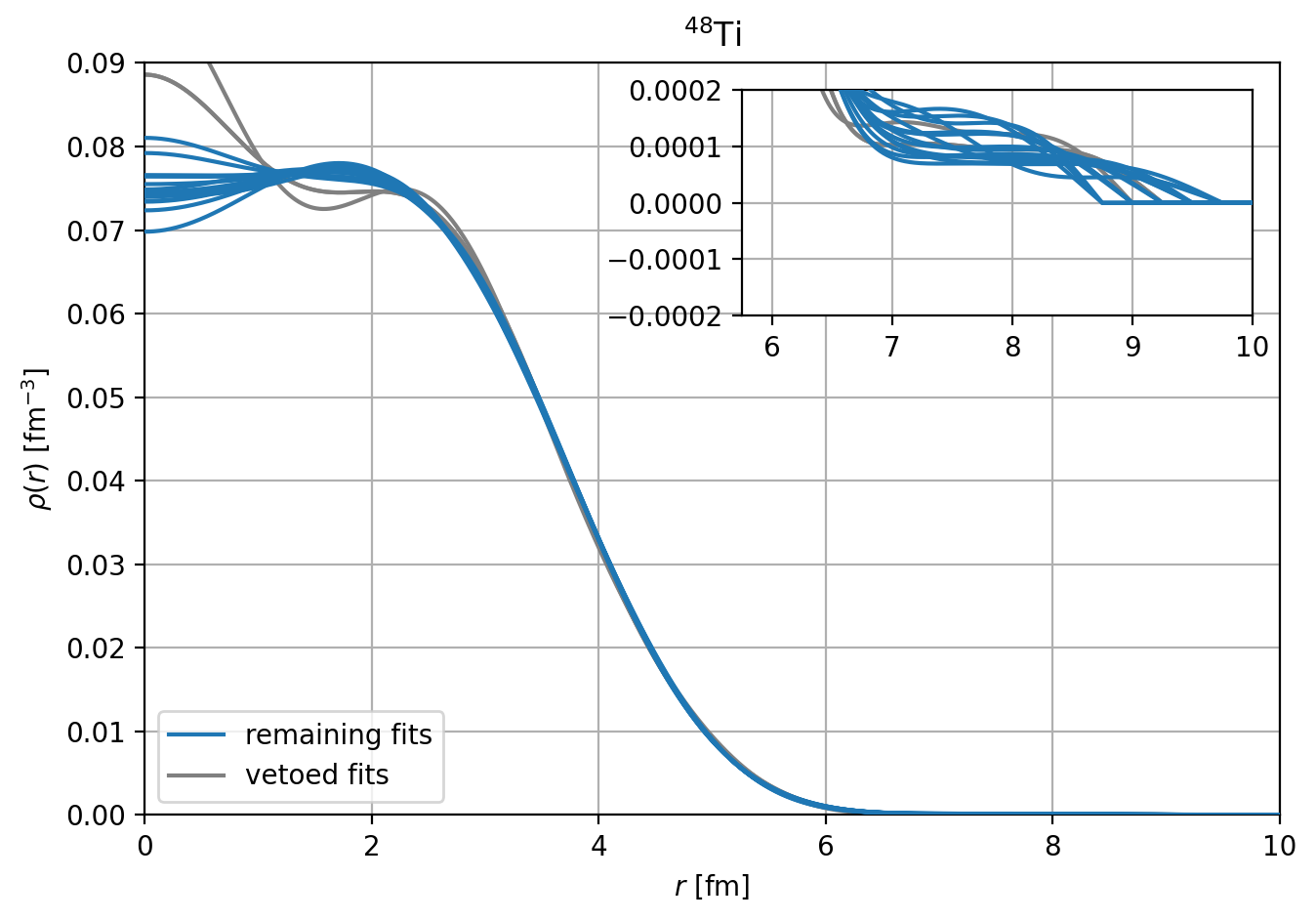}
    \hfil
    \includegraphics[width=0.49\textwidth]{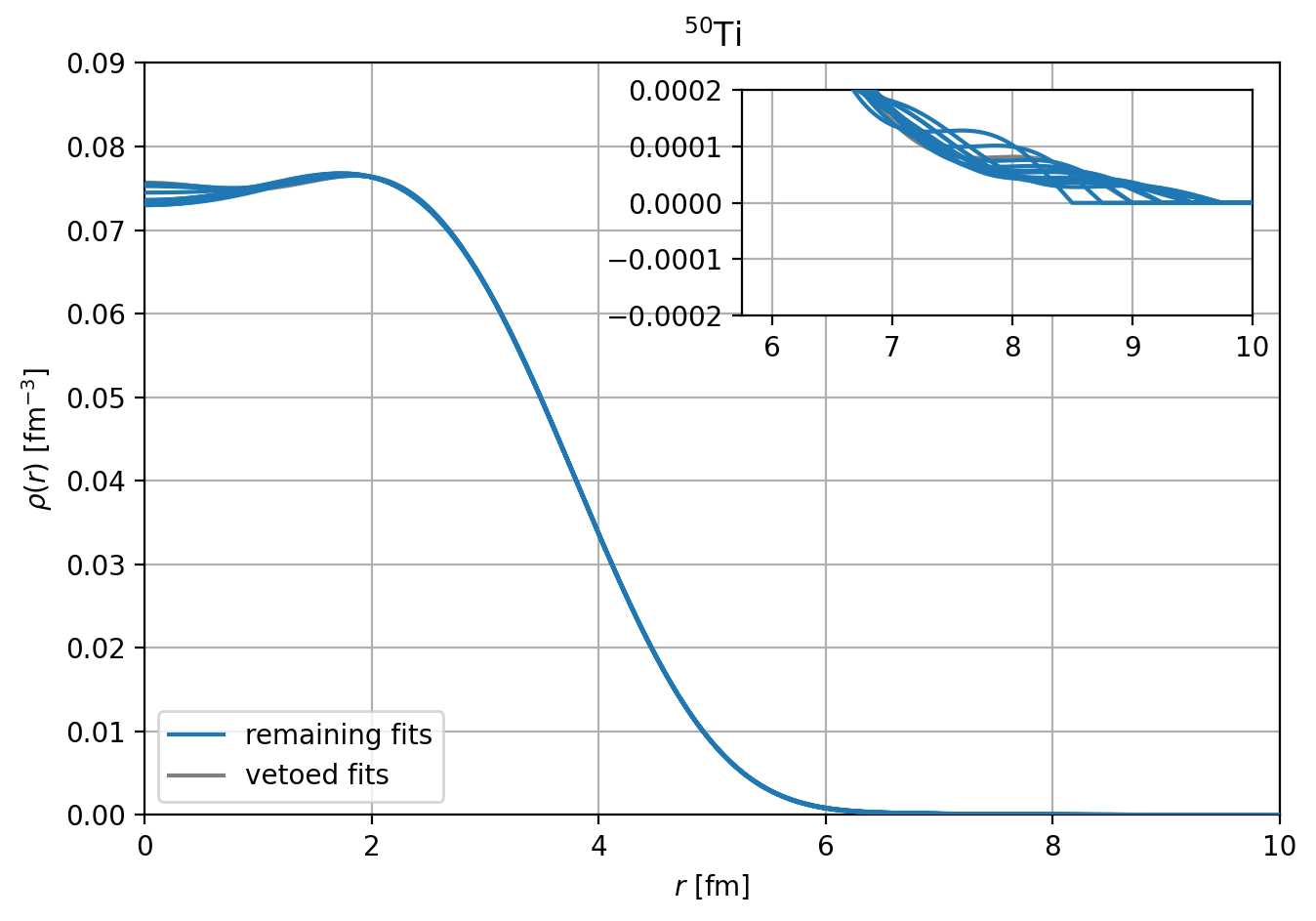}
    \hfil
    
    \hfil
    \includegraphics[width=0.49\textwidth]{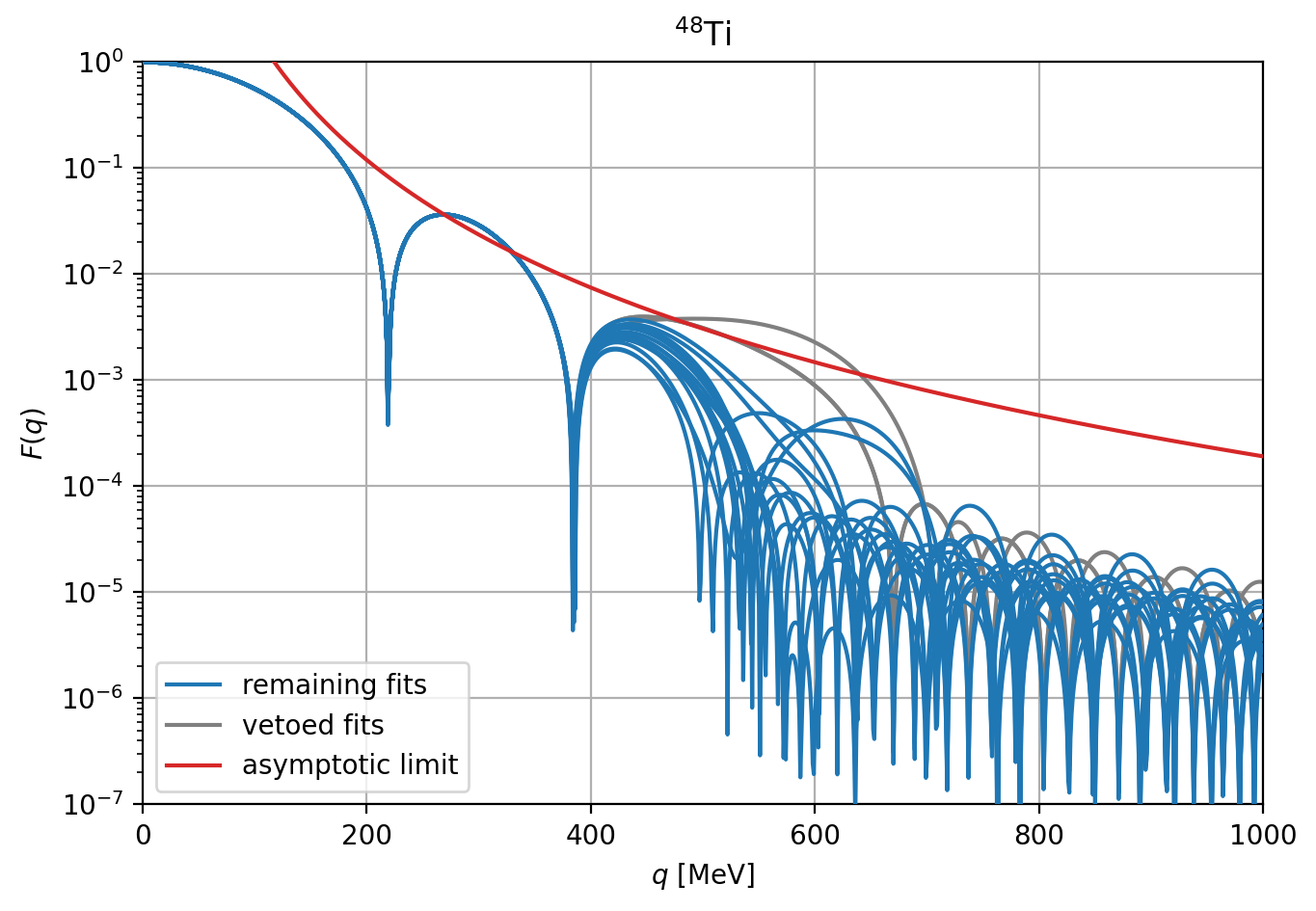}
    \hfil
    \includegraphics[width=0.49\textwidth]{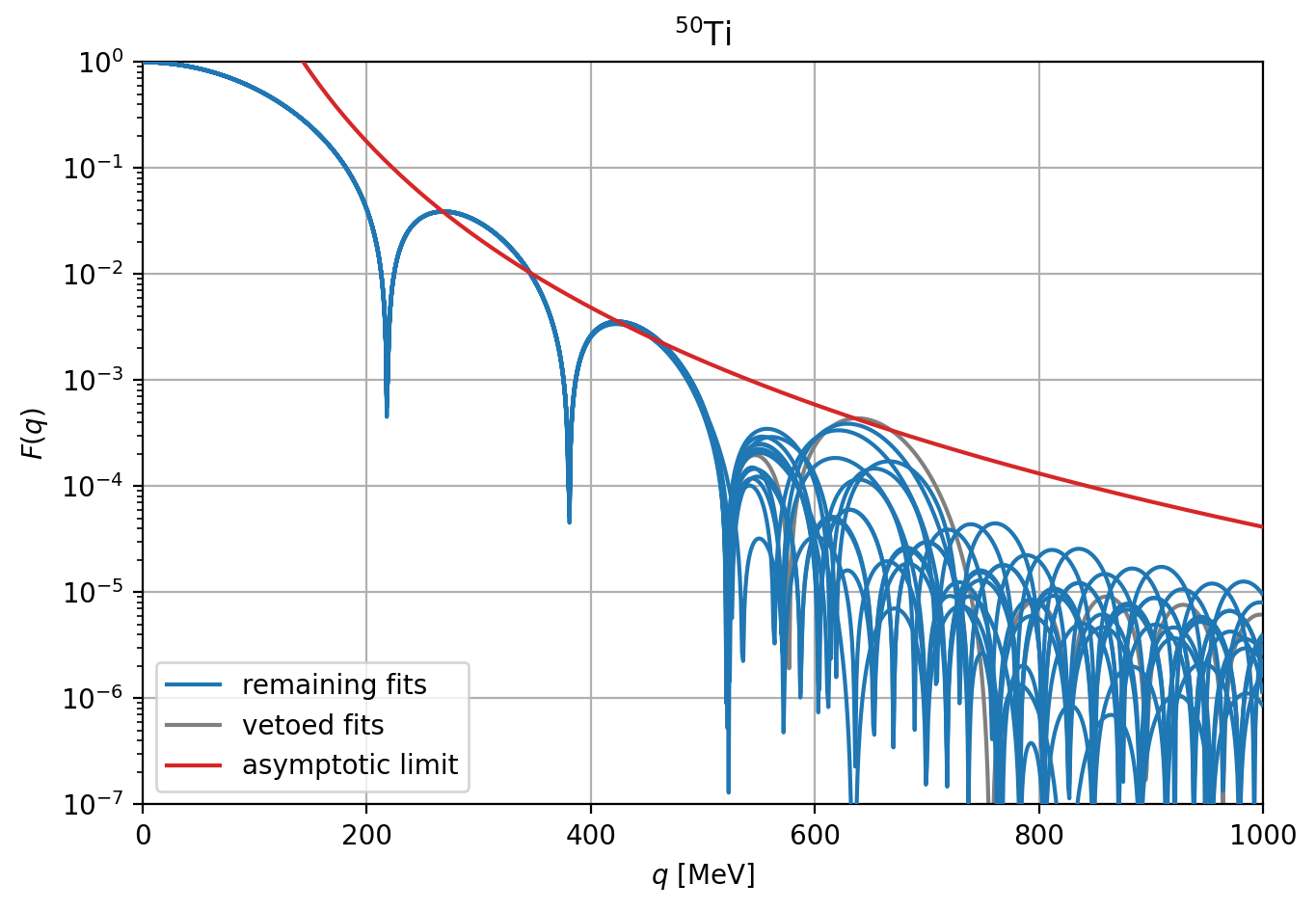}
    \hfil
    
    \hfil
    \resizebox{!}{0.115\textheight}{%
    \renewcommand{\arraystretch}{1.3}
    \begin{tabular}{ccccc}
        \toprule
        \multicolumn{5}{c}{$^{48}$Ti}\\ \midrule
        $R$\textbackslash{{}}$N_x$ & 6 &  7 &  8 &  9 \\ \midrule
        8.75 & 1.601 & 1.657 & 1.727 &  \\
        9.00 & 1.556 & 1.621 & 1.693 & {\color{gray} 1.742 } \\
        9.25 & 1.532 & 1.596 & {\color{gray} 1.640 } & {\color{gray} 1.718 } \\
        9.50 & 1.530 & 1.581 & 1.652 & 1.722 \\
        9.75 & 1.542 & 1.576 & 1.633 &  \\
    \bottomrule
    \end{tabular}
    \renewcommand{\arraystretch}{1.0}}%
    \hfil
    \resizebox{!}{0.115\textheight}{%
    \renewcommand{\arraystretch}{1.3}
    \begin{tabular}{cccccc}
        \toprule
        \multicolumn{6}{c}{$^{50}$Ti}\\ \midrule
        $R$\textbackslash{{}}$N_x$ & 6 &  7 &  8 &  9 \\ \midrule
        8.50 &  &  & 1.601 &  \\
        8.75 &  &  & 1.575 &  \\
        9.00 &  &  & 1.562 & {\color{gray} 1.591 } \\
        9.25 &  & 1.526 & 1.556 & 1.584 \\
        9.50 & 1.500 & 1.527 & 1.556 & 1.582 \\
        9.75 &  & 1.538 & 1.560 & 1.586 \\
        \bottomrule
    \end{tabular}
    \renewcommand{\arraystretch}{1.0}}%
    \hfil
    
    \caption{Same as Fig.~\ref{tab:Ti4850PEN}, but including the Barrett moment as an additional constraint. The tables show $\chi^2/\text{dof}$ ($\text{dof}=28+1+1-N_x$ for ${}^{48}$Ti, $\text{dof}=57+1+1-N_x$ for ${}^{50}$Ti). Solutions in gray are excluded for $p_\text{val}<0.25\%$ for ${}^{48}$Ti and $p_\text{val}<0.1\%$ for ${}^{50}$Ti or when violating the asymptotic limit (indicated in red).}
    \label{tab:Ti4850BAR}
\end{figure}


\begin{figure}[t]

    \hfil
    \includegraphics[width=0.49\textwidth]{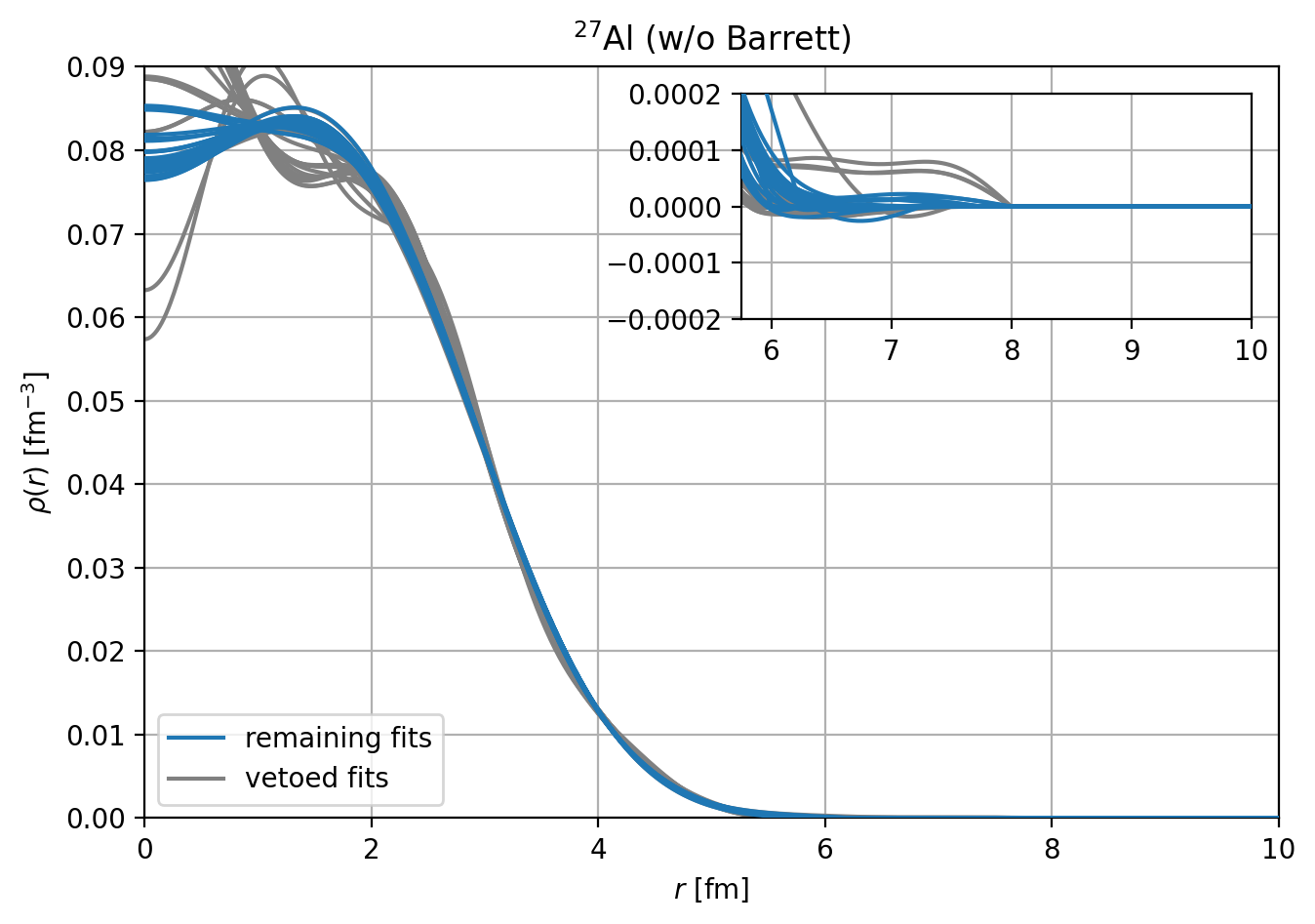}
    \hfil
    \includegraphics[width=0.49\textwidth]{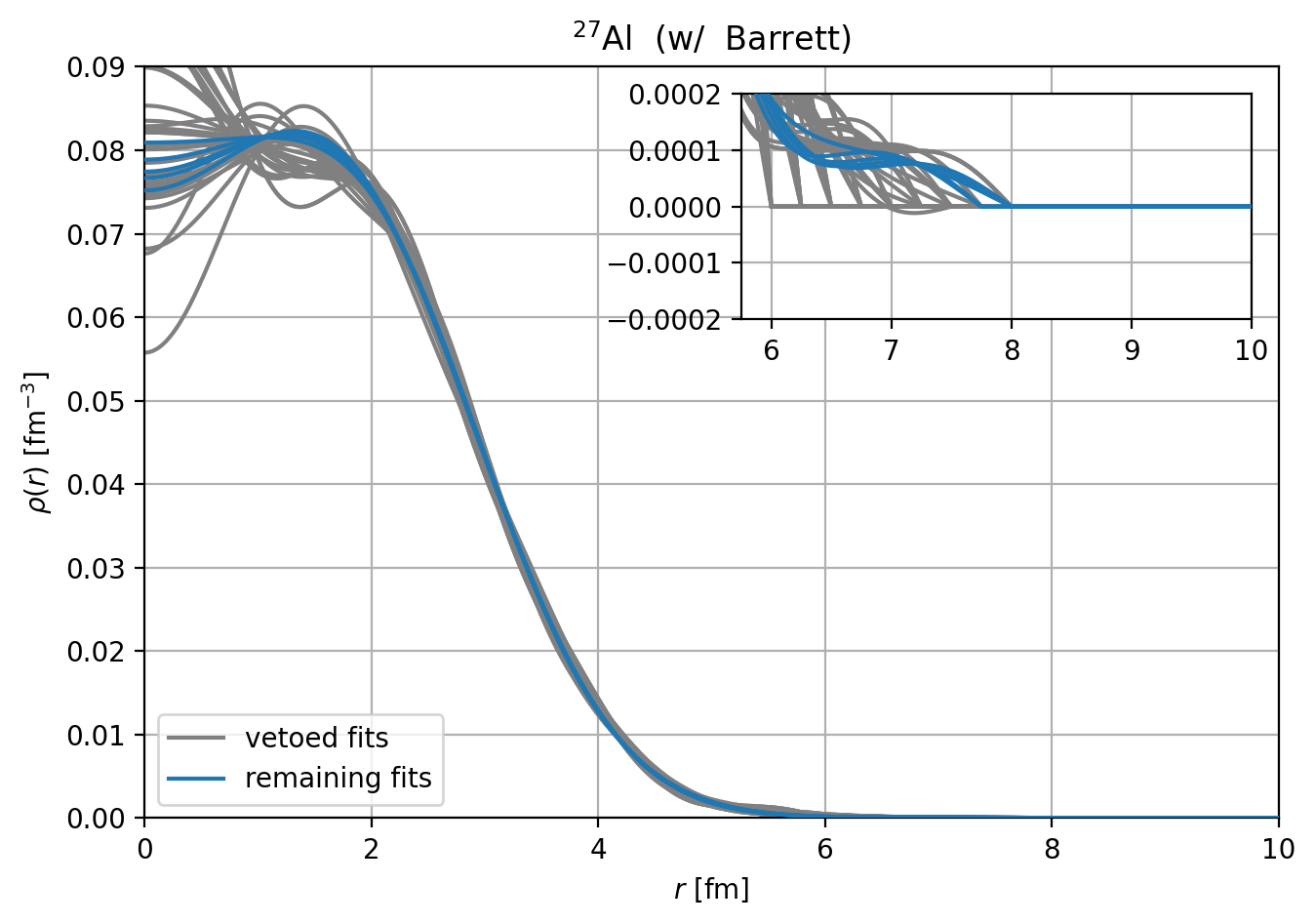} 
    \hfil
    
    \hfil
    \includegraphics[width=0.49\textwidth]{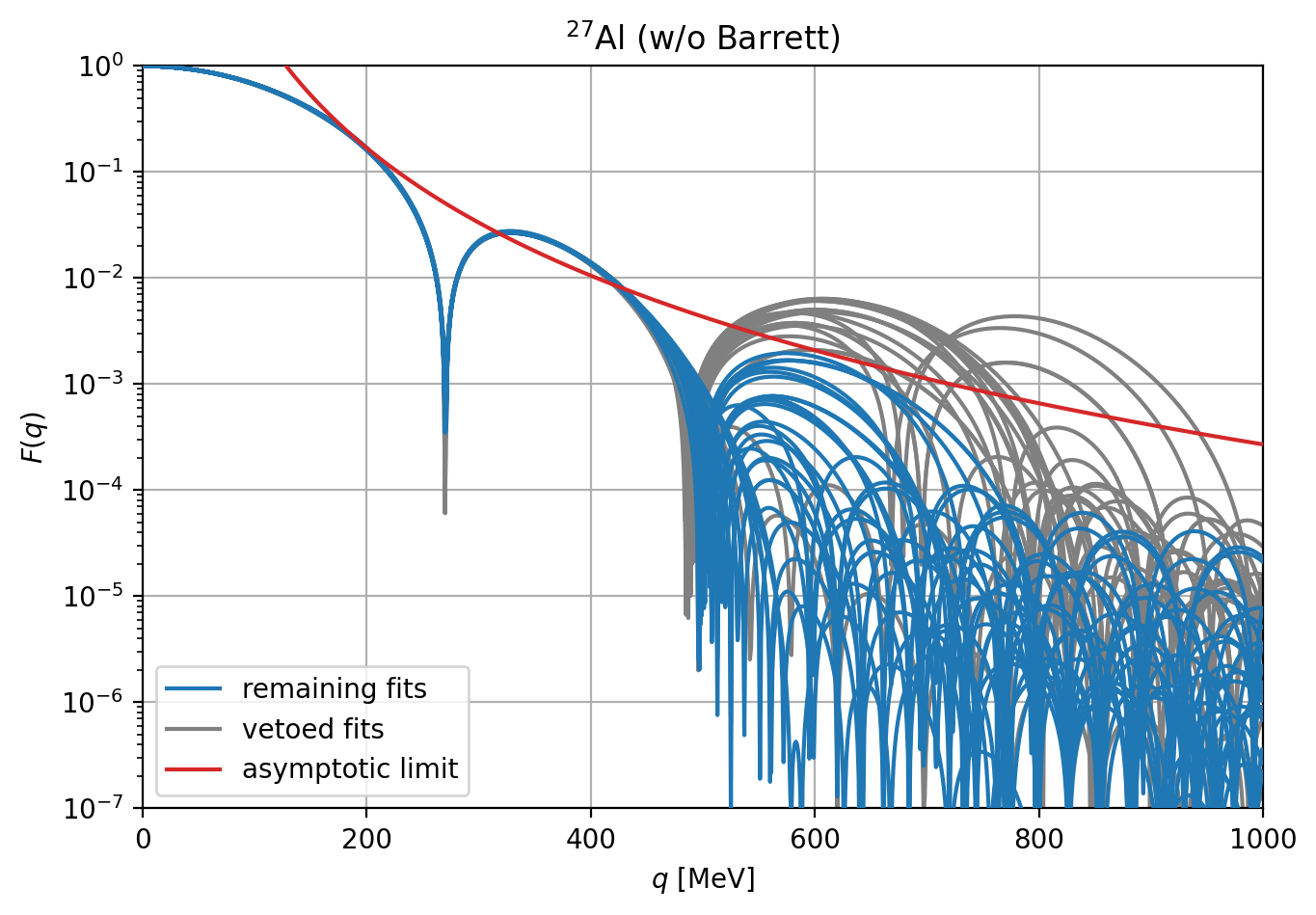}
    \hfil
    \includegraphics[width=0.49\textwidth]{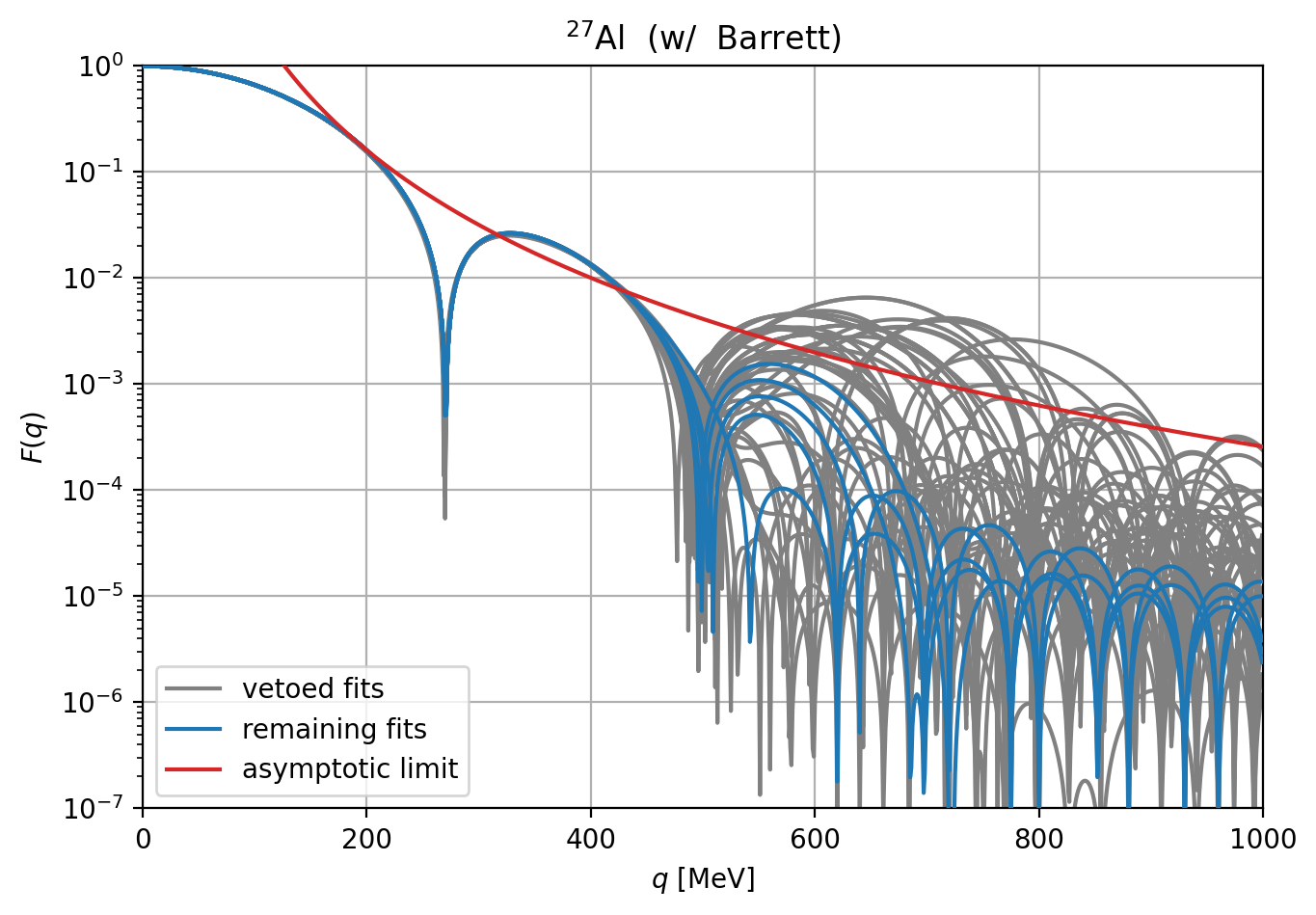} 
    \hfil
    
    \hfil
    \resizebox{!}{0.105\textheight}{%
    \renewcommand{\arraystretch}{1.3}
    \begin{tabular}{ccccccccc}
        \toprule
        \multicolumn{9}{c}{$^{27}$Al (w/o Barrett)}\\ \midrule
        $R$\textbackslash{{}}$N_x$ & 3 &  4 &  5 &  6 &  7 &  8 &  9 & 10 \\ \midrule
        6.00 & 1.951 & 1.994 & {\color{gray} 1.994 } & {\color{gray} 2.031 } & {\color{gray} 2.079 } & {\color{gray} 2.130 } &  &  \\
        6.25 & 2.103 & 2.019 & 2.013 & {\color{gray} 2.032 } & {\color{gray} 2.080 } & {\color{gray} 2.129 } &  &  \\
        6.50 &  & 2.026 & 2.032 & {\color{gray} 2.039 } & {\color{gray} 2.081 } & {\color{gray} 2.131 } &  &  \\
        6.75 &  & 2.021 & 2.048 & 2.054 & {\color{gray} 2.084 } & {\color{gray} 2.133 } & {\color{gray} 2.186 } &  \\
        7.00 &  & 2.013 & 2.059 & 2.070 & {\color{gray} 2.067 } & {\color{gray} 2.117 } & {\color{gray} 2.169 } &  \\
        7.25 &  & 2.113 & 2.069 & 2.092 & 2.110 & {\color{gray} 2.108 } & {\color{gray} 2.161 } &  \\
        7.50 &  & {\color{gray} 2.544 } & 2.080 & 2.104 & 2.131 & 2.183 & 2.238 &  \\
        7.75 &  &  & 2.175 & 2.109 & 2.149 & 2.196 & 2.251 &  \\
        8.00 &  &  & {\color{gray} 2.237 } & 2.121 & 2.154 & {\color{gray} 2.180 } & {\color{gray} 2.207 } & {\color{gray} 2.263 } \\
        \bottomrule
    \end{tabular}%
    \renewcommand{\arraystretch}{1.0}}%
    \hfil
    \resizebox{!}{0.105\textheight}{%
    \renewcommand{\arraystretch}{1.3}
    \begin{tabular}{ccccccccc}
        \toprule
        \multicolumn{9}{c}{$^{27}$Al  (w/  Barrett)}\\ \midrule
        $R$\textbackslash{{}}$N_x$ & 3 &  4 &  5 &  6 &  7 &  8 &  9 & 10 \\ \midrule
        6.00 & {\color{gray} 3.764 } & {\color{gray} 2.660 } & {\color{gray} 2.536 } & {\color{gray} 2.575 } & {\color{gray} 2.626 } & {\color{gray} 2.688 } &  &  \\
        6.25 & {\color{gray} 2.504 } & {\color{gray} 2.496 } & {\color{gray} 2.477 } & {\color{gray} 2.496 } & {\color{gray} 2.553 } & {\color{gray} 2.611 } &  &  \\
        6.50 &  & {\color{gray} 2.398 } & {\color{gray} 2.429 } & {\color{gray} 2.441 } & {\color{gray} 2.496 } & {\color{gray} 2.556 } &  &  \\
        6.75 &  & {\color{gray} 2.344 } & {\color{gray} 2.394 } & {\color{gray} 2.403 } & {\color{gray} 2.448 } & {\color{gray} 2.506 } & {\color{gray} 2.567 } &  \\
        7.00 &  & {\color{gray} 2.314 } & {\color{gray} 2.365 } & {\color{gray} 2.370 } & {\color{gray} 2.416 } & {\color{gray} 2.474 } & {\color{gray} 2.534 } &  \\
        7.25 &  & {\color{gray} 2.310 } & {\color{gray} 2.344 } & {\color{gray} 2.334 } & {\color{gray} 2.383 } & {\color{gray} 2.439 } & {\color{gray} 2.471 } &  \\
        7.50 &  & {\color{gray} 2.530 } & {\color{gray} 2.322 } & {\color{gray} 2.289 } & {\color{gray} 2.330 } & {\color{gray} 2.385 } & {\color{gray} 2.417 } &  \\
        7.75 &  &  & {\color{gray} 2.291 } & 2.230 & 2.270 & {\color{gray} 2.266 } & {\color{gray} 2.320 } &  \\
        8.00 &  &  & 2.214 & 2.158 & 2.197 & {\color{gray} 2.152 } & {\color{gray} 2.204 } & {\color{gray} 2.259 } \\
        \bottomrule
    \end{tabular}
    \renewcommand{\arraystretch}{1.0}}%
    \hfil
    
    \caption{Resulting charge densities and form factors after steps \ref{item:pen} and \ref{item:sel} for ${}^{27}$Al with and without Barrett moment. The tables show $\chi^2/\text{dof}$ ($\text{dof}=48+1-N_x$ without Barrett moment, $\text{dof}=48+1+1-N_x$ wit Barrett moment). Solutions in gray are excluded for $p_\text{val}<10^{-5}$ for fits without Barrett moment and $p_\text{val}<3\times10^{-6}$ for fits with Barrett moment or when violating the asymptotic limit (indicated in red).}
    \label{tab:Al27}
\end{figure}

Next, we show the resulting charge density, form factor, and reduced $\chi^2$ values after steps \ref{item:pen} and \ref{item:sel}, either with or without the inclusion of the corresponding Barrett moment. Here we again exclude fits based on the $p$-value, but, in addition,  remove fits based on their asymptotic behavior. The results are shown in Figs.~\ref{tab:Ca4048PEN}--\ref{tab:Al27}. As one can see, the limiting asymptotic behavior is crucial to suppress an extreme increase in the form factor for large momentum transfer, which otherwise results in extreme variations in the charge density for $r \to 0$.
The remaining solutions are the foundation for the central solutions and systematic uncertainty bands as presented in Sec.~\ref{sec:charge}.

On a more general note, one can see that the limiting factor in precision is generally the highest measured momentum transfer, as beyond this point without further external input the fit remains largely unconstrained. This then results in uncertainties for small and large distances. In particular for $^{40,48}$Ca, for which we observe significant tension between the input from muonic atoms in terms of Barrett moments and the pure electron scattering data, one sees that the fits including the Barrett moments modify the tail of the charge density to increase the value of the charge radius. In these cases, the charge density first decreases in a similar fashion as without the additional constraint, but then hovers just slightly above zero until the larger charge radius is reached. In this way, the shape of the charge density is not altered significantly, at least not in a way that would contradict the measured cross sections or any of the other constraints we imposed, while at the same time adjusting integrated quantities such as the charge radius or the Barrett moment.  

\begin{figure}[t]

    \hfil
    \includegraphics[width=0.49\textwidth]{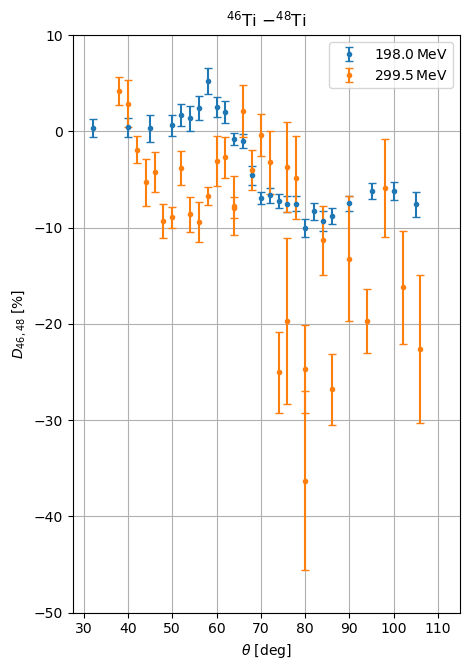}
    \hfil
    \includegraphics[width=0.49\textwidth]{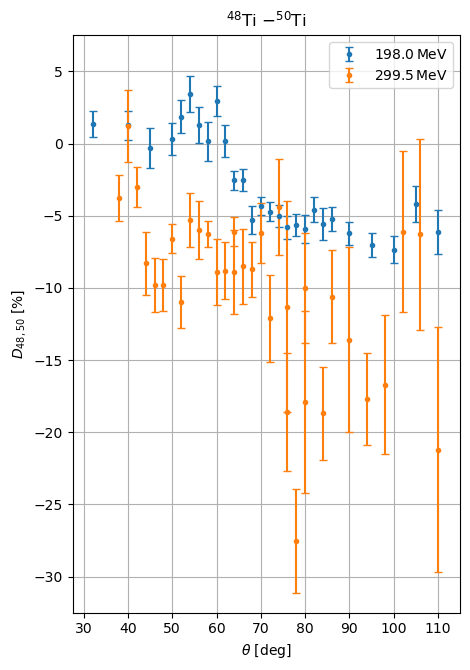}
    \hfil
    
\caption{Data of Ref.~\cite{Heisenberg:1972zza} for the differential-cross-section differences $D_{A,B}$ off target nuclei $A$ and $B$ (normalized to the sum).}
\label{fig:D4650}

\end{figure}

\subsection{Remarks on \texorpdfstring{$\boldsymbol{{}^{46,50}}$}{}Ti data}
\label{app:Ti4648}

Using the data of Ref.~\cite{Heisenberg:1972zza}, we attempted to also extract charge densities for both ${}^{46}$Ti and ${}^{50}$Ti. However, ultimately we had to conclude that at least for $^{46}$Ti inconsistencies in this data set were so severe that no meaningful charge distribution could be determined. 
Since the situation for $^{50}$Ti was still significantly better than for $^{46}$Ti, we 
included these results in this work, with the caveat that not all systematic tensions might be fully reflected by our uncertainty estimates. 
In Fig.~\ref{fig:D4650} we reproduce the original data from Ref.~\cite{Heisenberg:1972zza},
illustrating the inconsistencies alluded to above. 
In particular for $D_{46,48}$ there exist two (supposedly) identical measurements at $299.5\MeV$ and $76^\circ$, which are inconsistent at $>1.5\sigma$. Furthermore, the measurements at $299.5\MeV$ do not convey a clear structure that our fit could be expected to describe, as measurements very close in angular distance can deviate drastically. 

Moreover, the extracted charge density of $^{48}$Ti, necessary as input to calculate the cross sections of $^{46,50}$Ti, by itself is already afflicted with sizable uncertainties, which need to be propagated into the fit. 
Despite the large absolute value of these propagated uncertainties, surpassing the direct statistical ones, due to the strong correlations among cross sections at different angles, the direct statistical uncertainties appear to dominate the fit. This results in a fit with artificially small uncertainties, which are also not sufficiently enhanced by scale factors. Comparing uncertainties from $^{48}$Ti and $^{50}$Ti, it should be expected that the precision with which $^{48}$Ti can be extracted serves as a rough upper limit for the precision that can be achieved for $^{46,50}$Ti. This condition is, however, only approximately fulfilled for $^{50}$Ti, and strongly violated for $^{46}$Ti. Finally, the $p$-value for the best fits for $^{46}$Ti was at the $10^{-5}$ level, while for $^{50}$Ti we could reach $p$-values above $1\%$. This led to the conclusion that even if we could scale errors to better reflect the intrinsic uncertainties, for $^{46}$Ti we did not see a way to extract meaningful information beyond the $^{48}$Ti charge density, while for $^{50}$Ti the separate fits should still encode useful information.

\section{Fourier--Bessel parameter sets}
\label{app:parameterizations}

We tabulate the parameters for all our charge densities including statistic and systematic uncertainties as well as their correlations. We list the $x_i$ parameters including a statistical as well as an upper and lower systematic bound. The upper and lower bounds hereby refer to the side of the error band of the charge density they parameterize and do not necessarily refer to a direction in which the parameter itself may deviate. For easy accessibility we also list the $a_i$ parameters with a single uncertainty, defined by the combination of the individual components. We list these parameters in Tables~\ref{tab:Ca4048_params}--\ref{tab:Al27_params}, with the corresponding correlations provided in Tables~\ref{tab:Ca4048_corr}--\ref{tab:Al27_corr}. Note that by construction all uncertainty components of the $x_i$ have the same correlation, which would however not be the case anymore if we quoted individual uncertainty components for the $a_i$. 


\begin{table}[p]
    \centering

    \scriptsize
    
    \renewcommand{\arraystretch}{1.3}%
    \begin{tabular}{lllll}%
        \toprule
        & \multicolumn{4}{c}{$^{40}$Ca} \\ \midrule
        & \multicolumn{2}{c}{w/o Barrett} & \multicolumn{2}{c}{w/  Barrett} \\ \midrule
        $n$  & \multicolumn{1}{c}{$a_n$} & \multicolumn{1}{c}{$x_n$} & \multicolumn{1}{c}{$a_n$} & \multicolumn{1}{c}{$x_n$} \\ \midrule
        1 & \hphantom{$-$}$0.052028$(85) & $0.66089$(36)($\substack{ + 85 \\ - 50 }$)  & \hphantom{$-$}$0.0313412$(76) & $0.73243$(11)($\substack{ + 9 \\ - 11 }$)  \\
        2 & \hphantom{$-$}$0.05896$(30) & $0.46506$(57)($\substack{ + 101 \\ - 24 }$)  & \hphantom{$-$}$0.058045$(48) & $0.08136$(60)($\substack{ + 41 \\ - 66 }$)  \\
        3 & $-0.01332$(18) & $0.41163$(34)($\substack{ + 50 \\ - 15 }$)  & \hphantom{$-$}$0.022084$(92) & $0.64635$(87)($\substack{ + 70 \\ - 70 }$)  \\
        4 & $-0.02320$(15) & $0.66804$(53)($\substack{ + 80 \\ - 70 }$)  & $-0.019018$(69) & $0.80501$(76)($\substack{ + 70 \\ - 80 }$)  \\
        5 & \hphantom{$-$}$0.003893$(49) & $0.54160$(33)($\substack{ + 20 \\ - 40 }$)  & $-0.017467$(65) & $0.15359$(83)($\substack{ + 90 \\ - 60 }$)  \\
        6 & \hphantom{$-$}$0.00770$(20) & $0.40127$(65)($\substack{ + 250 \\ - 0 }$)  & \hphantom{$-$}$0.000966$(57) & $0.47675$(68)($\substack{ + 0 \\ - 120 }$)  \\
        7 & \hphantom{$-$}$0.001045$(35) & $0.51564$(53)($\substack{ + 0 \\ - 0 }$)  & \hphantom{$-$}$0.007210$(48) & $0.70235$(89)($\substack{ + 100 \\ - 100 }$)  \\
        8 & $-0.00075$(20) & $0.4977$(12)($\substack{ + 20 \\ - 10 }$)  & \hphantom{$-$}$0.003362$(68) & $0.39217$(90)($\substack{ + 0 \\ - 200 }$)  \\
        9 & \hphantom{$-$}$0.00012$(12) &  & $-0.000043$(39) & $0.49845$(98)($\substack{ + 0 \\ - 100 }$)  \\
        10 &  &   & $-0.00057$(12) & $0.4976$(13)($\substack{ + 20 \\ - 20 }$)  \\
        11 &  &   & \hphantom{$-$}$0.000055$(54) &  \\
        \bottomrule\\%
        \toprule
        & \multicolumn{4}{c}{$^{48}$Ca} \\ \midrule
        & \multicolumn{2}{c}{w/o Barrett} & \multicolumn{2}{c}{w/  Barrett} \\ \midrule
        $n$  & \multicolumn{1}{c}{$a_n$} & \multicolumn{1}{c}{$x_n$} & \multicolumn{1}{c}{$a_n$} & \multicolumn{1}{c}{$x_n$} \\ \midrule
        1 & \hphantom{$-$}$0.060470$(58) & $0.63198$(35)($\substack{ + 37 \\ - 38 }$)  & \hphantom{$-$}$0.029287$(26) & $0.74113$(12)($\substack{ + 57 \\ - 50 }$)  \\
        2 & \hphantom{$-$}$0.05005$(18) & $0.55505$(29)($\substack{ + 24 \\ - 30 }$)  & \hphantom{$-$}$0.056123$(56) & $0.03046$(35)($\substack{ + 10 \\ - 110 }$)  \\
        3 & $-0.03108$(18) & $0.41244$(47)($\substack{ + 60 \\ - 70 }$)  & \hphantom{$-$}$0.02222$(17) & $0.66040$(67)($\substack{ + 260 \\ - 250 }$)  \\
        4 & $-0.01544$(14) & $0.53997$(33)($\substack{ + 80 \\ - 20 }$)  & $-0.021143$(69) & $0.86326$(61)($\substack{ + 40 \\ - 0 }$)  \\
        5 & \hphantom{$-$}$0.012265$(89) & $0.56299$(54)($\substack{ + 40 \\ - 0 }$)  & $-0.02059$(36) & $0.06244$(47)($\substack{ + 400 \\ - 900 }$)  \\
        6 & \hphantom{$-$}$0.00493$(20) & $0.5151$(11)($\substack{ + 10 \\ - 16 }$)  & \hphantom{$-$}$0.000919$(48) & $0.47604$(74)($\substack{ + 100 \\ - 100 }$)  \\
        7 & $-0.00125$(16) &  & \hphantom{$-$}$0.009187$(36) & $0.7793$(11)($\substack{ + 0 \\ - 0 }$)  \\
        8 &  &   & \hphantom{$-$}$0.00448$(32) & $0.3571$(10)($\substack{ + 0 \\ - 0 }$)  \\
        9 &  &   & $-0.00019$(75) & $0.46486$(86)($\substack{ + 0 \\ - 0 }$)  \\
        10 &  &   & $-0.000809$(20) &  \\
        \bottomrule%
    \end{tabular}%
    \renewcommand{\arraystretch}{1.0}
    \caption{FB parameters for $^{40}$Ca and $^{48}$Ca.}
    \label{tab:Ca4048_params}
\end{table}


\begin{table}[p]
    \centering

    \scriptsize
    
    \renewcommand{\arraystretch}{1.3}%
    \begin{tabular}{lllll}%
        \toprule
        & \multicolumn{4}{c}{$^{48}$Ti} \\ \midrule
        & \multicolumn{2}{c}{w/o Barrett} & \multicolumn{2}{c}{w/  Barrett} \\ \midrule
        $n$  & \multicolumn{1}{c}{$a_n$} & \multicolumn{1}{c}{$x_n$} & \multicolumn{1}{c}{$a_n$} & \multicolumn{1}{c}{$x_n$} \\ \midrule
        1 & \hphantom{$-$}$0.03166$(28) & $0.7351$(20)($\substack{ + 0 \\ - 54 }$)  & \hphantom{$-$}$0.03392$(20) & $0.72798$(19)($\substack{ + 0 \\ - 380 }$)  \\
        2 & \hphantom{$-$}$0.0579$(10) & $0.1043$(51)($\substack{ + 0 \\ - 170 }$)  & \hphantom{$-$}$0.05913$(54) & $0.1639$(19)($\substack{ + 0 \\ - 80 }$)  \\
        3 & \hphantom{$-$}$0.01966$(48) & $0.5576$(28)($\substack{ + 0 \\ - 30 }$)  & \hphantom{$-$}$0.01547$(77) & $0.4950$(26)($\substack{ + 0 \\ - 0 }$)  \\
        4 & $-0.0232$(24) & $0.8333$(46)($\substack{ + 100 \\ - 460 }$)  & $-0.02550$(92) & $0.8333$(43)($\substack{ + 160 \\ - 90 }$)  \\
        5 & $-0.0178$(31) & $0.2312$(58)($\substack{ + 800 \\ - 0 }$)  & $-0.0152$(17) & $0.3217$(51)($\substack{ + 380 \\ - 100 }$)  \\
        6 & \hphantom{$-$}$0.0011$(51) & $0.395$(14)($\substack{ + 160 \\ - 21 }$)  & \hphantom{$-$}$0.0029$(37) & $0.406$(23)($\substack{ + 92 \\ - 68 }$)  \\
        7 & \hphantom{$-$}$0.0038$(58) &  & \hphantom{$-$}$0.0037$(37) &  \\
        \bottomrule\\%
        \toprule
        & \multicolumn{4}{c}{$^{50}$Ti} \\ \midrule
        & \multicolumn{2}{c}{w/o Barrett} & \multicolumn{2}{c}{w/  Barrett} \\ \midrule
        $n$  & \multicolumn{1}{c}{$a_n$} & \multicolumn{1}{c}{$x_n$} & \multicolumn{1}{c}{$a_n$} & \multicolumn{1}{c}{$x_n$} \\ \midrule
        1 & \hphantom{$-$}$0.03168$(15) & $0.7356$(11)($\substack{ + 29 \\ - 15 }$)  & \hphantom{$-$}$0.031816$(31) & $0.73837$(17)($\substack{ + 60 \\ - 20 }$)  \\
        2 & \hphantom{$-$}$0.05810$(58) & $0.1008$(38)($\substack{ + 94 \\ - 64 }$)  & \hphantom{$-$}$0.05854$(30) & $0.0933$(25)($\substack{ + 0 \\ - 52 }$)  \\
        3 & \hphantom{$-$}$0.01934$(46) & $0.5497$(27)($\substack{ + 27 \\ - 0 }$)  & \hphantom{$-$}$0.01935$(48) & $0.5518$(29)($\substack{ + 0 \\ - 40 }$)  \\
        4 & $-0.02511$(53) & $0.8681$(39)($\substack{ + 90 \\ - 0 }$)  & $-0.02471$(41) & $0.8612$(38)($\substack{ + 0 \\ - 63 }$)  \\
        5 & $-0.01908$(25) & $0.2008$(58)($\substack{ + 0 \\ - 0 }$)  & $-0.01882$(57) & $0.2090$(61)($\substack{ + 160 \\ - 20 }$)  \\
        6 & \hphantom{$-$}$0.00245$(64) & $0.3552$(73)($\substack{ + 140 \\ - 40 }$)  & \hphantom{$-$}$0.00243$(96) & $0.3692$(70)($\substack{ + 340 \\ - 10 }$)  \\
        7 & \hphantom{$-$}$0.00524$(57) &  & \hphantom{$-$}$0.0047$(13) &  \\
        \bottomrule%
    \end{tabular}%
    \renewcommand{\arraystretch}{1.0}
    \caption{FB parameters for $^{48}$Ti and $^{50}$Ti.}
    \label{tab:Ti4850_params}
\end{table}


\begin{table}[p]
    \centering

    \scriptsize
    
    \renewcommand{\arraystretch}{1.3}%
    \begin{tabular}{lllll}%
        \toprule
        & \multicolumn{4}{c}{$^{27}$Al} \\ \midrule
        & \multicolumn{2}{c}{w/o Barrett} & \multicolumn{2}{c}{w/  Barrett} \\ \midrule
        $n$  & \multicolumn{1}{c}{$a_n$} & \multicolumn{1}{c}{$x_n$} & \multicolumn{1}{c}{$a_n$} & \multicolumn{1}{c}{$x_n$} \\ \midrule
        1 & \hphantom{$-$}$0.06189$(74) & $0.6386$(17)($\substack{ + 54 \\ - 68 }$)  & \hphantom{$-$}$0.03125$(11) & $0.73357$(22)($\substack{ + 0 \\ - 230 }$)  \\
        2 & \hphantom{$-$}$0.0531$(25) & $0.5427$(10)($\substack{ + 50 \\ - 71 }$)  & \hphantom{$-$}$0.05727$(73) & $0.1042$(12)($\substack{ + 0 \\ - 137 }$)  \\
        3 & $-0.0212$(33) & $0.3895$(17)($\substack{ + 370 \\ - 70 }$)  & \hphantom{$-$}$0.02301$(71) & $0.6235$(14)($\substack{ + 0 \\ - 40 }$)  \\
        4 & $-0.0164$(55) &  & $-0.01565$(49) & $0.7163$(19)($\substack{ + 0 \\ - 0 }$)  \\
        5 &  &   & $-0.0163$(11) & $0.2457$(19)($\substack{ + 270 \\ - 0 }$)  \\
        6 &  &   & $-0.0039$(26) & $0.4723$(19)($\substack{ + 530 \\ - 150 }$)  \\
        7 &  &   & \hphantom{$-$}$0.0010$(19) &  \\
        \bottomrule
    \end{tabular}%
    \renewcommand{\arraystretch}{1.0}
    \caption{FB parameters for $^{27}$Al.}
    \label{tab:Al27_params}
\end{table}


\begin{table}[p]
    \centering
    \tiny
    \begin{tabular}{cccccccccc}
        \toprule 
        \multicolumn{10}{c}{$^{40}$Ca; w/o Barrett} \\ \midrule
        $a_i\backslash{}x_i$ & 1 & 2 & 3 & 4 & 5 & 6 & 7 & 8 \\ \midrule
        1 & $1.000$ & $-0.933$ & $-0.560$ & $0.777$ & $0.202$ & $-0.460$ & $0.023$ & $0.295$ & --  \\
        2 & $0.984$ & $1.000$ & $0.683$ & $-0.765$ & $-0.284$ & $0.338$ & $-0.087$ & $-0.153$ & --  \\
        3 & $-0.839$ & $-0.751$ & $1.000$ & $-0.755$ & $-0.434$ & $0.005$ & $-0.247$ & $0.194$ & --  \\
        4 & $-0.746$ & $-0.776$ & $0.699$ & $1.000$ & $0.138$ & $-0.466$ & $-0.015$ & $0.255$ & --  \\
        5 & $0.202$ & $0.245$ & $-0.153$ & $-0.252$ & $1.000$ & $-0.138$ & $0.020$ & $-0.017$ & --  \\
        6 & $0.460$ & $0.409$ & $-0.426$ & $-0.329$ & $0.138$ & $1.000$ & $-0.041$ & $-0.369$ & --  \\
        7 & $0.023$ & $0.055$ & $-0.013$ & $-0.077$ & $0.020$ & $0.041$ & $1.000$ & $-0.211$ & --  \\
        8 & $-0.407$ & $-0.380$ & $0.369$ & $0.327$ & $-0.232$ & $-0.915$ & $-0.272$ & $1.000$ & --  \\
        9 & $-0.295$ & $-0.231$ & $0.293$ & $0.110$ & $0.017$ & $-0.369$ & $0.211$ & $0.264$ & $1.000$  \\
        \bottomrule%
    \end{tabular}%
    \vspace{\baselineskip} 
    
    \begin{tabular}{cccccccccccc}
        \toprule 
        \multicolumn{12}{c}{$^{40}$Ca; w/  Barrett} \\ \midrule
        $a_i\backslash{}x_i$ & 1 & 2 & 3 & 4 & 5 & 6 & 7 & 8 & 9 & 10 \\ \midrule
        1 & $1.000$ & $-0.376$ & $-0.217$ & $0.054$ & $-0.180$ & $0.063$ & $0.083$ & $0.051$ & $0.110$ & $0.340$ & --  \\
        2 & $0.420$ & $1.000$ & $-0.622$ & $-0.294$ & $0.496$ & $-0.180$ & $-0.320$ & $-0.178$ & $-0.485$ & $-0.243$ & --  \\
        3 & $-0.233$ & $0.703$ & $1.000$ & $-0.278$ & $-0.371$ & $-0.074$ & $0.011$ & $0.050$ & $0.297$ & $0.092$ & --  \\
        4 & $-0.054$ & $-0.291$ & $0.079$ & $1.000$ & $-0.409$ & $0.171$ & $0.302$ & $0.215$ & $0.237$ & $-0.078$ & --  \\
        5 & $-0.189$ & $-0.431$ & $-0.352$ & $0.397$ & $1.000$ & $-0.173$ & $-0.191$ & $-0.352$ & $-0.403$ & $-0.027$ & --  \\
        6 & $-0.063$ & $-0.180$ & $-0.009$ & $0.171$ & $0.107$ & $1.000$ & $-0.010$ & $-0.053$ & $-0.018$ & $-0.139$ & --  \\
        7 & $0.083$ & $0.318$ & $0.117$ & $-0.302$ & $-0.134$ & $0.010$ & $1.000$ & $-0.022$ & $0.337$ & $0.372$ & --  \\
        8 & $-0.051$ & $-0.177$ & $-0.094$ & $0.215$ & $0.330$ & $-0.053$ & $0.022$ & $1.000$ & $0.046$ & $-0.046$ & --  \\
        9 & $0.110$ & $0.480$ & $0.368$ & $-0.237$ & $-0.360$ & $0.018$ & $0.337$ & $-0.046$ & $1.000$ & $0.280$ & --  \\
        10 & $0.036$ & $-0.499$ & $-0.714$ & $-0.250$ & $0.182$ & $-0.368$ & $-0.093$ & $-0.237$ & $-0.322$ & $1.000$ & --  \\
        11 & $-0.340$ & $-0.255$ & $-0.078$ & $-0.078$ & $0.023$ & $-0.139$ & $-0.372$ & $-0.046$ & $-0.280$ & $0.253$ & $1.000$  \\
        \bottomrule%
    \end{tabular}%
    \vspace{\baselineskip} 
    
    \begin{tabular}{cccccccc}
        \toprule 
        \multicolumn{8}{c}{$^{48}$Ca; w/o Barrett} \\ \midrule
        $a_i\backslash{}x_i$ & 1 & 2 & 3 & 4 & 5 & 6 \\ \midrule
        1 & $1.000$ & $0.776$ & $-0.775$ & $0.625$ & $0.443$ & $0.071$ & --  \\
        2 & $0.975$ & $1.000$ & $-0.714$ & $0.608$ & $0.426$ & $0.129$ & --  \\
        3 & $-0.828$ & $-0.731$ & $1.000$ & $-0.534$ & $-0.344$ & $0.021$ & --  \\
        4 & $-0.672$ & $-0.662$ & $0.759$ & $1.000$ & $0.492$ & $0.216$ & --  \\
        5 & $0.099$ & $0.159$ & $-0.226$ & $-0.098$ & $1.000$ & $0.295$ & --  \\
        6 & $0.374$ & $0.296$ & $-0.242$ & $-0.340$ & $-0.282$ & $1.000$ & --  \\
        7 & $-0.071$ & $-0.043$ & $0.028$ & $0.082$ & $0.048$ & $0.202$ & $1.000$  \\
        \bottomrule%
    \end{tabular}%
    \vspace{\baselineskip}

    \begin{tabular}{ccccccccccc}
        \toprule 
        \multicolumn{11}{c}{$^{48}$Ca; w/  Barrett} \\ \midrule
        $a_i\backslash{}x_i$ & 1 & 2 & 3 & 4 & 5 & 6 & 7 & 8 & 9 \\ \midrule
        1 & $1.000$ & $-0.453$ & $-0.157$ & $-0.352$ & $-0.088$ & $0.005$ & $-0.350$ & $0.141$ & $0.059$ & --  \\
        2 & $0.499$ & $1.000$ & $-0.466$ & $0.247$ & $0.172$ & $-0.309$ & $0.486$ & $-0.344$ & $-0.151$ & --  \\
        3 & $-0.465$ & $0.345$ & $1.000$ & $-0.513$ & $-0.350$ & $0.128$ & $-0.414$ & $0.287$ & $0.224$ & --  \\
        4 & $0.547$ & $0.338$ & $0.226$ & $1.000$ & $0.031$ & $-0.273$ & $0.479$ & $-0.368$ & $-0.167$ & --  \\
        5 & $-0.123$ & $-0.188$ & $-0.262$ & $-0.214$ & $1.000$ & $0.158$ & $0.042$ & $0.020$ & $-0.066$ & --  \\
        6 & $-0.005$ & $-0.301$ & $-0.168$ & $-0.133$ & $-0.161$ & $1.000$ & $-0.310$ & $0.369$ & $0.078$ & --  \\
        7 & $-0.350$ & $-0.493$ & $-0.266$ & $-0.519$ & $0.064$ & $0.310$ & $1.000$ & $-0.499$ & $-0.083$ & --  \\
        8 & $0.433$ & $0.089$ & $-0.219$ & $0.429$ & $0.701$ & $-0.107$ & $-0.238$ & $1.000$ & $0.046$ & --  \\
        9 & $-0.443$ & $-0.131$ & $0.187$ & $-0.461$ & $-0.692$ & $0.153$ & $0.299$ & $-0.991$ & $1.000$ & --  \\
        10 & $0.059$ & $0.150$ & $0.176$ & $0.188$ & $-0.074$ & $-0.078$ & $-0.083$ & $0.067$ & $-0.051$ & $1.000$  \\
        \bottomrule%
    \end{tabular}
    \caption{Correlations for $^{40}$Ca and $^{48}$Ca for $x_i$ (upper triangle) and $a_i$ (lower triangle).}
    \label{tab:Ca4048_corr}
\end{table}


\begin{table}[p]
    \centering
    \tiny
    
    \begin{tabular}{cccccccc}
        \toprule 
        \multicolumn{8}{c}{$^{48}$Ti; w/o Barrett} \\ \midrule
        $a_i\backslash{}x_i$ & 1 & 2 & 3 & 4 & 5 & 6 \\ \midrule
        1 & $1.000$ & $-0.932$ & $0.096$ & $-0.429$ & $0.151$ & $-0.075$ & --  \\
        2 & $0.947$ & $1.000$ & $-0.235$ & $0.423$ & $-0.285$ & $-0.038$ & --  \\
        3 & $-0.272$ & $0.025$ & $1.000$ & $-0.660$ & $0.196$ & $-0.127$ & --  \\
        4 & $0.444$ & $0.427$ & $0.162$ & $1.000$ & $-0.389$ & $-0.004$ & --  \\
        5 & $0.211$ & $0.333$ & $0.371$ & $0.551$ & $1.000$ & $0.303$ & --  \\
        6 & $-0.060$ & $-0.112$ & $-0.206$ & $-0.397$ & $-0.004$ & $1.000$ & --  \\
        7 & $0.075$ & $-0.025$ & $-0.174$ & $0.017$ & $-0.288$ & $0.695$ & $1.000$  \\
        \bottomrule%
    \end{tabular}%
    \vspace{\baselineskip} 
    
    \begin{tabular}{cccccccc}
        \toprule 
        \multicolumn{8}{c}{$^{48}$Ti; w/  Barrett} \\ \midrule
        $a_i\backslash{}x_i$ & 1 & 2 & 3 & 4 & 5 & 6 \\ \midrule
        1 & $1.000$ & $-0.424$ & $0.021$ & $-0.037$ & $0.089$ & $0.098$ & --  \\
        2 & $0.614$ & $1.000$ & $-0.271$ & $0.108$ & $-0.324$ & $-0.382$ & --  \\
        3 & $-0.696$ & $0.119$ & $1.000$ & $-0.708$ & $0.159$ & $-0.251$ & --  \\
        4 & $0.205$ & $0.093$ & $-0.035$ & $1.000$ & $-0.266$ & $-0.029$ & --  \\
        5 & $-0.143$ & $0.287$ & $0.460$ & $0.336$ & $1.000$ & $0.169$ & --  \\
        6 & $0.289$ & $-0.138$ & $-0.449$ & $-0.061$ & $0.053$ & $1.000$ & --  \\
        7 & $-0.098$ & $-0.357$ & $-0.120$ & $0.026$ & $-0.219$ & $0.752$ & $1.000$  \\
        \bottomrule%
    \end{tabular}
    \vspace{\baselineskip} 
    
    \begin{tabular}{cccccccc}
        \toprule 
        \multicolumn{8}{c}{$^{50}$Ti; w/o Barrett} \\ \midrule
        $a_i\backslash{}x_i$ & 1 & 2 & 3 & 4 & 5 & 6 \\ \midrule
        1 & $1.000$ & $-0.846$ & $0.158$ & $-0.262$ & $0.040$ & $0.402$ & --  \\
        2 & $0.875$ & $1.000$ & $-0.540$ & $0.060$ & $0.117$ & $-0.281$ & --  \\
        3 & $-0.166$ & $0.314$ & $1.000$ & $-0.242$ & $-0.165$ & $-0.042$ & --  \\
        4 & $0.310$ & $0.095$ & $-0.214$ & $1.000$ & $-0.522$ & $-0.032$ & --  \\
        5 & $-0.121$ & $-0.109$ & $0.076$ & $0.573$ & $1.000$ & $0.027$ & --  \\
        6 & $0.091$ & $-0.155$ & $-0.572$ & $-0.307$ & $-0.362$ & $1.000$ & --  \\
        7 & $-0.402$ & $-0.298$ & $0.182$ & $-0.066$ & $0.120$ & $0.329$ & $1.000$  \\
        \bottomrule%
    \end{tabular}%
    \vspace{\baselineskip} 
    
    \begin{tabular}{cccccccc}
        \toprule 
        \multicolumn{8}{c}{$^{50}$Ti; w/  Barrett} \\ \midrule
        $a_i\backslash{}x_i$ & 1 & 2 & 3 & 4 & 5 & 6 \\ \midrule
        1 & $1.000$ & $-0.460$ & $0.207$ & $0.076$ & $-0.077$ & $-0.082$ & --  \\
        2 & $0.489$ & $1.000$ & $-0.746$ & $-0.321$ & $0.313$ & $0.389$ & --  \\
        3 & $0.159$ & $0.887$ & $1.000$ & $-0.195$ & $-0.253$ & $-0.215$ & --  \\
        4 & $-0.056$ & $-0.355$ & $-0.115$ & $1.000$ & $-0.470$ & $-0.229$ & --  \\
        5 & $-0.106$ & $-0.256$ & $-0.238$ & $0.477$ & $1.000$ & $0.236$ & --  \\
        6 & $0.130$ & $0.019$ & $-0.079$ & $-0.085$ & $0.337$ & $1.000$ & --  \\
        7 & $0.082$ & $0.385$ & $0.332$ & $-0.253$ & $-0.200$ & $0.714$ & $1.000$  \\
        \bottomrule%
    \end{tabular}
    \caption{Correlations for $^{48}$Ti and $^{50}$Ti for $x_i$ (upper triangle) and $a_i$ (lower triangle).}
    \label{tab:Ti4850_corr}
\end{table}


\begin{table}[p]
    \centering
    \tiny
    \begin{tabular}{ccccc}
        \toprule 
        \multicolumn{5}{c}{$^{27}$Al; w/o Barrett} \\ \midrule
        $a_i\backslash{}x_i$ & 1 & 2 & 3 \\ \midrule
        1 & $1.000$ & $0.655$ & $-0.202$ & --  \\
        2 & $0.961$ & $1.000$ & $0.235$ & --  \\
        3 & $-0.595$ & $-0.570$ & $1.000$ & --  \\
        4 & $-0.202$ & $-0.328$ & $0.800$ & $1.000$  \\
        \bottomrule%
    \end{tabular}%
    \vspace{\baselineskip} 
    
    \begin{tabular}{cccccccc}
        \toprule 
        \multicolumn{8}{c}{$^{27}$Al; w/  Barrett} \\ \midrule
        $a_i\backslash{}x_i$ & 1 & 2 & 3 & 4 & 5 & 6 \\ \midrule
        1 & $1.000$ & $-0.335$ & $-0.052$ & $0.017$ & $-0.010$ & $-0.001$ & --  \\
        2 & $0.391$ & $1.000$ & $-0.460$ & $-0.323$ & $-0.205$ & $-0.031$ & --  \\
        3 & $-0.209$ & $0.787$ & $1.000$ & $-0.033$ & $0.397$ & $-0.127$ & --  \\
        4 & $0.221$ & $-0.766$ & $-0.861$ & $1.000$ & $-0.020$ & $0.586$ & --  \\
        5 & $-0.106$ & $0.463$ & $0.595$ & $-0.471$ & $1.000$ & $-0.088$ & --  \\
        6 & $0.176$ & $-0.449$ & $-0.467$ & $0.704$ & $0.172$ & $1.000$ & --  \\
        7 & $0.001$ & $-0.030$ & $0.011$ & $0.184$ & $0.072$ & $0.586$ & $1.000$  \\
        \bottomrule%
    \end{tabular}
    \caption{Correlations for $^{27}$Al for $x_i$ (upper triangle) and $a_i$ (lower triangle).}
    \label{tab:Al27_corr}
\end{table}

\section{Data tables}
\label{app:data}

Due to the complications of retrieving the original cross sections, we tabulate 
in this appendix
the data we used in our analysis.
Table~\ref{tab:Ca_data} shows the data for calcium, Table~\ref{tab:Ti_data} the data for titanium, and Table~\ref{tab:Al_data} the data for aluminum. 

\begin{table}[tp]%
\hfil
    \renewcommand{\arraystretch}{1.3}
    \resizebox{!}{0.474\textheight}{%
    \begin{tabular}{cccccrrl}
    \toprule
    \multicolumn{8}{c}{$^{40}$Ca} \\
    \midrule
     $E$ [MeV] & $\theta$ [deg] &&& $\diff\sigma/\diff\Omega$ & (stat) & (sys) & [$\text{fm}^2$/sr] \\
    \midrule $100$&$40$&&&$8.698$&$(74)$&$(107)$&$\times ~ 10^{-1}$ \\
    &$45$&&&$4.673$&$(40)$&$(58)$&$\times ~ 10^{-1}$ \\
    &$50$&&&$2.542$&$(17)$&$(31)$&$\times ~ 10^{-1}$ \\
    &$55$&&&$1.462$&$(13)$&$(18)$&$\times ~ 10^{-1}$ \\
    &$60$&&&$8.317$&$(58)$&$(103)$&$\times ~ 10^{-2}$ \\
    &$65$&&&$4.873$&$(34)$&$(60)$&$\times ~ 10^{-2}$ \\
    &$70$&&&$2.840$&$(20)$&$(35)$&$\times ~ 10^{-2}$ \\
    &$75$&&&$1.679$&$(11)$&$(21)$&$\times ~ 10^{-2}$ \\
    &$80$&&&$9.856$&$(67)$&$(122)$&$\times ~ 10^{-3}$ \\
    &$85$&&&$5.846$&$(40)$&$(72)$&$\times ~ 10^{-3}$ \\
    &$90$&&&$3.433$&$(29)$&$(42)$&$\times ~ 10^{-3}$ \\
    &$95$&&&$2.007$&$(14)$&$(25)$&$\times ~ 10^{-3}$ \\
    &$100$&&&$1.173$&$(10)$&$(14)$&$\times ~ 10^{-3}$ \\
    &$105$&&&$6.776$&$(58)$&$(84)$&$\times ~ 10^{-4}$ \\
    &$110$&&&$3.897$&$(33)$&$(48)$&$\times ~ 10^{-4}$ \\
    &$115$&&&$2.223$&$(19)$&$(27)$&$\times ~ 10^{-4}$ \\
    \midrule $210$&$45$&&&$7.703$&$(73)$&$(182)$&$\times ~ 10^{-3}$ \\
    &$50$&&&$1.976$&$(18)$&$(47)$&$\times ~ 10^{-3}$ \\
    &$52.5$&&&$9.434$&$(85)$&$(223)$&$\times ~ 10^{-4}$ \\
    &$55$&&&$4.274$&$(39)$&$(101)$&$\times ~ 10^{-4}$ \\
    &$57.5$&&&$1.847$&$(17)$&$(44)$&$\times ~ 10^{-4}$ \\
    &$60$&&&$8.293$&$(77)$&$(196)$&$\times ~ 10^{-5}$ \\
    &$62.5$&&&$4.571$&$(39)$&$(108)$&$\times ~ 10^{-5}$ \\
    &$65$&&&$3.644$&$(31)$&$(86)$&$\times ~ 10^{-5}$ \\
    &$67.5$&&&$3.599$&$(31)$&$(85)$&$\times ~ 10^{-5}$ \\
    &$70$&&&$3.731$&$(32)$&$(88)$&$\times ~ 10^{-5}$ \\
    &$72.5$&&&$3.675$&$(32)$&$(87)$&$\times ~ 10^{-5}$ \\
    &$75$&&&$3.468$&$(29)$&$(82)$&$\times ~ 10^{-5}$ \\
    &$80$&&&$2.658$&$(22)$&$(63)$&$\times ~ 10^{-5}$ \\
    &$85$&&&$1.748$&$(15)$&$(41)$&$\times ~ 10^{-5}$ \\
    &$90$&&&$9.997$&$(90)$&$(236)$&$\times ~ 10^{-6}$ \\
    &$95$&&&$5.250$&$(52)$&$(124)$&$\times ~ 10^{-6}$ \\
    &$100$&&&$2.565$&$(28)$&$(61)$&$\times ~ 10^{-6}$ \\
    \midrule $320$&$45$&&&$9.678$&$(93)$&$(267)$&$\times ~ 10^{-5}$ \\
    &$50$&&&$7.860$&$(76)$&$(217)$&$\times ~ 10^{-5}$ \\
    &$55$&&&$3.872$&$(38)$&$(107)$&$\times ~ 10^{-5}$ \\
    &$60$&&&$1.339$&$(13)$&$(37)$&$\times ~ 10^{-5}$ \\
    &$65$&&&$3.056$&$(32)$&$(84)$&$\times ~ 10^{-6}$ \\
    &$67.5$&&&$1.206$&$(14)$&$(33)$&$\times ~ 10^{-6}$ \\
    &$70$&&&$3.983$&$(58)$&$(110)$&$\times ~ 10^{-7}$ \\
    &$72$&&&$1.449$&$(27)$&$(40)$&$\times ~ 10^{-7}$ \\
    &$74$&&&$6.653$&$(136)$&$(184)$&$\times ~ 10^{-8}$ \\
    &$76$&&&$6.423$&$(133)$&$(177)$&$\times ~ 10^{-8}$ \\
    &$78$&&&$8.751$&$(157)$&$(242)$&$\times ~ 10^{-8}$ \\
    &$80$&&&$1.133$&$(14)$&$(31)$&$\times ~ 10^{-7}$ \\
    &$85$&&&$1.218$&$(22)$&$(34)$&$\times ~ 10^{-7}$ \\
    &$90$&&&$9.064$&$(172)$&$(250)$&$\times ~ 10^{-8}$ \\
    &$95$&&&$4.596$&$(123)$&$(127)$&$\times ~ 10^{-8}$ \\
    \midrule $400^*$&$63.528$&&&$2.240$&$(79)$&$(55)$&$\times ~ 10^{-7}$ \\
    &$66.196$&&&$2.110$&$(75)$&$(52)$&$\times ~ 10^{-7}$ \\
    &$68.882$&&&$1.740$&$(62)$&$(43)$&$\times ~ 10^{-7}$ \\
    &$71.588$&&&$1.200$&$(48)$&$(30)$&$\times ~ 10^{-7}$ \\
    &$74.316$&&&$6.640$&$(281)$&$(163)$&$\times ~ 10^{-8}$ \\
    &$77.067$&&&$3.940$&$(139)$&$(97)$&$\times ~ 10^{-8}$ \\
    &$79.845$&&&$1.750$&$(62)$&$(43)$&$\times ~ 10^{-8}$ \\
    &$82.65$&&&$7.110$&$(360)$&$(175)$&$\times ~ 10^{-9}$ \\
    &$85.485$&&&$2.480$&$(151)$&$(61)$&$\times ~ 10^{-9}$ \\
    &$88.355$&&&$7.210$&$(488)$&$(177)$&$\times ~ 10^{-10}$ \\
    &$91.189$&&&$1.770$&$(191)$&$(44)$&$\times ~ 10^{-10}$ \\
    &$95.671$&&&$6.680$&$(660)$&$(164)$&$\times ~ 10^{-11}$ \\
    &$98.72$&&&$7.330$&$(977)$&$(180)$&$\times ~ 10^{-11}$ \\
    &$103.398$&&&$8.280$&$(842)$&$(204)$&$\times ~ 10^{-11}$ \\
    &$109.758$&&&$3.570$&$(434)$&$(88)$&$\times ~ 10^{-11}$ \\
    &$116.507$&&&$8.360$&$(21  54)$&$(206)$&$\times ~ 10^{-12}$ \\
    &$123.719$&&&$5.110$&$(2991)$&$(126)$&$\times ~ 10^{-13}$ \\
    \bottomrule
    \renewcommand{\arraystretch}{1.0}
\end{tabular}}%
\hfil
\resizebox{!}{0.474\textheight}{%
    \renewcommand{\arraystretch}{1.3}
    \begin{tabular}{cccccrrl}
    \toprule
    \multicolumn{8}{c}{$^{48}$Ca} \\
    \midrule
     $E$ [MeV] & $\theta$ [deg] &&& $\diff\sigma/\diff\Omega$ & (stat) & (sys) & [$\text{fm}^2$/sr] \\
    \midrule $100$&$40$&&&$8.694$&$(75)$&$(97)$&$\times ~ 10^{-1}$ \\
    &$45$&&&$4.636$&$(40)$&$(52)$&$\times ~ 10^{-1}$ \\
    &$50$&&&$2.515$&$(17)$&$(28)$&$\times ~ 10^{-1}$ \\
    &$55$&&&$1.441$&$(12)$&$(16)$&$\times ~ 10^{-1}$ \\
    &$60$&&&$8.220$&$(57)$&$(92)$&$\times ~ 10^{-2}$ \\
    &$65$&&&$4.828$&$(33)$&$(54)$&$\times ~ 10^{-2}$ \\
    &$70$&&&$2.783$&$(19)$&$(31)$&$\times ~ 10^{-2}$ \\
    &$75$&&&$1.645$&$(11)$&$(18)$&$\times ~ 10^{-2}$ \\
    &$80$&&&$9.552$&$(65)$&$(106)$&$\times ~ 10^{-3}$ \\
    &$85$&&&$5.635$&$(37)$&$(63)$&$\times ~ 10^{-3}$ \\
    &$90$&&&$3.266$&$(26)$&$(36)$&$\times ~ 10^{-3}$ \\
    &$95$&&&$1.888$&$(13)$&$(21)$&$\times ~ 10^{-3}$ \\
    &$100$&&&$1.081$&$(9)$&$(12)$&$\times ~ 10^{-3}$ \\
    &$105$&&&$6.170$&$(52)$&$(69)$&$\times ~ 10^{-4}$ \\
    &$110$&&&$3.480$&$(29)$&$(39)$&$\times ~ 10^{-4}$ \\
    &$115$&&&$1.945$&$(17)$&$(22)$&$\times ~ 10^{-4}$ \\
    \midrule $210$&$45$&&&$7.138$&$(67)$&$(156)$&$\times ~ 10^{-3}$ \\
    &$50$&&&$1.707$&$(16)$&$(37)$&$\times ~ 10^{-3}$ \\
    &$52.5$&&&$7.744$&$(72)$&$(170)$&$\times ~ 10^{-4}$ \\
    &$55$&&&$3.312$&$(31)$&$(73)$&$\times ~ 10^{-4}$ \\
    &$57.5$&&&$1.401$&$(13)$&$(31)$&$\times ~ 10^{-4}$ \\
    &$60$&&&$6.975$&$(65)$&$(153)$&$\times ~ 10^{-5}$ \\
    &$62.5$&&&$5.034$&$(42)$&$(110)$&$\times ~ 10^{-5}$ \\
    &$65$&&&$5.085$&$(41)$&$(111)$&$\times ~ 10^{-5}$ \\
    &$67.5$&&&$5.437$&$(43)$&$(119)$&$\times ~ 10^{-5}$ \\
    &$70$&&&$5.599$&$(45)$&$(123)$&$\times ~ 10^{-5}$ \\
    &$72.5$&&&$5.352$&$(43)$&$(117)$&$\times ~ 10^{-5}$ \\
    &$75$&&&$4.904$&$(39)$&$(107)$&$\times ~ 10^{-5}$ \\
    &$80$&&&$3.565$&$(28)$&$(78)$&$\times ~ 10^{-5}$ \\
    &$85$&&&$2.219$&$(17)$&$(49)$&$\times ~ 10^{-5}$ \\
    &$90$&&&$1.210$&$(10)$&$(26)$&$\times ~ 10^{-5}$ \\
    &$95$&&&$6.124$&$(59)$&$(134)$&$\times ~ 10^{-6}$ \\
    &$100$&&&$2.732$&$(27)$&$(60)$&$\times ~ 10^{-6}$ \\
    \midrule $320$&$45$&&&$1.450$&$(14)$&$(39)$&$\times ~ 10^{-4}$ \\
    &$50$&&&$1.076$&$(10)$&$(29)$&$\times ~ 10^{-4}$ \\
    &$55$&&&$4.837$&$(46)$&$(130)$&$\times ~ 10^{-5}$ \\
    &$60$&&&$1.512$&$(15)$&$(41)$&$\times ~ 10^{-5}$ \\
    &$65$&&&$2.866$&$(30)$&$(77)$&$\times ~ 10^{-6}$ \\
    &$67.5$&&&$9.757$&$(108)$&$(262)$&$\times ~ 10^{-7}$ \\
    &$70$&&&$2.618$&$(39)$&$(70)$&$\times ~ 10^{-7}$ \\
    &$72$&&&$1.137$&$(20)$&$(31)$&$\times ~ 10^{-7}$ \\
    &$74$&&&$1.182$&$(26)$&$(32)$&$\times ~ 10^{-7}$ \\
    &$76$&&&$1.732$&$(34)$&$(47)$&$\times ~ 10^{-7}$ \\
    &$78$&&&$2.232$&$(39)$&$(60)$&$\times ~ 10^{-7}$ \\
    &$80$&&&$2.513$&$(48)$&$(68)$&$\times ~ 10^{-7}$ \\
    &$85$&&&$2.139$&$(37)$&$(58)$&$\times ~ 10^{-7}$ \\
    &$90$&&&$1.385$&$(26)$&$(37)$&$\times ~ 10^{-7}$ \\
    &$95$&&&$6.482$&$(169)$&$(174)$&$\times ~ 10^{-8}$ \\
    \midrule $502$&$38$&&&$3.321$&$(114)$&$(56)$&$\times ~ 10^{-5}$ \\
    &$40$&&&$1.183$&$(41)$&$(20)$&$\times ~ 10^{-5}$ \\
    &$42$&&&$2.759$&$(111)$&$(46)$&$\times ~ 10^{-6}$ \\
    &$44$&&&$4.614$&$(188)$&$(78)$&$\times ~ 10^{-7}$ \\
    &$46$&&&$3.797$&$(173)$&$(64)$&$\times ~ 10^{-7}$ \\
    &$48$&&&$6.558$&$(289)$&$(110)$&$\times ~ 10^{-7}$ \\
    &$50$&&&$7.631$&$(298)$&$(128)$&$\times ~ 10^{-7}$ \\
    &$52$&&&$6.586$&$(253)$&$(111)$&$\times ~ 10^{-7}$ \\
    &$54$&&&$4.972$&$(186)$&$(84)$&$\times ~ 10^{-7}$ \\
    &$56$&&&$3.040$&$(111)$&$(51)$&$\times ~ 10^{-7}$ \\
    &$60$&&&$7.259$&$(408)$&$(122)$&$\times ~ 10^{-8}$ \\
    &$64$&&&$1.102$&$(81)$&$(19)$&$\times ~ 10^{-8}$ \\
    &$68$&&&$9.921$&$(2049)$&$(167)$&$\times ~ 10^{-10}$ \\
    &$73$&&&$6.231$&$(1342)$&$(105)$&$\times ~ 10^{-10}$ \\
    &$75$&&&$5.457$&$(966)$&$(92)$&$\times ~ 10^{-10}$ \\
    &$78$&&&$3.607$&$(681)$&$(61)$&$\times ~ 10^{-10}$ \\
    &$86$&&&$1.217$&$(865)$&$(20)$&$\times ~ 10^{-11}$ \\
     \bottomrule
     \renewcommand{\arraystretch}{1.0}
\end{tabular}}%
\hfil
\caption{Electron scattering off $^{40}$Ca and $^{48}$Ca cross-section data from Ref.~\cite{Emrich:1983}. The $400^*\MeV$ data for $^{40}$Ca are adjusted (initially measured at $502\MeV$) from Ref.~\cite{Sick:1979bkt}, as used in Ref.~\cite{Emrich:1983}.}
\label{tab:Ca_data}
\end{table}%

\begin{table}[tp]%
\hfil
\resizebox{!}{0.438\textheight}{
    \renewcommand{\arraystretch}{1.3}
    \begin{tabular}{cccccrl}
    \toprule
    \multicolumn{7}{c}{$^{48}\text{Ti}-{}^{50}\text{Ti}$} \\
    \midrule
    $E$ [MeV] & $\theta$ [deg] &&& $D_{48,50}$ & (stat) & \\
    \midrule $198$&$32$&&&$1.36$&$(92)$&$\%$ \\
    &$40$&&&$1.26$&$(100)$&$\%$ \\
    &$45$&&&$$-0.32$$&$(140)$&$\%$ \\
    &$50$&&&$0.32$&$(110)$&$\%$ \\
    &$52$&&&$1.86$&$(115)$&$\%$ \\
    &$54$&&&$3.45$&$(125)$&$\%$ \\
    &$56$&&&$1.27$&$(125)$&$\%$ \\
    &$58$&&&$0.17$&$(135)$&$\%$ \\
    &$60$&&&$2.96$&$(105)$&$\%$ \\
    &$62$&&&$0.16$&$(110)$&$\%$ \\
    &$64$&&&$-2.55$&$(65)$&$\%$ \\
    &$66$&&&$-2.51$&$(75)$&$\%$ \\
    &$68$&&&$-5.30$&$(100)$&$\%$ \\
    &$70$&&&$-4.33$&$(65)$&$\%$ \\
    &$72$&&&$-4.72$&$(65)$&$\%$ \\
    &$74$&&&$-5.00$&$(77)$&$\%$ \\
    &$76$&&&$-5.81$&$(79)$&$\%$ \\
    &$78$&&&$-5.63$&$(76)$&$\%$ \\
    &$80$&&&$-5.91$&$(98)$&$\%$ \\
    &$82$&&&$-4.60$&$(86)$&$\%$ \\
    &$84$&&&$-5.56$&$(111)$&$\%$ \\
    &$86$&&&$-5.24$&$(81)$&$\%$ \\
    &$90$&&&$-6.22$&$(80)$&$\%$ \\
    &$95$&&&$-7.02$&$(82)$&$\%$ \\
    &$100$&&&$-7.35$&$(95)$&$\%$ \\
    &$105$&&&$-4.21$&$(124)$&$\%$ \\
    &$110$&&&$-6.14$&$(150)$&$\%$ \\
    \midrule $299.5$&$38$&&&$-3.8$&$(16)$&$\%$ \\
    &$40$&&&$1.2$&$(25)$&$\%$ \\
    &$42$&&&$-3.0$&$(14)$&$\%$ \\
    &$44$&&&$-8.3$&$(22)$&$\%$ \\
    &$46$&&&$-9.8$&$(19)$&$\%$ \\
    &$48$&&&$-9.8$&$(18)$&$\%$ \\
    &$50$&&&$-6.6$&$(10)$&$\%$ \\
    &$52$&&&$-11.0$&$(18)$&$\%$ \\
    &$54$&&&$-5.3$&$(19)$&$\%$ \\
    &$56$&&&$-6.0$&$(20)$&$\%$ \\
    &$58$&&&$-6.3$&$(9)$&$\%$ \\
    &$60$&&&$-8.9$&$(23)$&$\%$ \\
    &$62$&&&$-8.8$&$(20)$&$\%$ \\
    &$64$&&&$-7.5$&$(29)$&$\%$ \\
    &$66$&&&$-8.5$&$(26)$&$\%$ \\
    &$68$&&&$-8.7$&$(19)$&$\%$ \\
    &$70$&&&$-6.2$&$(21)$&$\%$ \\
    &$72$&&&$-12.1$&$(30)$&$\%$ \\
    &$74$&&&$-4.4$&$(33)$&$\%$ \\
    &$76$&&&$-15.0$&$(73)$&$\%$ \\
    &$78$&&&$-27.5$&$(36)$&$\%$ \\
    &$80$&&&$-14.0$&$(63)$&$\%$ \\
    &$84$&&&$-18.7$&$(32)$&$\%$ \\
    &$86$&&&$-10.6$&$(32)$&$\%$ \\
    &$90$&&&$-13.6$&$(64)$&$\%$ \\
    &$94$&&&$-17.7$&$(32)$&$\%$ \\
    &$98$&&&$-16.7$&$(48)$&$\%$ \\
    &$102$&&&$-6.1$&$(56)$&$\%$ \\
    &$106$&&&$-6.3$&$(66)$&$\%$ \\
    &$110$&&&$-21.2$&$(85)$&$\%$ \\
     \bottomrule
     \renewcommand{\arraystretch}{1.0}
\end{tabular}}%
\hfil
\resizebox{!}{0.438\textheight}{
    \renewcommand{\arraystretch}{1.3}
    \begin{tabular}{cccccrl}
    \toprule
    \multicolumn{7}{c}{$^{46}\text{Ti}-{}^{48}\text{Ti}$} \\
    \midrule
    $E$ [MeV] & $\theta$ [deg] &&& $D_{46,48}$ & (stat) & \\
    \midrule $198$&$32$&&&$0.33$&$(92)$&$\%$ \\
    &$40$&&&$0.44$&$(100)$&$\%$ \\
    &$45$&&&$0.32$&$(140)$&$\%$ \\
    &$50$&&&$0.63$&$(110)$&$\%$ \\
    &$52$&&&$1.75$&$(115)$&$\%$ \\
    &$54$&&&$1.35$&$(125)$&$\%$ \\
    &$56$&&&$2.46$&$(125)$&$\%$ \\
    &$58$&&&$5.20$&$(135)$&$\%$ \\
    &$60$&&&$2.51$&$(105)$&$\%$ \\
    &$62$&&&$2.02$&$(110)$&$\%$ \\
    &$64$&&&$-0.76$&$(65)$&$\%$ \\
    &$66$&&&$-0.99$&$(75)$&$\%$ \\
    &$68$&&&$-4.57$&$(100)$&$\%$ \\
    &$70$&&&$-6.90$&$(65)$&$\%$ \\
    &$72$&&&$-6.63$&$(75)$&$\%$ \\
    &$74$&&&$-7.23$&$(71)$&$\%$ \\
    &$76$&&&$-7.56$&$(82)$&$\%$ \\
    &$78$&&&$-7.50$&$(75)$&$\%$ \\
    &$80$&&&$-10.00$&$(94)$&$\%$ \\
    &$82$&&&$-8.30$&$(86)$&$\%$ \\
    &$84$&&&$-9.30$&$(104)$&$\%$ \\
    &$86$&&&$-8.80$&$(86)$&$\%$ \\
    &$90$&&&$-7.45$&$(78)$&$\%$ \\
    &$95$&&&$-6.20$&$(84)$&$\%$ \\
    &$100$&&&$-6.20$&$(97)$&$\%$ \\
    &$105$&&&$-7.58$&$(129)$&$\%$ \\ \\
     \midrule $299.5$&$38$&&&$4.2$&$(15)$&$\%$ \\
    &$40$&&&$2.9$&$(24)$&$\%$ \\
    &$42$&&&$-1.9$&$(14)$&$\%$ \\
    &$44$&&&$-5.3$&$(24)$&$\%$ \\
    &$46$&&&$-4.2$&$(21)$&$\%$ \\
    &$48$&&&$-9.3$&$(18)$&$\%$ \\
    &$50$&&&$-8.9$&$(11)$&$\%$ \\
    &$52$&&&$-3.8$&$(18)$&$\%$ \\
    &$54$&&&$-8.6$&$(18)$&$\%$ \\
    &$56$&&&$-9.4$&$(21)$&$\%$ \\
    &$58$&&&$-6.7$&$(9)$&$\%$ \\
    &$60$&&&$-3.1$&$(26)$&$\%$ \\
    &$62$&&&$-2.7$&$(21)$&$\%$ \\
    &$64$&&&$-7.8$&$(31)$&$\%$ \\
    &$66$&&&$2.1$&$(27)$&$\%$ \\
    &$68$&&&$-4.0$&$(21)$&$\%$ \\
    &$70$&&&$-0.4$&$(22)$&$\%$ \\
    &$72$&&&$-3.2$&$(33)$&$\%$ \\
    &$74$&&&$-25.0$&$(42)$&$\%$ \\
    &$76$&&&$-11.7$&$(86)$&$\%$ \\
    &$78$&&&$-4.8$&$(43)$&$\%$ \\
    &$80$&&&$-30.5$&$(93)$&$\%$ \\
    &$84$&&&$-11.3$&$(36)$&$\%$ \\
    &$86$&&&$-26.8$&$(37)$&$\%$ \\
    &$90$&&&$-13.2$&$(65)$&$\%$ \\
    &$94$&&&$-19.7$&$(33)$&$\%$ \\
    &$98$&&&$-5.9$&$(51)$&$\%$ \\
    &$102$&&&$-16.2$&$(59)$&$\%$ \\
    &$106$&&&$-22.6$&$(77)$&$\%$ \\ \\
    \bottomrule
     \renewcommand{\arraystretch}{1.0}
\end{tabular}}%
\hfil
\resizebox{!}{0.438\textheight}{%
    \renewcommand{\arraystretch}{1.3}
    \begin{tabular}{cccccrl}
    \toprule
    \multicolumn{7}{c}{$^{48}\text{Ca} - {}^{48}\text{Ti}$} \\
    \midrule
     $E$ [MeV] & $\theta$ [deg] &&& $D_{\text{Ca},\text{Ti}}$ & (stat) & \\
    \midrule $149.5$&$32$&&&$-7.9$&$(12)$&$\%$ \\
     $174.5$&$32$&&&$-5.7$&$(16)$&$\%$ \\
     $199.5$&$32$&&&$-5.2$&$(18)$&$\%$ \\
    \midrule $249.5$&$32$&&&$0.9$&$(12)$&$\%$ \\
    &$35$&&&$1.2$&$(15)$&$\%$ \\
    &$37.5$&&&$3.8$&$(10)$&$\%$ \\
    &$40$&&&$4.3$&$(7)$&$\%$ \\
    &$42$&&&$5.0$&$(10)$&$\%$ \\
    &$44$&&&$5.4$&$(8)$&$\%$ \\
    &$46$&&&$6.1$&$(10)$&$\%$ \\
    &$48$&&&$3.4$&$(10)$&$\%$ \\
    &$50$&&&$-7.2$&$(7)$&$\%$ \\
    &$52$&&&$-16.1$&$(15)$&$\%$ \\
    &$54$&&&$-19.3$&$(15)$&$\%$ \\
    &$56$&&&$-19.9$&$(15)$&$\%$ \\
    &$58$&&&$-16.1$&$(14)$&$\%$ \\
    &$60$&&&$-10.5$&$(12)$&$\%$ \\
    &$62$&&&$-7.4$&$(12)$&$\%$ \\
    &$64$&&&$-4.8$&$(12)$&$\%$ \\
    &$66$&&&$-1.1$&$(15)$&$\%$ \\
    &$68$&&&$1.8$&$(14)$&$\%$ \\
    &$70$&&&$4.0$&$(14)$&$\%$ \\
    &$72$&&&$5.5$&$(14)$&$\%$ \\
    &$74$&&&$11.4$&$(20)$&$\%$ \\
    &$76$&&&$8.6$&$(14)$&$\%$ \\
    &$80$&&&$17.5$&$(14)$&$\%$ \\
    &$85$&&&$20.9$&$(24)$&$\%$ \\
    &$90$&&&$34.9$&$(30)$&$\%$ \\
    \bottomrule
    \\ \\ \\ \\ \\ \\ \\ \\ \\ \\ \\ \\ \\ \\ \\ \\ \\ \\ \\ \\ \\ \\ \\ \\ \\ \\ \\ \\ \\ 
    \renewcommand{\arraystretch}{1.0}
\end{tabular}}%
\hfil
\caption{Electron scattering cross-section differences $D_{A,B}$ between  $^{48}$Ca, $^{48}$Ti, $^{50}$Ti, and $^{46}$Ti from Refs.~\cite{Heisenberg:1972zza,Frosch:1968zz}.} 
\label{tab:Ti_data}
\end{table}%

\begin{table}[tp]%
\hfil
    \resizebox{!}{0.454\textheight}{
    \renewcommand{\arraystretch}{1.3}
    \begin{tabular}{cccccrrl}
    \toprule
    \multicolumn{8}{c}{$^{27}$Al} \\
    \midrule
     $E$ [MeV] & $\theta$ [deg] &&& $\diff\sigma/\diff\Omega$ & (stat) & (sys) & [$\text{fm}^2$/sr] \\
    \midrule $250$&$34$&&&$2.78$&$(8)$&$(8)$&$\times ~ 10^{-2}$ \\
    &$36$&&&$1.79$&$(5)$&$(5)$&$\times ~ 10^{-2}$ \\
    &$38$&&&$1.16$&$(3)$&$(3)$&$\times ~ 10^{-2}$ \\
    &$40$&&&$7.05$&$(21)$&$(21)$&$\times ~ 10^{-3}$ \\
    &$42$&&&$4.37$&$(13)$&$(13)$&$\times ~ 10^{-3}$ \\
    &$44$&&&$2.65$&$(8)$&$(8)$&$\times ~ 10^{-3}$ \\
    &$46$&&&$1.55$&$(5)$&$(5)$&$\times ~ 10^{-3}$ \\
    &$48$&&&$8.76$&$(26)$&$(26)$&$\times ~ 10^{-4}$ \\
    &$50$&&&$5.15$&$(15)$&$(15)$&$\times ~ 10^{-4}$ \\
    &$52$&&&$2.86$&$(9)$&$(9)$&$\times ~ 10^{-4}$ \\
    &$54$&&&$1.53$&$(5)$&$(5)$&$\times ~ 10^{-4}$ \\
    &$56$&&&$7.70$&$(23)$&$(23)$&$\times ~ 10^{-5}$ \\
    &$58$&&&$4.12$&$(12)$&$(12)$&$\times ~ 10^{-5}$ \\
    &$60$&&&$2.36$&$(7)$&$(7)$&$\times ~ 10^{-5}$ \\
    &$62$&&&$1.39$&$(4)$&$(4)$&$\times ~ 10^{-5}$ \\
    &$64$&&&$1.01$&$(3)$&$(3)$&$\times ~ 10^{-5}$ \\
    &$66$&&&$8.50$&$(26)$&$(26)$&$\times ~ 10^{-6}$ \\
    &$68$&&&$8.62$&$(26)$&$(26)$&$\times ~ 10^{-6}$ \\
    &$70$&&&$8.35$&$(25)$&$(25)$&$\times ~ 10^{-6}$ \\
    &$72$&&&$8.20$&$(25)$&$(25)$&$\times ~ 10^{-6}$ \\
    &$74$&&&$7.95$&$(24)$&$(24)$&$\times ~ 10^{-6}$ \\
    &$76$&&&$7.10$&$(21)$&$(21)$&$\times ~ 10^{-6}$ \\
    &$78$&&&$6.13$&$(18)$&$(18)$&$\times ~ 10^{-6}$ \\
    &$80$&&&$5.42$&$(16)$&$(16)$&$\times ~ 10^{-6}$ \\
    &$82$&&&$4.64$&$(14)$&$(14)$&$\times ~ 10^{-6}$ \\
    &$84$&&&$3.74$&$(12)$&$(11)$&$\times ~ 10^{-6}$ \\
    &$86$&&&$3.14$&$(9)$&$(9)$&$\times ~ 10^{-6}$ \\
    &$88$&&&$2.60$&$(8)$&$(8)$&$\times ~ 10^{-6}$ \\
    &$90$&&&$2.19$&$(7)$&$(7)$&$\times ~ 10^{-6}$ \\
    &$92$&&&$1.62$&$(5)$&$(5)$&$\times ~ 10^{-6}$ \\
    &$94$&&&$1.44$&$(4)$&$(4)$&$\times ~ 10^{-6}$ \\
    &$98$&&&$7.76$&$(23)$&$(23)$&$\times ~ 10^{-7}$ \\
    &$102$&&&$5.26$&$(16)$&$(16)$&$\times ~ 10^{-7}$ \\
    &$106$&&&$3.26$&$(11)$&$(10)$&$\times ~ 10^{-7}$ \\
    &$110$&&&$2.04$&$(67)$&$(6)$&$\times ~ 10^{-7}$ \\
    &$114$&&&$1.24$&$(38)$&$(4)$&$\times ~ 10^{-7}$ \\
    &$118$&&&$8.03$&$(32)$&$(24)$&$\times ~ 10^{-8}$ \\
    &$122$&&&$6.55$&$(34)$&$(20)$&$\times ~ 10^{-8}$ \\
    &$126$&&&$4.00$&$(19)$&$(12)$&$\times ~ 10^{-8}$ \\
    \midrule $500$&$34$&&&$3.64$&$(12)$&$(11)$&$\times ~ 10^{-5}$ \\
    &$35$&&&$3.75$&$(11)$&$(11)$&$\times ~ 10^{-5}$ \\
    &$36$&&&$3.54$&$(11)$&$(11)$&$\times ~ 10^{-5}$ \\
    &$38$&&&$2.70$&$(8)$&$(8)$&$\times ~ 10^{-5}$ \\
    &$40$&&&$1.90$&$(6)$&$(6)$&$\times ~ 10^{-5}$ \\
    &$42$&&&$1.25$&$(4)$&$(4)$&$\times ~ 10^{-5}$ \\
    &$44$&&&$7.52$&$(20)$&$(23)$&$\times ~ 10^{-6}$ \\
    &$46$&&&$4.45$&$(13)$&$(13)$&$\times ~ 10^{-6}$ \\
    &$48$&&&$2.04$&$(6)$&$(6)$&$\times ~ 10^{-6}$ \\
    &$50$&&&$1.08$&$(8)$&$(3)$&$\times ~ 10^{-6}$ \\
    &$52$&&&$4.99$&$(49)$&$(15)$&$\times ~ 10^{-7}$ \\
    &$54$&&&$2.55$&$(26)$&$(8)$&$\times ~ 10^{-7}$ \\
    &$56$&&&$1.56$&$(16)$&$(5)$&$\times ~ 10^{-7}$ \\
    &$58$&&&$8.87$&$(89)$&$(27)$&$\times ~ 10^{-8}$ \\
    &$60$&&&$5.10$&$(51)$&$(15)$&$\times ~ 10^{-8}$ \\
    &$62$&&&$3.26$&$(33)$&$(10)$&$\times ~ 10^{-8}$ \\
    &$64$&&&$1.73$&$(17)$&$(5)$&$\times ~ 10^{-8}$ \\
    &$66$&&&$1.60$&$(16)$&$(5)$&$\times ~ 10^{-8}$ \\
    &$68$&&&$7.78$&$(78)$&$(23)$&$\times ~ 10^{-9}$ \\
    &$70$&&&$2.85$&$(41)$&$(9)$&$\times ~ 10^{-9}$ \\
    \bottomrule
    \renewcommand{\arraystretch}{1.0}
\end{tabular}}%
\hfil
\resizebox{!}{0.454\textheight}{%
    \renewcommand{\arraystretch}{1.3}
    \begin{tabular}{cccccrcl}
    \toprule
    \multicolumn{8}{c}{$^{27}$Al} \\
    \midrule
    $E$ [MeV] & $\theta$ [deg] &&& $\diff\sigma/\diff\Omega$ & (stat) & (sys) & [$\text{fm}^2$/sr] \\
    \midrule $170$ & $135$ &&& $8.05$ & $(80)$ & $(24)$ & $\cdot ~ 10^{-7}$ \\
    $188.5$ & $135$ &&& $4.50$ & $(45)$ & $(13)$ & $\cdot ~ 10^{-7}$ \\
    $195$ & $135$ &&& $3.32$ & $(20)$ & $(10)$ & $\cdot ~ 10^{-7}$ \\
    $221$ & $135$ &&& $9.15$ & $(92)$ & $(27)$ & $\cdot ~ 10^{-8}$ \\
    $254$ & $135$ &&& $2.00$ & $(20)$ & $( 6)$ & $\cdot ~ 10^{-8}$ \\
    $270$ & $135$ &&& $1.14$ & $(11)$ & $( 3)$ & $\cdot ~ 10^{-8}$ \\
    $285.5$ & $135$ &&& $1.28$ & $(13)$ & $( 4)$ & $\cdot ~ 10^{-8}$ \\
    \midrule \\ \\
    \toprule
    \multicolumn{8}{c}{$^{27}$Al} \\
    \midrule
    $E$ [MeV] & $\theta$ [deg] &&& $\diff\sigma/\diff\Omega$ & (stat) & (sys) & [$\text{fm}^2$/sr] \\
    \midrule $47.34$&$70$&&&$20.71$&$(50)$&--&$\times ~ 10^{-2}$ \\
    $61.45$&$70$&&&$9.87$&$(22)$&--&$\times ~ 10^{-2}$ \\
    $72.34$&$70$&&&$6.06$&$(12)$&--&$\times ~ 10^{-2}$ \\
    \midrule $53.86$&$90$&&&$3.97$&$(8)$&--&$\times ~ 10^{-2}$ \\
    $75.31$&$90$&&&$1.14$&$(3)$&--&$\times ~ 10^{-2}$ \\
    $79.08$&$90$&&&$1.020$&$(28)$&--&$\times ~ 10^{-2}$ \\
    \midrule $28.7$&$110$&&&$7.32$&$(17)$&--&$\times ~ 10^{-2}$ \\
    $37.35$&$110$&&&$3.60$&$(8)$&--&$\times ~ 10^{-2}$ \\
    $55.24$&$110$&&&$1.050$&$(28)$&--&$\times ~ 10^{-2}$ \\
    \bottomrule 
    \\ \\ \\ \\ \\ \\ \\ \\ \\ \\ \\ \\ \\ \\ \\ \\ \\ \\ \\ \\ \\ \\ \\ \\ \\ \\ \\ \\ \\ \\ \\ \\ \\ \\ \\ \\ \\ \\ \\ \\
    \renewcommand{\arraystretch}{1.0}
\end{tabular}}
\hfil
\caption{Electron scattering off $^{27}$Al from Ref.~\cite{Li:1974vj} including measurements at large angles, without fixed energy and Ref.~\cite{Dolbilkin:1983} for low momentum transfer; the systematic error from Ref.~\cite{Li:1974vj} corresponds to $3\%$ of the cross section.}
\label{tab:Al_data}
\end{table}

\bibliographystyle{apsrev4-1_mod_2}
\bibliography{ref}
	
\end{document}